\documentclass[twocolumn]{autart}    

\usepackage{graphicx}          
\usepackage{amsmath}
\usepackage{amsfonts}
\usepackage{mathrsfs}
\usepackage{amssymb}
\usepackage{multirow}
\usepackage{latexsym}
\usepackage{psfrag}
\usepackage{graphicx}
\usepackage[round,sort&compress,square,comma,authoryear]{natbib}
\usepackage[colorlinks=true,urlcolor=blue,citecolor=blue,linkcolor=red,bookmarks=true]{hyperref}
\setcitestyle{round}
\setcitestyle{citesep={;}}
\usepackage{color}
\usepackage{times,balance}
\usepackage{algorithm,multirow,xcolor}
\usepackage{algorithmicx}
\usepackage{algpseudocode}
\usepackage{longtable}
\usepackage{booktabs}
\usepackage{colortbl}
\usepackage{dsfont}
\usepackage{etoolbox}
\definecolor{mygray}{gray}{.8}
\newtheorem{theorem}{Theorem}
\newtheorem{definition}{Definition}

\newtheorem{lemma}{Lemma}
\newtheorem{remark}{Remark}

\allowdisplaybreaks[4]

\begin{document}

\begin{frontmatter}

\title{Aperiodic-sampled neural network controllers with closed-loop stability verifications (extended version)} 

\small{
\thanks{This paper was supported in part by the National Key Research and Development Program of China under Grant 
2023YFE02099 00, the 
National Natural Science Foundation
of China under Grants 62033005, 62320106001, and 62273208, 
the Natural Science Fou- ndation of Heilongjiang Province under Grant LH2024F026, and 
the Natural Science Foundation of Chongqing Municipality under Grant CSTB2023NSCQ-MSX0625. }
}
\small{\thanks{ This paper was not presented at any IFAC
meeting. Corresponding author: Ligang Wu.}}

\author[RM1,RM2]{Renjie Ma}
\ead{renjiema@hit.edu.cn},   
\author[ZH]{Zhijian Hu}
\ead{zhijian.hu@ntu.edu.sg},               
\author[RY]{Rongni Yang}
\ead{rnyang@sdu.edu.cn},
\author[LW]{Ligang Wu}
\ead{ligangwu@hit.edu.cn}  

\address[RM1]{State Key Laboratory of Robotics and Systems, Harbin Institute of Technology, Harbin 150001, China}
\address[RM2]{Chongqing Research Institute, Harbin Institute of Technology, Chongqing 401120, China}
\address[ZH]{LAAS-CNRS, University of Toulouse, CNRS, Toulouse 31077, France}             
\address[RY]{School of Control Science and Engineering, Shandong University, Jinan 250061, China}        
\address[LW]{School of Astronautics, Harbin Institute of Technology, Harbin 150001, China}


\begin{abstract}                          
In this paper, 
we synthesize two aperiodic-sampled deep neural network (DNN) control schemes, based on the closed-loop 
tracking stability guarantees.
By means of the integral quadratic constraint coping with the input-output behaviour of system uncertainties/nonlinearities and the convex relaxations of nonlinear DNN activations leveraging their
local sector-bounded attributes, we establish conditions to design the event- and self-triggered logics and to compute the ellipsoidal inner approximations of region of attraction, respectively. Finally, 
we perform a numerical example of an inverted pendulum to illustrate the effectiveness of the
proposed aperiodic-sampled DNN control schemes.
\end{abstract}

\end{frontmatter}

\section{Introduction}


The rapid developments of imitation learning motivate new perspectives of control designs for cyber-physical systems.
Apart from minimizing a loss function based on exploration-exploitation tradeoffs \citep*{Jin2018Control}, learning 
a control strategy to obtain the prescribed system
performances and to imitate the expert demonstrations has obtained   
eye-catching achievements in recent years \citep*{Yin2022Imitation}.
Within this context, deep neural networks (DNNs) can be utilized to displace
computation-expensive control policies, such as model predictive control,
and to present better adaptivity to uncertainties
and less conservatism \citep*{Karg2020Efficient}. 
However, providing a theoretical guarantee of learned behaviour is challenging, due to the nonlinear and large-scale attributes of DNN,
which limits practical applications of neural-feedback loops. 

\subsection{Related Literature}
 The numerous hidden layers and neurons, 
as well as nonlinear DNN activation functions
bring  challenges for system analysis \citep*{Bella2025Regional,Nino2025Online}. Thus, the convex relaxations of DNNs are
crucial for deriving theoretical guarantees of neural-feedback
loops.
For example, the Lipschitz continuity of DNNs \citep*{Pauli2021Training,Fazlyab2019Efficient,Amico2023An}
is a common assumption,
and the Lipschitz constant provides an upper bound on
the degrees of DNN output changes, which can be incorporated into the Lyapunov-based system analysis \citep*{Talukder2023Robust}.
Particularly, the
\textrm{LipSDP} algorithm \citep*{Fazlyab2019Efficient} computes tight estimations of upper bounds on the Lipschitz constant of DNNs, based on the slope-restricted attribute of activation functions.
Then, the work \citep*{Pauli2021Training} proposes an optimization framework to train a DNN and to promote its robustness with a small Lipschitz constant. 
Note that the specific nonlinear attributes of
activation functions are crucial for relaxing DNNs, 
such as linear programming \citep*{Wong2018Provable} and semidefinite programming \citep*{Raghunathan2018Semi}
methods for \texttt{ReLU} DNNs. 
In general,
slope restrictions and sector bounds are commonly utilized
to abstract nonlinear DNN activation functions by virtue
of quadratic constraints \citep*{Fazlyab2022Safety,Ma2024Deep}, 
which lead to sufficient conditions of stability guarantee
based on linear matrix inequalities.
Accordingly,
robust ellipsoidal inner approximations of region of attraction (RoA) can be obtained for uncertain 
neural-feedback loops with perturbations formulated by integral quadratic constraints (IQCs)
in \citep*{Yin2022Stability} and \citep*{Pauli2021Offset}. 
Besides, quadratic constraints of DNNs
can also be used for analyzing the forward reachability of neural-feedback loops by semidefinite programming in both
centralized \citep*{Hu2020Reach} and distributed \citep*{Gates2023Scalable} scenarios. 
Moreover, interval arithmetic frameworks of DNNs can be incorporated
into a simulation-guided reachability analysis \citep*{Xiang2021Reachable} or be involved in a mixed-integer linear programming to obtain the closed-loop stability guarantees \citep*{Fabiani2022Reliability}.

Periodic time-triggered sampling schemes are not resource-
efficient in the sense of networked control
 \citep*{Heemels2012An}. 
Enabling communication transmissions only when
necessary can refine 
unnecessary waste of communication resources, which
motivates the developments of
event- and self-triggered control. 
Based on continuous monitoring of system states, 
an event trigger can check whether the state
value has changed and satisfied the specific logic, and can authorize the communication transmissions to update control inputs \citep*{Seuret2012A,Hu2025Resilient}.
The core of event-triggered control (ETC) lies in specifying a triggering logic incorporated into a control strategy such that the desired closed-loop performances can be guaranteed.
Normally, a triggering logic
relies on the Lyapunov function \citep*{Ghodrat2023A}
or the error signal
\citep*{Tripathy2017Stabilization,Zhao2022Event,Seifullaev2022Event,Girard2015Dynamic,Liu2015Event,Ye2023Global,Coutinho2022Codesign}. Compared with static ETC, both
the state trajectory and the auxiliary variable need to be converged synchronously for dynamic ETC, which potentially enlarges the minimum inter-execution time  \citep*{Girard2015Dynamic}. 
For a self-triggered control (STC) scheme,
the crux is to compute the upcoming transmission instant,
by virtue of the transmitted signals, in advance \citep*{Akashi2018Self}. 
Therefore, a continuous event detector for monitoring system state is unnecessary, and the communication transmission is shut until the upcoming triggering instant.
Extensive applications of STC are widely emerged in quantized control \citep*{Wakaiki2023Self}, resilient control against communication delays and packet dropouts \citep*{Peng2016On}, nonlinear control \citep*{Li2022Stability}, switched control \citep*{Cao2023Self,Wan2021Dynamic}, and data-driven control \citep*{Wang2022Model,Wildhagen2023Data}, where both closed-loop stability guarantees and STC syntheses are performed together.
Fewer results have explored DNN-based ETC and STC 
schemes, compared to time-triggered DNN controllers \citep*{Yin2022Stability,Pauli2021Offset,Hu2020Reach,Gates2023Scalable}, which acts as the first motivation of this paper.

The uncertainty impacts, such as unmodeled dynamics, time delays, and adversarial attacks/faults \citep*{Ma2021Sparse},
inevitably degrade closed-loop performances of neural-feedback loops, in view of the intrinsic vulnerability of DNNs \citep*{Fazlyab2022Safety}. Enhancing resilience of neural-feedback loops in the sense of robust control is critically important.
Integral quadratic constraints (IQCs) can describe the local
input-output behaviour of a general uncertainty, which interconnects in feedback with a nominal system \citep*{Seiler2015Stability}. 
IQCs cover a multitude of uncertainties and nonlinearities in the time domain \citep*{Lessard2016Analysis} and perform less conservatism for deriving 
closed-loop performance guarantees \citep*{Schwenkel2022Model}.
Due to the nonlinearity attribute of neural-feedback loops, the stabilization control
is normally not in the global sense. 
The size of region of attraction (RoA), acting as a stability metric, can be utilized to assess the effectiveness of specified control policy. 
In practice, computing an exact RoA to obtain its complicated analytical expression is difficult, therefore, many algorithms
have been developed to obtain inner approximations of RoA
in view of sublevel sets of Lyapunov functions \citep*{Valmorbida2017Region,Iannelli2019Region}, in which nonrestrictive IQCs can
be well involved. 
However, fewer results have revealed inner approximations of RoA for neural-feedback loops with
aperiodic-sampled communication between sensor and controller, which acts as the second motivation of this paper.

\subsection{Technical Contributions}
In this paper, we aim at filling the gap between aperiodic-sampled communication transmissions and neural-feedback loops.
Note that the technical results herein are different from the existing literature. First, establishing the relationships between the triggering parameters and the size of robust
inner approximation of RoA, were not revealed in \citep*{Yin2022Stability}. Besides,
the triggering mechanisms herein
are theoretically different from the counterpart in \citep*{Souza2023Event}, where the event triggers are located 
within the interior of a DNN. Although the work \citep*{Wang2022Model} provides promising tools for aperiodic-sampled controller
designs for linear systems, it cannot be used directly toward neural-feedback loops,
due to intrinsically nonlinear DNN activations. In addition, 
the stability-guaranteed STC synthesis, connecting self-triggered logics with closed-loop performances, was not developed therein.
The main contributions of this paper 
are summarized by the following points.\newline
\noindent $\bullet$ The resource-efficient event- and self-triggered communication 
schemes are deployed with the channel between the sampler and the DNN controller, 
such that the triggering time instants can be autonomously determined by monitoring 
the state trajectories of neural-feedback loops, rather than the preactivation and activation of each DNN layer, 
which reduces communication and computation burdens. \newline
\noindent $\bullet$ Based on the quadratic constraint of DNN induced by the local attributes of repeated activation
function and the auxiliary looped functions, the sufficient conditions on ensuring the setpoint tracking stability of 
aperiodic-sampled neural-feedback loops
with less conservatism can be derived. \newline
\noindent $\bullet$ The size of ellipsoidal inner approximation of RoA in the sense of robust control
is extracted as a stability metric, which closely associates with the triggering parameters. By solving the proposed semidefinite programming problems,
the triggering parameters can be  determined on the premise of enlarging the RoA inner approximation. 

\subsection{Outline of This Paper}
The problem formulation is stated in Section \ref{sec2}. Then the
event-triggered and self-triggered DNN controller designs together with the closed-loop stability analyses are put forward in
Sections \ref{sec3} and \ref{sec4}, respectively. Section \ref{sec5} exhibits an example of an inverted pendulum
to validate the effectiveness of proposed theoretical results. We conclude this paper  in Section \ref{sec6}.	

\subsection{Applicatory Notations}
$\mathds{S}^{n}\hspace*{-0.2em}$ denotes the set of $n$-by-$n$ symmetric matrices. $\mathds{RL}_{\infty}^{m\hspace*{-0.1em}\times\hspace*{-0.1em} p}\hspace*{-0.1em}$ denotes
the set of $m$-by-$p$ real-rational and proper transfer matrices, with no poles on the unit circle. Its subset
$\hspace*{-0.1em}\mathds{RH}_{\infty}^{m\hspace*{-0.1em}\times\hspace*{-0.1em} p}\hspace*{-0.1em}$ contains functions
which are analytic outside the closed unit disk.
We define the set of sequence in $\mathds{R}^{n}$ by $\ell_{2e}^{n}=\{(v(k)),k\hspace*{-0.2em}\in\hspace*{-0.2em}\mathds{N}|
v(k)\in\mathds{R}^{n}\}$. 
$\hspace*{-0.1em}\mathds{N}_{[r_{1},r_{2}]}\hspace*{-0.2em}\subseteq\hspace*{-0.2em}\mathds{N}$ denotes
the nonnegative integer set $\hspace*{-0.1em}\{\hspace*{-0.1em}r_{1},\hspace*{-0.1em}r_{1}\hspace*{-0.2em}+\hspace*{-0.2em}1,
\hspace*{-0.1em}\cdots\hspace*{-0.2em},\hspace*{-0.1em}r_{2}\}$ with $r_{2}\hspace*{-0.2em}>\hspace*{-0.2em}r_{1}\hspace*{-0.2em}
\geq\hspace*{-0.2em} 0$. We denote the positive (semi-)
definite and symmetric matrix by
$\hspace*{-0.1em}Q\hspace*{-0.2em}\succ\hspace*{-0.2em}\mathbf{0}(\hspace*{-0.1em}\succeq\hspace*{-0.2em}\mathbf{0})\hspace*{-0.1em}$, and
$\hspace*{-0.1em}\|v_{k}\hspace*{-0.1em}\|_{Q}^{2}\hspace*{-0.1em}$ equals to $\hspace*{-0.1em}v_{k}^{\top}\hspace*{-0.1em}Qv_{k}\hspace*{-0.1em}$. $\hspace*{-0.1em}\mathrm{Det}(\hspace*{-0.1em}R)\hspace*{-0.1em}$ captures \hspace*{-0.1em}the \hspace*{-0.1em}determinant \hspace*{-0.1em}of symmetric matrix $R$. The column-arranged matrix/vector
$\hspace*{-0.1em}[\hspace*{-0.1em}L_{1}^{\top}\hspace*{-0.2em},\hspace*{-0.1em}\cdots\hspace*{-0.2em},
\hspace*{-0.1em}L_{r_{1}}^{\top}]^{\hspace*{-0.1em}\top}
\hspace*{-0.1em}$ is depicted by
$\hspace*{-0.1em}\mathrm{Col}(\hspace*{-0.1em}L_{1}\hspace*{-0.1em},\hspace*{-0.1em}\cdots\hspace*{-0.2em},
\hspace*{-0.1em}L_{\hspace*{-0.05em}r_{1}}\hspace*{-0.1em})$.
$\hspace*{-0.1em}\mathrm{Diag}(\hspace*{-0.1em}H_{1}\hspace*{-0.1em},\hspace*{-0.1em}\cdots\hspace*{-0.2em},
\hspace*{-0.1em}H_{\hspace*{-0.05em}r_{2}}\hspace*{-0.1em})\hspace*{-0.1em}$
denotes the symmetric matrix whose
diagonal elements are $\hspace*{-0.1em}H_{\imath},\hspace*{-0.1em}\imath\hspace*{-0.3em}\in\hspace*{-0.3em}\mathds{N}_{[1,r_{2}]}$.

\vspace{-0.6em}

\section{Problem Formulation}
\label{sec2}

\vspace{-0.1em}	

We consider a discrete-time control system $\digamma(\Gamma,\Theta)$ with an interconnection of a nominal plant $\Gamma$ and an uncertainty
part $\hspace*{-0.1em}\Theta$ \citep*{Hu2016Robustness}.  For sampling time  instant $\hspace*{-0.1em}k\hspace*{-0.3em}\in\hspace*{-0.2em}\mathds{N}$,  the nominal plant $ \hspace*{-0.1em}\Gamma\hspace*{-0.1em}$ can be described by
\begin{equation}
\label{sys}
\begin{array}{rcl}
x(k\hspace*{-0.2em}+\hspace*{-0.2em}1)& \hspace*{-0.3em}=\hspace*{-0.3em} &
A_{\Gamma}x(k)\hspace*{-0.2em}+\hspace*{-0.2em}B_{\Gamma}u(k)\hspace*{-0.2em}+\hspace*{-0.2em}F_{\Gamma}\omega(k),\\
\nu(k)& \hspace*{-0.3em}=\hspace*{-0.3em} &
C_{\Gamma}x(k)\hspace*{-0.2em}+\hspace*{-0.2em}D_{\Gamma}u(k)\hspace*{-0.2em}+\hspace*{-0.2em}G_{\Gamma}\omega(k),
\end{array}
\end{equation}
where $x\hspace*{-0.2em}\in\hspace*{-0.2em}\mathds{R}^{n}$ denotes the system state,
$u\hspace*{-0.2em}\in\hspace*{-0.2em}\mathds{R}^{m}$
denotes the control input, $\nu\hspace*{-0.2em}\in\hspace*{-0.2em}\mathds{R}^{v}$ and
$\omega\hspace*{-0.2em}\in\hspace*{-0.2em}\mathds{R}^{w}$ represent the input and output of the uncertainty part $\Theta$, respectively.
We assume that the system matrices of \eqref{sys} are known with suitable dimensions.

The nominal plant $\Gamma$ is interconnected in feedback with a bounded and casual operator
$\Theta\hspace*{-0.2em}:\hspace*{-0.2em}\ell_{2e}^{v}\hspace*{-0.2em}\mapsto\hspace*{-0.2em}\ell_{2e}^{w}$ formulated by
\begin{equation}
\label{un}
\begin{array}{rcl}
\omega(k)& \hspace*{-0.3em}=\hspace*{-0.3em} & \Theta(\nu(k)).
\end{array}
\end{equation}
We assume that the interconnection of $\Gamma$ and $\Theta$ is well-posed, namely,
there exist the unique response $x\hspace*{-0.2em}\in\hspace*{-0.2em}\ell_{2e}^{n}$,
$\omega\hspace*{-0.2em}\in\hspace*{-0.2em}\ell_{2e}^{w}$, $\nu\hspace*{-0.2em}\in\hspace*{-0.2em}\ell_{2e}^{v}$ for each
$u\hspace*{-0.2em}\in\hspace*{-0.2em}\ell_{2e}^{m}$, which guarantees that the control system $\digamma\hspace*{-0.1em}(\Gamma,\Theta)$ has the unique solution, and there is no algebraic loop for the interconnection of $\Gamma$ and $\Theta$  \citep*{Schwenkel2022Model}.
Besides, all state measurements $x$ are available for feedback control.

IQCs exhibit powerful effectiveness for analyzing the interconnection in feedback between $\Gamma$ and $\Theta$, 
which is extended
to time-domain cases by dissipation inequalities \citep*{Seiler2015Stability}. The keypoint of characterizing the input-output relationship of
uncertainty operator $\Theta$ with IQC lies in constructing a virtual filter $\Phi_{\Theta}$, whose state $\xi$ relates to
the input $\nu$ and the output $\omega$ of $\Theta$, and output $r$ leads to a quadratic constraint. The
filter $\Phi_{\Theta}$ can be described with the form 
\begin{equation}
\label{fil}
\begin{array}{rcl}
\xi(k\hspace*{-0.2em}+\hspace*{-0.2em}1)& \hspace*{-0.3em}=\hspace*{-0.3em} &
A_{\Phi}\xi(k)\hspace*{-0.2em}+\hspace*{-0.2em}B_{\Phi}\nu(k)\hspace*{-0.2em}+\hspace*{-0.2em}F_{\Phi}\omega(k),\\
r(k)& \hspace*{-0.8em}=\hspace*{-0.8em} &
C_{\Phi}\xi(k)\hspace*{-0.2em}+\hspace*{-0.2em}D_{\Phi}\nu(k)\hspace*{-0.2em}+\hspace*{-0.2em}G_{\Phi}\omega(k),
\end{array}
\end{equation}
where $\xi\hspace*{-0.2em}\in\hspace*{-0.2em}\mathds{R}^{\psi}$ with $\xi(0)\hspace*{-0.2em}=\hspace*{-0.2em}0$ and
$r\hspace*{-0.2em}\in\hspace*{-0.2em}\mathds{R}^{\kappa}$. It is intuitive that the filter parameters in \eqref{fil} are with proper dimensions.
$A_{\Phi}$ implies a Schur matrix, whose spectrum is contained in the open unit disk in the complex plane, 
ensuring that \eqref{fil} has a unique solution $(\xi_{*},r_{*})$ for any choice of $(\nu_{*},\omega_{*})$, see \citep*[Subsection 3.1]{Lessard2016Analysis} for more technical details. 
The definition of time-domain $\rho$-hard IQC is recapped in what follows.
\begin{definition}
\citep*{Lessard2016Analysis} 
For a scalar $\rho\hspace*{-0.2em}\in\hspace*{-0.2em}\mathds{R}_{(0,1]}$, a matrix
$M_{\Theta}\hspace*{-0.2em}\in\hspace*{-0.2em}\mathds{S}^{\kappa}$, as well as a function
$\Phi_{\Theta}\hspace*{-0.2em}\in\hspace*{-0.2em}\mathds{RH}^{\kappa\hspace*{-0.1em}\times\hspace*{-0.1em}
(v\hspace*{-0.1em}+\hspace*{-0.1em}w)}_{\infty}$,
the bounded and casual operator $\Theta$ defined in \eqref{un} satisfies the $\rho$-hard IQC described by $(\Phi_{\Theta},M_{\Theta})$, if for
all $\nu\hspace*{-0.2em}\in\hspace*{-0.2em}\ell_{2e}^{v}$ with $\omega\hspace*{-0.2em}=\hspace*{-0.2em}\Theta(\nu)$, and for all
$\mathcal{I}\hspace*{-0.2em}\in\hspace*{-0.2em}\mathds{N}_{+}$, we have
\begin{equation}
\label{iqc}
\begin{array}{rcl}
\sum\limits_{k=0}^{\mathcal{I}}\rho^{-2k}r^{\top}(k)M_{\Theta}r(k)& \hspace*{-0.3em}\geq\hspace*{-0.3em} & 0.
\end{array}
\end{equation}
\end{definition}

We denote the augmented state by $\eta\hspace*{-0.2em}=\hspace*{-0.2em}\textrm{Col}(x,\xi)$ with
$\eta\hspace*{-0.2em}\in\hspace*{-0.2em}\mathds{R}^{z}$ and
$z\hspace*{-0.2em}=\hspace*{-0.2em}n\hspace*{-0.2em}+\hspace*{-0.2em}\psi$.
Then, in terms of \eqref{sys} and \eqref{fil}, we can obtain the augmented dynamics with the form  
\begin{equation}
\label{aug}
\begin{array}{rcl}
\eta(k\hspace*{-0.2em}+\hspace*{-0.2em}1)& \hspace*{-0.3em}=\hspace*{-0.3em} &
A\eta(k)\hspace*{-0.2em}+\hspace*{-0.2em}B\bar{u}(k),\\
r(k)& \hspace*{-0.8em}=\hspace*{-0.8em} &
C\eta(k)\hspace*{-0.2em}+\hspace*{-0.2em}D\bar{u}(k),
\end{array}
\end{equation}
where $\bar{u}\hspace*{-0.2em}=\hspace*{-0.2em}\mathrm{Col}(u,\omega)$ and
\begin{eqnarray*}
A&\hspace*{-0.2em}\triangleq\hspace*{-0.2em}& \left[
\begin{array}{cc}
\hspace*{-0.2em}A_{\Gamma} & \mathbf{0}\hspace*{-0.2em}\\
\hspace*{-0.2em}B_{\Phi}C_{\Gamma} &A_{\Phi}\hspace*{-0.2em}
\end{array}
\right], B\hspace*{-0.2em}\triangleq\hspace*{-0.2em}\left[\hspace*{-0.2em}
\begin{array}{cc}
\hspace*{-0.2em}B_{\Gamma}& F_{\Gamma}\hspace*{-0.2em}\\
\hspace*{-0.2em}B_{\Phi}D_{\Gamma} & B_{\Phi}G_{\Gamma}\hspace*{-0.2em}+\hspace*{-0.2em}
F_{\Phi}\hspace*{-0.2em}
\end{array}
\hspace*{-0.2em}\right], \\
C&\hspace*{-0.8em}\triangleq\hspace*{-0.8em}& \left[
\begin{array}{cc}
\hspace*{-0.2em}D_{\Phi}C_{\Gamma} &C_{\Phi}\hspace*{-0.1em}
\end{array}
\right], D\hspace*{-0.2em}\triangleq\hspace*{-0.2em}\left[\hspace*{-0.2em}
\begin{array}{cc}
\hspace*{-0.2em}D_{\Phi}D_{\Gamma} & D_{\Phi}G_{\Gamma}\hspace*{-0.2em}+\hspace*{-0.2em}
G_{\Phi}\hspace*{-0.2em}
\end{array}
\hspace*{-0.2em}\right].
\end{eqnarray*}
The stable equilibrium of augmented system \eqref{aug} is denoted by
$\eta^{*}\hspace*{-0.2em}=\hspace*{-0.2em}\textrm{Col}(x^{*},\xi^{*})$. We let $\chi(k,x_{0},\Theta)$ denote the state response of
neural-feedback loops $\digamma(\Gamma,\Theta)$ starting from the initial state
$x_{0}\hspace*{-0.2em}=\hspace*{-0.2em}x(0)$. It is unlikely to have a global stabilizing DNN controller, and mostly, local stability guarantees corresponding to the specific data region are sufficient \citep*{Pauli2021Offset}. Hence,
we assume that $\chi(k,x_{0},\Theta)$ can converge to the stable equilibrium  $x^{*}$ locally, which
yields the definition of robust RoA.
\begin{definition}
\label{def2}
Let $\mathcal{S}$ denote the set of uncertainty impact $\Theta$, that is, $\Theta\hspace*{-0.2em}\in\hspace*{-0.2em}\mathcal{S}$,
then the robust RoA of DNN-controlled system \eqref{aug} is defined by
\begin{equation}
\label{rroa}
\begin{array}{rcl}
\mathcal{R}_{F}^{x}& \hspace*{-0.3em}=\hspace*{-0.3em} & \left\{x_{0}\hspace*{-0.2em}\in\hspace*{-0.2em}\mathds{R}^{n}|
\lim_{k\rightarrow\infty}\chi(k,x_{0},\Theta)\hspace*{-0.2em}=\hspace*{-0.2em}x^{*}, \forall
\Theta\hspace*{-0.2em}\in\hspace*{-0.2em}\mathcal{S}\right\}.
\end{array}
\end{equation}
\end{definition}

The control input $u$ can be obtained by an $l$-layer feed-forward DNN $\pi_{\mathrm{DNN}}(x)\hspace*{-0.2em}:
\hspace*{-0.2em}\mathds{R}^{n}\hspace*{-0.2em}\mapsto\hspace*{-0.2em}\mathds{R}^{m}$ with the form
\begin{equation}
\label{dnn}
\begin{array}{rcl}
m_{0}(k)& \hspace*{-0.3em}=\hspace*{-0.3em} & x(k), \\
m_{i}(k)& \hspace*{-0.8em}=\hspace*{-0.8em} &\varphi_{i}(W_{i}m_{i-1}(k)\hspace*{-0.2em}+\hspace*{-0.2em}b_{i}),
i\hspace*{-0.2em}\in\hspace*{-0.2em}\mathds{N}_{[1,l]}, \\
u(k)& \hspace*{-0.8em}=\hspace*{-0.8em} & W_{l+1}m_{l}(k)+b_{l+1},
\end{array}
\end{equation}
where $m_{i}\hspace*{-0.2em}\in\hspace*{-0.2em}\mathds{R}^{a_{i}}$ denotes the activation of $i$-th layer with
$a_{0}\hspace*{-0.2em}=\hspace*{-0.2em}n$, and the perception of $i$-th layer is attained by the weight matrix
$W_{i}$ and the bias $b_{i}$. The activation function $\varphi_{i}$ is element-wise, that is,
\begin{equation}
\label{ela}
\begin{array}{rcl}
\varphi_{i}(\mathbf{v})& \hspace*{-0.3em}=\hspace*{-0.3em} &
\mathrm{Col}\left(\mathbf{s}(\mathbf{v}_{1}),\mathbf{s}(\mathbf{v}_{2}),\cdots,\mathbf{s}(\mathbf{v}_{a_{i}})\right),
\end{array}
\end{equation}
where $\mathbf{s}$ implies a scalar nonlinear function satisfying $\mathbf{s}(0)=0$, which can be selected as ReLU, sigmoid, and softmax. Normally, $\psi$ is
assumed to be identical for all layers without loss of generality.

We consider that the system state is sampled and transmitted to the controller at specific time instant
$k_{q}\hspace*{-0.2em}\in\hspace*{-0.2em}\mathds{N}$, where $k_{0}\hspace*{-0.2em}=\hspace*{-0.2em}0$ and
$k_{q+1}\hspace*{-0.2em}-\hspace*{-0.2em}k_{q}\hspace*{-0.2em}\geq\hspace*{-0.2em}1$ hold for
$q\hspace*{-0.2em}\in\hspace*{-0.2em}\mathds{N}$. The available sampled state $x(k_{q})$
is transmitted into $\pi_{\mathrm{DNN}}$ and leads to the control input
$u(k)\hspace*{-0.2em}=\hspace*{-0.2em}\pi_{\mathrm{DNN}}(x(k_{q}))$ for 
$k\hspace*{-0.2em}\in\hspace*{-0.2em}\mathds{N}_{[k_{q},k_{q+1}-1]}$, in view of the zero-order hold (ZOH).
To improve the communication efficiency, we will design ETC and STC schemes for neural-feedback loops \eqref{aug}, on the premise of the closed-loop stability guarantees.

\begin{figure}[htbp]
\centerline{
\includegraphics[width=8.5cm]{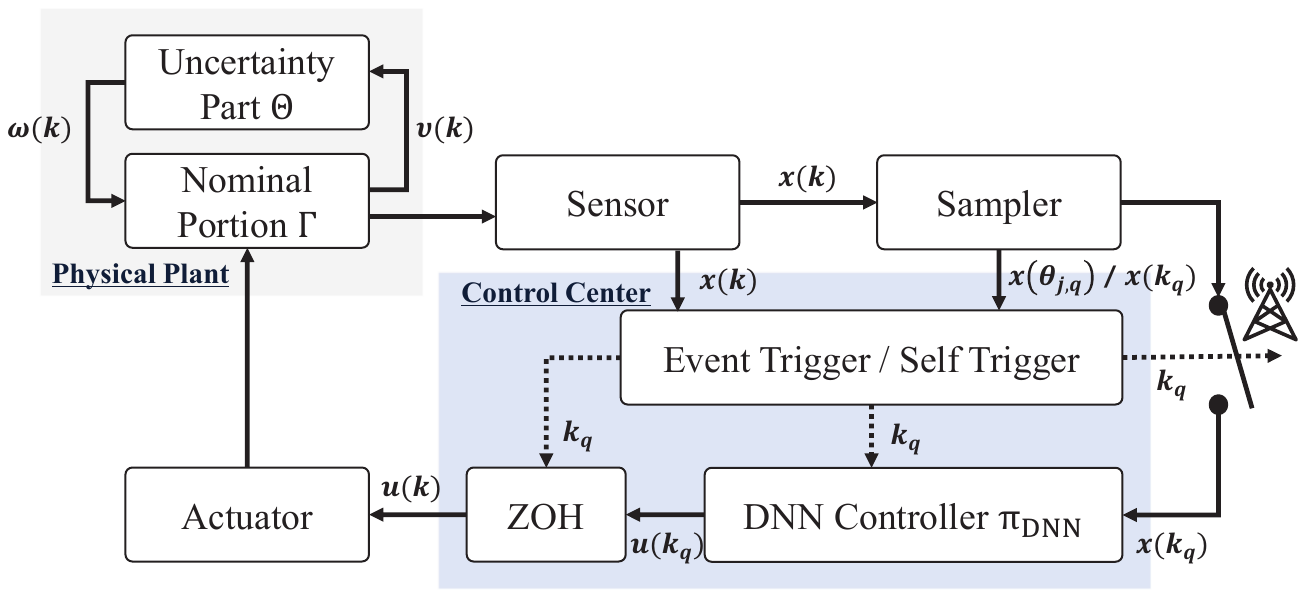} \vspace{-1ex}}
\caption{The design schematic of aperiodic-sampled neural-feedback loops.}
\label{F1}
\end{figure}
We illustrate the schematic of aperiodic-sampled
neural-feedback loops $\digamma(\Gamma,\Theta)$ in Fig. \ref{F1}, based on which, the input $\bar{u}(k)$ of 
neural-feedback loops \eqref{aug} is formulated by
\begin{equation}
\label{aaug}
\begin{array}{rcl}
\bar{u}(k)& \hspace*{-0.3em}=\hspace*{-0.3em} &
\textrm{Col}(\pi_{\mathrm{DNN}}(x(k_{q})),\omega(k)), k\hspace*{-0.2em}\in\hspace*{-0.2em}\mathds{N}_{[k_{q},k_{q+1}-1]}.
\end{array}
\end{equation}

Hence, the aperiodic-sampled DNN controller \eqref{aaug} depends on both the DNN architecture \eqref{dnn} and the
adaptive triggering time instant $k_{q}$. 
According to Fig. \ref{F1}, the transmission switch is located in the channel between the sampler and the controller, leading to aperiodic-sampled DNN control schemes \eqref{aaug}, and the triggering time instants are autonomously determined by the state trajectory of neural-feedback loops via an event/self trigger. In contrast, the work \citep*{Souza2023Event} deploys an event-triggered transmission scheme in the channel between two coterminous DNN layers, and the triggering logic is based on the quadratic constraint with regard to the preactivation and activation of each layer. The number of event triggers equals to that of DNN layers therein, which implies a more complex structure than our results.
The problems of interests of this paper are 
summarized by the following aspects. \newline
\noindent $\bullet$ We deploy two effective aperiodic-sampled transmission schemes to autonomously determine the triggering instants
in the channel between the sampler and the DNN controller, and to mitigate communication and computation burdens. \newline
\noindent $\bullet$ We derive the closed-loop stability guarantees of aperiodic-sampled neural-feedback loops, in the light of the convex
relaxations of DNN and the auxiliary looped functions combined with Lyapunov functions. \newline
\noindent $\bullet$ We assess the stability metric by virtue of maximizing the inner approximation of robust RoA influenced by
aperiodic-sampled schemes, and determine the triggering parameters based on the feasibility of proposed convex programming.

\section{Event-Triggered Neural Control Policy}
\label{sec3}

\subsection{Event-Triggered Scheme}
\label{301}

The blue rectangle within Fig. \ref{F1} depicts an event-triggered DNN controller module, where the communication time instants are
captured by $\{k_{q}\}_{q\in\mathds{N}}$. For ETC, the state trajectory is sampled periodically at discrete time instants
$\{k_{q}\hspace*{-0.2em}+\hspace*{-0.2em}j\vartheta\}$ with $j\hspace*{-0.2em}\in\hspace*{-0.2em}\mathds{N}$ and the constant sampling
interval $\vartheta\hspace*{-0.2em}\in\hspace*{-0.2em}\mathds{R}_{[\vartheta_{l},\vartheta_{u}]}$, in which $\vartheta_{l}$ and
$\vartheta_{u}\hspace*{-0.2em}\in\hspace*{-0.2em}\mathds{N}_{\geq 1}$ are the lower and upper bounds, respectively.
The following triggering rule is assessed in terms of the sampled state
$x(k_{q}\hspace*{-0.2em}+\hspace*{-0.2em}j\vartheta)$
\begin{equation}
\label{asl}
\begin{array}{rcl}
\alpha(\theta_{j,q})\hspace*{-0.2em}+\hspace*{-0.2em}g
\beta(\theta_{j,q})& \hspace*{-0.3em}<\hspace*{-0.3em} & 0,
\end{array}
\end{equation}
where $g\hspace*{-0.2em}\in\hspace*{-0.2em}\mathds{R}_{\geq0}$ is a parameter to be determined, $\theta_{j,q}\hspace*{-0.2em}\triangleq\hspace*{-0.2em}
k_{q}\hspace*{-0.2em}+\hspace*{-0.2em}j\vartheta$ holds for $j\hspace*{-0.2em}\in\hspace*{-0.2em}\mathds{N}_{[0,h(q)]}$ with
$h(q)\hspace*{-0.2em}\triangleq\hspace*{-0.2em}(k_{q+1}\hspace*{-0.2em}-\hspace*{-0.2em}k_{q})\hspace*{-0.1em}/
\hspace*{-0.1em}\vartheta\hspace*{-0.2em}-\hspace*{-0.2em}1$, $\beta(\theta_{j,q})$ defines a quadratic function around $x^{*}$
with the form
\begin{equation}
\label{as2}
\begin{array}{rcl}
\hspace*{-0.8em}\beta(\theta_{j,q})& \hspace*{-0.3em}=\hspace*{-0.3em} & \epsilon_{1}\|x(\theta_{j,q})\hspace*{-0.3em}-\hspace*{-0.3em}x^{*}\|_{\Xi_{1}}^{2}
\hspace*{-0.3em}+\hspace*{-0.3em}\epsilon_{2}
\|x(k_{q})
\hspace*{-0.3em}-\hspace*{-0.3em}x^{*}\|_{\Xi_{1}}^{2} \\
\hspace*{-1.6em}& \hspace*{-0.9em}\hspace*{-0.9em} &
-\|e(\theta_{j,q})\|_{\Xi_{2}}^{2},
\end{array}
\end{equation}
where $\epsilon_{1},\epsilon_{2}\hspace*{-0.2em}\in\hspace*{-0.2em}\mathds{R}_{[0,1]}$ are triggering discount parameters, and
$\Xi_{1},\Xi_{2}\hspace*{-0.2em}\in\hspace*{-0.2em}\mathds{S}^{n}_{\succ0}$ are weight matrices. The error between
the state trajectory $x(\theta_{j,q})$ at the current time instant and the counterpart $x(k_{q})$ at the latest transmission instant is
defined by $e(\theta_{j,q})$, that is, $e(\theta_{j,q})\hspace*{-0.2em}\triangleq\hspace*{-0.2em}x(\theta_{j,q})\hspace*{-0.2em}-\hspace*{-0.2em}
x(k_{q})$. Moreover, $\alpha(t)$ formulates a dynamic parameter with $\alpha(k)\hspace*{-0.2em}=\hspace*{-0.2em}\alpha(\theta_{j,q})$ and
\begin{equation}
\label{as3}
\begin{array}{rcl}
\alpha(\theta_{j+1,q})& \hspace*{-0.3em}=\hspace*{-0.3em} & (1\hspace*{-0.2em}-\hspace*{-0.2em}\mu)\alpha(\theta_{j,q})
\hspace*{-0.2em}+\hspace*{-0.2em}\beta(\theta_{j,q}),
\end{array}
\end{equation}
where $\alpha(0)\hspace*{-0.2em}\in\hspace*{-0.2em}\mathds{R}_{\geq0}$ and
$\mu\hspace*{-0.2em}\in\hspace*{-0.2em}\mathds{R}_{[0,1)}$ are prescribed coefficients.

The sampled state $x(\theta_{j,q})$ can be transmitted into the DNN controller $\pi_{\mathrm{DNN}}$ once the event-triggered logic
\eqref{asl} is violated. We can obtain the newly computed DNN control input held by a ZOH for the
subsequent triggering interval $\hspace*{-0.1em}[k_{q+1},k_{q+2}\hspace*{-0.2em}-\hspace*{-0.2em}1]$,
where the latest transmitted state is used to update the event-triggered DNN control input, and accordingly, to determine
the next sampled state. Thus, the dynamic event-triggered
logic is formulated by
\begin{equation}
\label{ets1}
\begin{array}{rcl}
k_{q+1}& \hspace*{-0.3em}=\hspace*{-0.3em} & k_{q}\hspace*{-0.2em}+\hspace*{-0.2em}\vartheta\min_{j}
\{j\hspace*{-0.2em}\in\hspace*{-0.2em}\mathds{N}_{+}| \alpha(\theta_{j,q})\hspace*{-0.2em}+\hspace*{-0.2em}g
\beta(\theta_{j,q})\hspace*{-0.2em}<\hspace*{-0.2em}0 \}.
\end{array}
\end{equation}
\begin{remark}
\label{r1}
Once the event-triggered condition \eqref{asl} is violated, we have $\alpha(\theta_{j,q})+g
\beta(\theta_{j,q})\hspace*{-0.2em}\geq\hspace*{-0.2em} 0$. For $g\hspace*{-0.2em}=\hspace*{-0.2em}0$,  $\alpha(\theta_{j,q})\hspace*{-0.2em}\geq\hspace*{-0.2em} 0$ holds.
While for $g\hspace*{-0.2em}\in\hspace*{-0.2em}\mathds{R}_{>0}$,
$\beta(\theta_{j,q})\hspace*{-0.2em}\geq-\alpha(\theta_{j,q})/g$ holds, then based on which, \eqref{as3}
leads to $\alpha(\theta_{j+1,q})\hspace*{-0.2em}\geq\hspace*{-0.2em}(1\hspace*{-0.2em}-\hspace*{-0.2em}
\mu\hspace*{-0.2em}-\hspace*{-0.2em}g^{-1})\alpha(\theta_{j,q})$. Thus, $\alpha(\theta_{j,q})\hspace*{-0.2em}\geq\hspace*{-0.2em}0$ holds
for any $j\hspace*{-0.2em}\in\hspace*{-0.2em}\mathds{N}_{[0,h(q)]}$ on the premise of both the initial condition
$\alpha(0)\hspace*{-0.2em}\geq\hspace*{-0.2em}0$ and the prerequisite constraint
$1\hspace*{-0.3em}-\hspace*{-0.3em}\mu\hspace*{-0.3em}-\hspace*{-0.3em}g^{-1}
\hspace*{-0.3em}\geq\hspace*{-0.3em}0$ \citep*{Wang2022Model}.
Hence, the nonnegativity of $\alpha(\theta_{j,q})$ is satisfied when deploying the event-triggered scheme \eqref{ets1}.
\end{remark}
\begin{remark}
\label{r302}
In essence, the deployed event-triggered scheme \eqref{ets1} is more general. For example, it can boil down
to the dynamic triggering logic established in \citep*{Girard2015Dynamic}, if
$\vartheta\hspace*{-0.3em}=\hspace*{-0.3em}g\hspace*{-0.3em}=\hspace*{-0.3em}1$,
$\epsilon_{1}\hspace*{-0.3em}=\hspace*{-0.3em}1$, and $\epsilon_{2}\hspace*{-0.3em}=\hspace*{-0.3em}0$
hold for \eqref{as2}. In addition, it can further cover the
static triggering logics \citep*{Tripathy2017Stabilization,Zhao2022Event,Seifullaev2022Event,Ghodrat2023A} without the consideration of
the non-negative $\alpha(\theta_{j,q})$. Besides, if we set
$\epsilon_{1}\hspace*{-0.3em}=\hspace*{-0.3em}\epsilon_{2}\hspace*{-0.3em}=\hspace*{-0.3em}0$,
then the condition \eqref{asl} leads to the
triggering scheme developed by \citep*{Liu2015Event}. Moreover, compared to \citep*{Wang2022Model}, different weight parameters $\Xi_{1}$ and
$\Xi_{2}$ embedded in $\beta(\theta_{j,q})$ can incorporate more general triggering conditions in
\citep*{Girard2015Dynamic,Tripathy2017Stabilization} better. Hence, the conducted triggering scheme \eqref{ets1} potentially unifies and generalizes most of existing event-triggered mechanisms.
\end{remark}

\subsection{Convex Relaxation of Deep Neural Network}
\label{302}

We define the DNN preactivations by
$p_{i}(k)\hspace*{-0.2em}=\hspace*{-0.2em}W_{i}m_{i-1}(k)\hspace*{-0.2em}+\hspace*{-0.2em}b_{i}$, such that  
$m_{i}(k)\hspace*{-0.2em}=\hspace*{-0.2em}\varphi_{i}(p_{i}(k))$ holds for $i\hspace*{-0.2em}\in\hspace*{-0.2em}\mathds{N}_{[1,l]}$. Hence, by
aggregating the inputs and outputs of the activation layers with
$\mathbf{g}\hspace*{-0.2em}=\hspace*{-0.2em}\mathrm{Col}(\mathbf{g}_{1},\hspace*{-0.1em}\cdots\hspace*{-0.2em},\mathbf{g}_{l})
\hspace*{-0.2em}\in\hspace*{-0.2em}\mathds{R}^{a},a\hspace*{-0.2em}=\hspace*{-0.2em}\sum_{i=1}^{l}a_{i}$, in which $\mathbf{g}$ can be $p$ and $m$,
respectively, the condition $m\hspace*{-0.2em}=\hspace*{-0.2em}\varphi(p)$ holds element-wisely for repeated nonlinearities \eqref{ela}.
The nonlinearities can be isolated based on \eqref{dnn}
combined with the superposition of linear operator \citep*{Fazlyab2022Safety}, which yields
\begin{eqnarray}
\left[\hspace*{-0.2em}
\begin{array}{c}
\hspace*{-0.1em}u(k)\hspace*{-0.1em}  \\
\hspace*{-0.1em}p(k)\hspace*{-0.1em}
\end{array}%
\hspace*{-0.2em}\right] & \hspace*{-0.3em}=\hspace*{-0.3em} &%
\left[
\begin{array}{ccc}
\hspace*{-0.1em}\Pi_{ux}&\Pi_{um}&\Pi_{u1} \hspace*{-0.3em}\\
\hspace*{-0.1em}\Pi_{px}&\Pi_{pm}&\Pi_{p1}\hspace*{-0.3em}
\end{array}%
\right]\hspace*{-0.3em}\left[
\begin{array}{c}
\hspace*{-0.1em}x(k_{q})\hspace*{-0.1em}  \\
\hspace*{-0.1em}m(k)\hspace*{-0.1em}\\
\hspace*{-0.1em}1\hspace*{-0.1em}
\end{array}%
\right]\hspace*{-0.2em}, m(k)\hspace*{-0.2em}=\hspace*{-0.2em}\varphi(p(k)), \label{dnn3}
\end{eqnarray}
for the time instant $k\hspace*{-0.2em}\in\hspace*{-0.2em}\mathds{N}_{[k_{q},k_{q+1}-1]}$, and
\begin{eqnarray}
\Pi_{ux}\hspace*{-0.1em} & \hspace*{-0.1em}\triangleq\hspace*{-0.1em} &%
\hspace*{-0.1em} \mathbf{0}_{m\hspace*{-0.1em}\times\hspace*{-0.1em} n},
\Pi_{um}\hspace*{-0.3em}\triangleq\hspace*{-0.3em}\left[\hspace*{-0.1em}
\begin{array}{cc}
\hspace*{-0.1em}\mathbf{0}_{m\hspace*{-0.1em}\times\hspace*{-0.1em}\mathcal{Z}(a_{1},a_{l\hspace*{-0.1em}-\hspace*{-0.1em}1})}
\hspace*{-0.1em}&\hspace*{-0.1em} W_{l\hspace*{-0.1em}+\hspace*{-0.1em}1}\hspace*{-0.1em}
\end{array}%
\hspace*{-0.2em}\right]\hspace*{-0.2em}, \Pi_{u1}\hspace*{-0.3em}\triangleq\hspace*{-0.3em}b_{l\hspace*{-0.1em}+\hspace*{-0.1em}1},
\notag\\
\Pi_{px}\hspace*{-0.1em} & \hspace*{-0.1em}\triangleq\hspace*{-0.1em} &
\hspace*{-0.1em}\mathrm{Col}(W_{1},\mathbf{0}_{\mathcal{Z}(a_{2},a_{l})\hspace*{-0.1em}\times\hspace*{-0.1em}n}),
\Pi_{p1}\hspace*{-0.3em}\triangleq\hspace*{-0.3em}\mathrm{Col}(b_{1},b_{2},\cdots,b_{l}), \notag \\
\Pi_{pm}\hspace*{-0.1em} & \hspace*{-0.1em}\triangleq\hspace*{-0.1em}
&\hspace*{-0.1em}\mathrm{Col}(\mathbf{0}_{a_{1}\hspace*{-0.1em}\times\hspace*{-0.1em}\mathcal{Z}(a_{1},a_{l})},\tilde{\Pi}_{pm}),
\tilde{\Pi}_{pm}\hspace*{-0.3em}\triangleq\hspace*{-0.3em}\left[
\begin{array}{cc}
\hspace*{-0.1em}\tilde{\Pi}_{pm}^{(1)}
& \mathbf{0}_{\mathcal{Z}(a_{2},a_{l})\hspace*{-0.1em}\times\hspace*{-0.1em}a_{l}}\hspace*{-0.3em}
\end{array}%
\right]\hspace*{-0.2em}, \notag \\
\tilde{\Pi}_{pm}^{(1)}\hspace*{-0.1em} & \hspace*{-0.1em}\triangleq\hspace*{-0.1em} &
\hspace*{-0.1em}\mathrm{Diag}(W_{2},W_{3},\cdots,W_{l}),
\label{OE10}
\end{eqnarray}
holding with $\mathcal{Z}(a_{i},a_{j})\hspace*{-0.3em}=\hspace*{-0.3em}\sum_{\kappa\hspace*{-0.1em}=\hspace*{-0.1em}i}^{j}a_{\kappa}$.
The repeated nonlinearities \eqref{ela} normally satisfy the sector-bounded or/and slope-restricted properties,
which can be potentially utilized to
relax $\pi_{\mathrm{DNN}}$ by performing quadratic constraints based on \eqref{dnn3}.

Note that the global sector bounds of an activation function are too conservative to assess the closed-loop stability in a more exact way. Thus, the local sector bounds are preferred when formulating a tighter approximation of
$\pi_{\mathrm{DNN}}$, which yields the \emph{largest} inner approximation of
RoA\footnote{Based on the formulation of RoA \citep*[Proposition 1]{Persis2023Learning}, any Lyapunov-based sublevel set
$\mathcal{E}_{\hspace*{-0.1em}P_{1}}^{\Im}(x^{*}\hspace*{-0.1em})\hspace*{-0.2em}\triangleq\hspace*{-0.2em}
\{x\hspace*{-0.2em}\in\hspace*{-0.2em}\mathds{R}^{n}|\|x\hspace*{-0.2em}-\hspace*{-0.2em}x^{*}\|^{2}_{P_{1}}
\hspace*{-0.3em}\leq\hspace*{-0.3em}\Im\}$ contained in the region of $\mathcal{Z}\hspace*{-0.1em}\cup\hspace*{-0.1em}\{x^{*}\}$, with
$\mathcal{Z}\hspace*{-0.2em}\triangleq\hspace*{-0.2em}
\{x\hspace*{-0.2em}\in\hspace*{-0.2em}\mathds{R}^{n}|\Delta\mathcal{V}(k)\hspace*{-0.3em}<\hspace*{-0.3em}0\}$ and $\Delta\mathcal{V}(k)$ specified as Lemma \ref{lf}, is a positive invariant set for the closed loop and implies an RoA estimation.
Herein, the $1$-level set $\mathcal{E}_{\hspace*{-0.1em}P_{1}}(x^{*}\hspace*{-0.1em})$ with $\Im\hspace*{-0.3em}=\hspace*{-0.3em}1$ as  \citep*{Hindi1998Analysis} is considered as a metric for assessing stability, that is, the \emph{largest} RoA inner approximation is in the
sense of $1$-level set. We make this statement to avoid readers' confusion.
}
in the sense of robust control. We consider that each scalar activation function satisfies local sector-bounded 
restrictions, namely,
$\varphi_{i}\hspace*{-0.2em}\in\hspace*{-0.2em}\mathrm{Sec}[\rho_{i},\sigma_{i}]$ holds around the equilibrium $p_{i}^{*}$ with an input
$p_{i}\hspace*{-0.3em}\in\hspace*{-0.3em}[\underline{p}_{i}\hspace*{-0.1em},\hspace*{-0.1em}\overline{p}_{i}]$, 
$i\hspace*{-0.3em}\in\hspace*{-0.3em}\mathds{N}_{[1,l]}$. The local sector bounds are aggregated into vectors
$\rho,\sigma,\underline{p},\overline{p},p^{*}\in\mathds{R}^{a}$, which yields the following quadratic constraint \citep*{Fazlyab2022Safety}.
\begin{lemma}
\label{lem1}
The activation function $\varphi$ satisfies the local sector bound $\mathrm{Sec}[\rho,\sigma]$ around the point $(p^{*},m^{*}\hspace*{-0.1em})$ for $p,p^{*}\in[\underline{p},\overline{p}]$.
Then, there exists $\gamma\hspace*{-0.3em}=\hspace*{-0.3em}\mathrm{Col}(\gamma_{1},\cdots,\gamma_{a})$ with
$\gamma_{i}\hspace*{-0.3em}\geq\hspace*{-0.3em}0$, such that
\begin{eqnarray}
\left[
\begin{array}{c}
\hspace*{-0.3em}\Delta p\hspace*{-0.3em}  \\
\hspace*{-0.3em}\Delta m\hspace*{-0.3em}
\end{array}%
\right]^{\hspace*{-0.2em}\top}\hspace*{-0.4em}
\underbrace{\left[
\begin{array}{cc}
\hspace*{-0.3em}M_{\sigma}  & -I_{a}\hspace*{-0.3em}\\
\hspace*{-0.3em}-M_{\rho}  & I_{a}\hspace*{-0.3em}
\end{array}%
\right]^{\hspace*{-0.2em}\top}\hspace*{-0.4em}\left[
\begin{array}{cc}
\hspace*{-0.3em}\mathbf{0}_{a}  & M_{\gamma}\hspace*{-0.3em}\\
\hspace*{-0.3em}M_{\gamma}  & \mathbf{0}_{a}\hspace*{-0.3em}
\end{array}%
\right]\hspace*{-0.4em}\left[
\begin{array}{cc}
\hspace*{-0.3em}M_{\sigma}  &-I_{a}\hspace*{-0.3em}\\
\hspace*{-0.3em}-M_{\rho}  & I_{a}\hspace*{-0.3em}
\end{array}%
\right]}_{\triangleq\mathcal{M}_{\mathrm{DNN}}}\hspace*{-0.4em}\left[
\begin{array}{c}
\hspace*{-0.3em}\Delta p\hspace*{-0.3em}  \\
\hspace*{-0.3em}\Delta m\hspace*{-0.3em}
\end{array}%
\right] & \hspace*{-0.1em}\geq\hspace*{-0.1em} & 0 \label{dnnqc}
\end{eqnarray}
holds for $m\hspace*{-0.2em}=\hspace*{-0.2em}\varphi(p)$,
$\rho,\sigma,\underline{p},\overline{p},p^{*}\hspace*{-0.2em}\in\hspace*{-0.2em}\mathds{R}^{a}$ with element-wise
$\rho\hspace*{-0.2em}\leq\hspace*{-0.2em}\sigma$ and $\underline{p}\hspace*{-0.2em}\leq\hspace*{-0.2em}\overline{p}$, where
$\Delta p\hspace*{-0.2em}\triangleq\hspace*{-0.2em}p\hspace*{-0.2em}-\hspace*{-0.2em}p^{*}$,
$\Delta m\hspace*{-0.2em}\triangleq\hspace*{-0.2em}m\hspace*{-0.2em}-\hspace*{-0.2em}m^{*}$, and
\begin{eqnarray}
M_{\sigma}\hspace*{-0.1em}& \hspace*{-0.1em}\triangleq\hspace*{-0.1em} &\hspace*{-0.1em}
\mathrm{Diag}(\sigma_{1},\sigma_{2},\cdots,\sigma_{a}),
M_{\rho}\hspace*{-0.3em}\triangleq\hspace*{-0.3em}
\mathrm{Diag}(\rho_{1},\rho_{2},\cdots,\rho_{a}), \notag\\
M_{\gamma}\hspace*{-0.1em}& \hspace*{-0.1em}\triangleq\hspace*{-0.1em} &\hspace*{-0.1em}
\mathrm{Diag}(\gamma_{1},\gamma_{2},\cdots,\gamma_{a}).
\end{eqnarray}
\end{lemma}
\textbf{Proof}\ Based on the definitions of $\Delta p$ and $\Delta m$, the offset local sector bound $\mathrm{Sec}[\rho,\sigma]$ for the activation
nonlinearity $\varphi$ implies that there exist scalars  $\gamma_{i}\hspace*{-0.2em}\geq\hspace*{-0.2em}0,i\hspace*{-0.3em}\in\hspace*{-0.3em}\mathds{N}_{[1,a]}$, which yields the
condition $\gamma_{i}(\Delta m_{i}\hspace*{-0.2em}-\hspace*{-0.2em}\rho_{i}\Delta p_{i})(\sigma_{i}\Delta
p_{i}\hspace*{-0.2em}-\hspace*{-0.2em}\Delta m_{i})\hspace*{-0.2em}\geq\hspace*{-0.2em}0$. Then summing up this inequality from
$i\hspace*{-0.2em}=\hspace*{-0.2em}1$ to $i\hspace*{-0.2em}=\hspace*{-0.2em}a$ leads to \eqref{dnnqc}. The proof is
along the same line as \citep*{Fazlyab2022Safety}, and more details are omitted.
\begin{remark}
\label{ibp}
Before incorporating the quadratic constraint of $\pi_{\mathrm{DNN}}\hspace*{-0.2em}$ in \eqref{dnnqc} into the closed-loop stability analysis, leading to an inner approximation of robust RoA, we need to achieve the bounds $\underline{p},\overline{p}\hspace*{-0.2em}\in\hspace*{-0.2em}\mathds{R}^{a}$ of $p$, starting from the counterpart of the first layer
$p_{1}\hspace*{-0.3em}\in\hspace*{-0.3em}[\underline{p}_{1}\hspace*{-0.1em},\hspace*{-0.1em}\overline{p}_{1}]$. Then, the bounds
$\underline{m}_{1}$ and $\overline{m}_{1}$ can be derived by $m_{1}(k)\hspace*{-0.3em}=\hspace*{-0.3em}\varphi_{1}(p_{1}(k))$, hence to calculate
the bounds $\hspace*{-0.2em}[\underline{p}_{2}\hspace*{-0.1em},\hspace*{-0.1em}\overline{p}_{2}]\hspace*{-0.2em}$ of the preactivation $\hspace*{-0.1em}p_{2}$. By the interval bound propagation (IBP)
\citep*[Section 3]{Gowai2018On}, such calculation process is propagated through all layers of a trained $\pi_{\mathrm{DNN}}$. More explicitly, the interval bounds on $p_{1}$ can be symmetrical about $p_{1}^{*}$ with
$\underline{p}_{1}\hspace*{-0.4em}=\hspace*{-0.3em}2p_{1}^{*}\hspace*{-0.2em}-\hspace*{-0.2em}\overline{p}_{1}$, which are taken into account
in what follows.
\end{remark}
We assume that the DNN controller is with zero-bias terms, such that the augmented state
$\eta$ can converge to the (local) equilibrium point $\eta^{*}$, which implies that the setpoint tracking error $\eta\hspace*{-0.2em}-\hspace*{-0.2em}\eta^{*}$ can 
converge to the original point. Within this context, zero bias in conjunction with $\mu(0)=0$ can suit for the special case of
$\bar{u}^{*}\hspace*{-0.2em}=\hspace*{-0.2em}\mathbf{0}_{m}$ and $\eta^{*}\hspace*{-0.2em}=\hspace*{-0.2em}\mathbf{0}_{\bar{n}}$, as discussed in
\citep*[Assumption 1]{Yin2022Imitation}.

\subsection{Closed-Loop Stability Analysis}
\label{sec33}

For a sequence of time instants $\{\theta_{j,q}\}$, a function $H(\eta,k)$ mapping $\mathds{R}^{\bar{n}}\hspace*{-0.2em}\times\hspace*{-0.2em}
\mathds{N}_{[\theta_{j,q},\theta_{j\hspace*{-0.1em}+\hspace*{-0.1em}1,q}\hspace*{-0.1em}-1]}$ to $\mathds{R}$ is called a looped function, if
$H(\eta(\theta_{j\hspace*{-0.1em}+\hspace*{-0.1em}1,q}),\theta_{j\hspace*{-0.1em}+\hspace*{-0.1em}1,q})
\hspace*{-0.3em}=\hspace*{-0.3em}H(\eta(\theta_{j,q}),\theta_{j,q})$ holds for $\hspace*{-0.1em}j\in\mathds{N}_{[0,h(q)]}$. Based on the
looped function defined in discrete time, the following lemma plays an important role in deriving the closed-loop stability guarantees in this subsection.
\begin{lemma}
\label{lf}
\hspace*{-0.4em}
For scalars $\mu_{1}\hspace*{-0.2em}<\hspace*{-0.2em}\mu_{2}$ with $\mu_{1},\mu_{2}\hspace*{-0.2em}\in\hspace*{-0.2em}\mathds{R}_{>0}$, and a
differentiable value function $V(\eta):\mathds{R}^{\varsigma}\rightarrow\mathds{R}_{\geq0}$
for $\eta\hspace*{-0.2em}\in\hspace*{-0.2em}\mathds{R}^{\varsigma}$, satisfying $\mu_{1}\|\eta\|^{2}_{2}\hspace*{-0.2em}\leq\hspace*{-0.2em}
V(\eta)\hspace*{-0.2em}\leq\hspace*{-0.2em}\mu_{2}\|\eta\|^{2}_{2}$, the following two statements are equivalent. \newline
(i) The existence of looped function $H(\eta(k),k)$ ensures that
$\Delta\mathcal{V}(k)\hspace*{-0.2em}\triangleq\hspace*{-0.2em}
\mathcal{V}(k\hspace*{-0.2em}+\hspace*{-0.2em}1)\hspace*{-0.2em}-\hspace*{-0.2em}\mathcal{V}(k)
\hspace*{-0.2em}<\hspace*{-0.2em}0$ holds for
$k\hspace*{-0.3em}\in\hspace*{-0.3em}[\theta_{j,q},\theta_{j\hspace*{-0.1em}+\hspace*{-0.1em}1,q}\hspace*{-0.2em}-\hspace*{-0.2em}1]$, non-zero
$\eta(\theta_{j,q})$ and $\{\theta_{j,q}\}$, where  $\mathcal{V}\hspace*{-0.1em}(k)\hspace*{-0.2em}\triangleq\hspace*{-0.2em}V\hspace*{-0.1em}(\eta(k))\hspace*{-0.2em}+\hspace*{-0.2em}H(\eta(k),k)$. \newline
(ii) The increment of $V(\eta)$ satisfies $\Delta V\hspace*{-0.3em}\triangleq\hspace*{-0.3em}V(\eta(\hspace*{-0.1em}\theta_{j\hspace*{-0.1em}+\hspace*{-0.1em}1,q}\hspace*{-0.1em})\hspace*{-0.1em})
\hspace*{-0.3em}-\hspace*{-0.3em}V(\eta(\hspace*{-0.1em}\theta_{j,q}\hspace*{-0.1em})\hspace*{-0.1em})\hspace*{-0.3em}<\hspace*{-0.3em}0$ with non-zero
$\eta(\theta_{j,q})$ for a sequence $\{\theta_{j,q}\}$.
\end{lemma}
\textbf{Proof}\
We
sum up $\Delta\mathcal{V}(k)$ over the interval $[\theta_{j,q},\theta_{j\hspace*{-0.1em}+\hspace*{-0.1em}1,q}\hspace*{-0.2em}-1]$ and take advantage of the definition of looped function as aforementioned. Allowing for $\Delta\mathcal{V}(k)<0$ satisfying the first statement, we obtain
\begin{eqnarray}
\sum_{k=\theta_{j,q}}^{\theta_{j+1,q}-1}\Delta\mathcal{V}(k)&\hspace*{-0.2em}=\hspace*{-0.2em}&
\sum_{k=\theta_{j,q}}^{\theta_{j+1,q}-1}\left(V(\eta(k\hspace*{-0.2em}+\hspace*{-0.2em}1)\hspace*{-0.2em}-\hspace*{-0.2em}V(\eta(k))\right)+ \notag \\
&\hspace*{-0.2em}\hspace*{-0.2em}& \sum_{k=\theta_{j,q}}^{\theta_{j+1,q}-1}\left(H(x(k\hspace*{-0.2em}+\hspace*{-0.2em}1),k\hspace*{-0.2em}+\hspace*{-0.2em}1)\hspace*{-0.2em}-\hspace*{-0.2em}H(x(k),k))\right) \notag \\
&\hspace*{-0.2em}=\hspace*{-0.2em}&
V(\eta(\hspace*{-0.1em}\theta_{j\hspace*{-0.1em}+\hspace*{-0.1em}1,q}\hspace*{-0.1em})\hspace*{-0.1em})
\hspace*{-0.3em}-\hspace*{-0.3em}V(\eta(\hspace*{-0.1em}\theta_{j,q}\hspace*{-0.1em}))\hspace*{-0.2em}=\hspace*{-0.2em}\Delta V\hspace*{-0.2em}<\hspace*{-0.2em}0, \notag
\end{eqnarray}
which can prove that the second statement holds. \newline
In contrast, we suppose the second statement is satisfied and construct the looped function
$H(\eta(k),k)=-V(\eta(k))+k\Delta V/(\theta_{j+1,q}-\theta_{j,q})$ for
$k\hspace*{-0.2em}\in\hspace*{-0.2em}[\theta_{j,q},\theta_{j\hspace*{-0.1em}+\hspace*{-0.1em}1,q}\hspace*{-0.2em}-\hspace*{-0.2em}1]$, which satisfies $H(\eta(\theta_{j\hspace*{-0.1em}+\hspace*{-0.1em}1,q}),\theta_{j\hspace*{-0.1em}+\hspace*{-0.1em}1,q})
\hspace*{-0.3em}=\hspace*{-0.3em}H(\eta(\theta_{j,q}),\theta_{j,q})$. Hence, we calculate
\begin{eqnarray}
\Delta\mathcal{V}(k)=\left(V(\eta(\hspace*{-0.1em}\theta_{j\hspace*{-0.1em}+\hspace*{-0.1em}1,q}\hspace*{-0.1em})\hspace*{-0.1em})
\hspace*{-0.3em}-\hspace*{-0.3em}V(\eta(\hspace*{-0.1em}\theta_{j,q}\hspace*{-0.1em})\right)/(\theta_{j+1,q}-\theta_{j,q})<0, \notag
\end{eqnarray}
which can prove that the first statement holds.

\begin{remark}
\label{loop}
Lemma \ref{lf} provides a methodology of verifying the closed-loop stability of the event-triggered neural-feedback loops \eqref{aug} based on a looped function
$H(x(k),k)$, the value of which is not restricted to be positively definite, implying less conservatism of stability
analysis. Within this context, the Lyapunov tracking loss $V(\eta)$ should only decrease at the communication instant $k_{q}$, rather than
at each time instant $k$. Thus, the core of verifying closed-loop stability lies in selecting proper looped functions combined with the
specified event-triggered logic, which motivates the following theorem.
\end{remark}
\begin{theorem}
\label{th1}
For specified scalars $1\hspace*{-0.2em}\leq\hspace*{-0.2em}\vartheta_{l}\hspace*{-0.2em}<\hspace*{-0.2em}\vartheta_{u}$,
$\epsilon_{1},\epsilon_{2}\hspace*{-0.2em}\in\hspace*{-0.2em}\mathds{R}_{[0,1]}$, and $\mu,g\hspace*{-0.2em}\geq\hspace*{-0.2em}0$ satisfying
$1\hspace*{-0.3em}-\hspace*{-0.3em}\mu\hspace*{-0.3em}-\hspace*{-0.3em}g^{-1}\hspace*{-0.3em}\geq\hspace*{-0.3em}0$,
the neural-feedback loops \eqref{aug} asymptotically converge to the equilibrium point
$(\eta^{*}\hspace*{-0.1em},\hspace*{-0.1em}u^{*}\hspace*{-0.1em},\hspace*{-0.1em}\omega^{*}\hspace*{-0.1em},
\hspace*{-0.1em}\nu^{*}\hspace*{-0.1em},\hspace*{-0.1em}p^{*}\hspace*{-0.1em},\hspace*{-0.1em}m^{*})$ under the event-triggered communication scheme  \eqref{ets1}, where $\alpha(\theta_{j,q})$ converges to the origin with $\alpha(0)\hspace*{-0.2em}\geq\hspace*{-0.2em}0$,
if there exist matrices $P\hspace*{-0.2em}\succ\hspace*{-0.2em}0$, $T_{\imath}\hspace*{-0.2em}\succ\hspace*{-0.2em}0$,
$\Xi_{\imath}\hspace*{-0.2em}\succ\hspace*{-0.2em}0$, $R,N_{\imath}$, and a vector $\gamma\hspace*{-0.2em}\in\hspace*{-0.2em}\mathds{R}^{a}$ with
element-wise $\gamma_{i}\hspace*{-0.3em}\geq\hspace*{-0.3em}0$ as decision variables, such that the following conditions
\begin{eqnarray}
\left[
\begin{array}{cc}
\hspace*{-0.1em}\mathcal{G}\hspace*{-0.2em}+\hspace*{-0.2em}\vartheta\amalg_{\imath}&\vartheta N_{\imath}\hspace*{-0.1em}\\
\hspace*{-0.1em}\star\hspace*{-0.1em}&\hspace*{-0.1em}-\vartheta \mathcal{T}_{\imath}\hspace*{-0.1em}
\end{array}%
\right]\hspace*{-0.1em} & \hspace*{-0.1em}\prec\hspace*{-0.1em} &%
\hspace*{-0.1em}0, \hspace*{-0.3em}\left[
\begin{array}{cc}
\hspace*{-0.1em}(\bar{p}_{1,\jmath}\hspace*{-0.2em}-\hspace*{-0.2em}p^{*}_{1,\jmath})^{2}&\left[
\begin{array}{cc}
\hspace*{-0.1em}W_{1,\jmath}
&\mathbf{0}_{1\times\psi}\hspace*{-0.1em}
\end{array}%
\right]\hspace*{-0.2em}\\
\hspace*{-0.3em}\star\hspace*{-0.2em}&\hspace*{-0.2em}P\hspace*{-0.2em}
\end{array}%
\hspace*{-0.2em}\right]\hspace*{-0.4em}\succeq
\hspace*{-0.2em}0,\label{ma1}
\end{eqnarray}
hold for
$\underline{p}_{1}\hspace*{-0.2em}=\hspace*{-0.2em}2p_{1}^{*}\hspace*{-0.2em}-\hspace*{-0.2em}\overline{p}_{1}$,
$\imath\hspace*{-0.2em}\in\hspace*{-0.2em}\mathds{N}_{[1,2]}$, $\jmath\hspace*{-0.2em}\in\hspace*{-0.2em}\mathds{N}_{[1,a_{1}]}$, $i\hspace*{-0.2em}\in\hspace*{-0.2em}\mathds{N}_{[1,a]}$, and
$\vartheta_{l}\hspace*{-0.2em}\leq\hspace*{-0.2em}\vartheta\hspace*{-0.2em}\leq\hspace*{-0.2em}\vartheta_{u}$,
where $\mathcal{T}_{\imath}\hspace*{-0.2em}\triangleq\hspace*{-0.2em}\mathrm{Diag}(T_{\imath},3T_{\imath})$,  
\begin{eqnarray*}
\mathcal{G}\hspace*{-0.1em} & \hspace*{-0.1em}\triangleq\hspace*{-0.1em} &\hspace*{-0.1em}
(AL_{1}\hspace*{-0.3em}+\hspace*{-0.3em}B\Sigma_{1}\Lambda_{1})^{\hspace*{-0.2em}\top}\hspace*{-0.2em}
P(AL_{1}\hspace*{-0.3em}+\hspace*{-0.3em}B\Sigma_{1}\Lambda_{1})
+\mathcal{G}_{1}^
{\hspace*{-0.2em}\top}(T_{2}\hspace*{-0.2em}-\hspace*{-0.2em}T_{1})\mathcal{G}_{1} \\
\hspace*{-0.1em} & \hspace*{-0.1em}\hspace*{-0.1em} &\hspace*{-0.1em}+\mathcal{G}_{2}^
{\hspace*{-0.2em}\top}M_{\Theta}\mathcal{G}_{2}\hspace*{-0.2em}+\hspace*{-0.2em}\mathcal{G}_{3}^
{\hspace*{-0.2em}\top}\mathcal{M}_{\mathrm{DNN}}\mathcal{G}_{3}\hspace*{-0.2em}+\hspace*{-0.2em}\mathrm{Sym}( R_{1}^{\hspace*{-0.2em}\top}
RR_{2}\hspace*{-0.2em}-\hspace*{-0.2em} R_{5}^{\hspace*{-0.2em}\top}
RR_{6}\\
\hspace*{-0.1em} & \hspace*{-0.1em}\hspace*{-0.1em} &\hspace*{-0.1em}+N_{1} R_{9}\hspace*{-0.2em}+\hspace*{-0.2em}
N_{2} R_{10})\hspace*{-0.2em}-\hspace*{-0.2em}L_{1}\hspace*{-0.2em}^{\hspace*{-0.2em}\top}PL_{1}
\hspace*{-0.2em}+\hspace*{-0.2em}\mathcal{Q},
\\
\mathcal{Q}\hspace*{-0.1em} & \hspace*{-0.1em}\triangleq\hspace*{-0.1em} &\hspace*{-0.1em}
\epsilon_{1}L_{4}\hspace*{-0.2em}^{\hspace*{-0.2em}\top}\Xi_{1}L_{4}\hspace*{-0.2em}+\hspace*{-0.2em}
\epsilon_{2}L_{8}\hspace*{-0.2em}^{\hspace*{-0.2em}\top}\Xi_{1}L_{8}\hspace*{-0.2em}-\hspace*{-0.2em}
(L_{4}\hspace*{-0.2em}-\hspace*{-0.2em}L_{8})^{\hspace*{-0.2em}\top}\Xi_{2}(L_{4}\hspace*{-0.2em}-\hspace*{-0.2em}L_{8}),
\\
\amalg_{1}\hspace*{-0.1em} & \hspace*{-0.1em}\triangleq\hspace*{-0.1em} &\hspace*{-0.1em}
\mathcal{G}_{1}^{\hspace*{-0.2em}\top}\hspace*{-0.2em}T_{2}\mathcal{G}_{1}\hspace*{-0.3em}+\hspace*{-0.3em}\mathrm{Sym}( R_{3}
\hspace*{-0.2em}^{\hspace*{-0.2em}\top}\hspace*{-0.2em}R R_{7}),
\amalg_{2} \hspace*{-0.2em}\triangleq\hspace*{-0.2em}
\mathcal{G}_{1}^{\hspace*{-0.2em}\top}\hspace*{-0.2em}T_{1}\mathcal{G}_{1}\hspace*{-0.3em}+\hspace*{-0.3em}\mathrm{Sym}( R_{8}
\hspace*{-0.2em}^{\hspace*{-0.2em}\top}\hspace*{-0.2em}R R_{4}),
\\
\hspace*{-0.6em}\Sigma_{1}\hspace*{-0.1em} & \hspace*{-0.1em}\triangleq\hspace*{-0.1em} &\hspace*{-0.1em}
\mathrm{Col}\hspace*{-0.2em}\left(\hspace*{-0.1em}\left[
\begin{array}{ccc}
\hspace*{-0.3em}\Pi_{ux}\hspace*{-0.1em}
&\hspace*{-0.1em}\Pi_{um}\hspace*{-0.1em}
&\hspace*{-0.1em} \mathbf{0}_{m\times w}\hspace*{-0.3em}
\end{array}%
\right]\hspace*{-0.1em},\hspace*{-0.1em}\left[
\begin{array}{ccc}
\hspace*{-0.3em}\mathbf{0}_{w\times n}\hspace*{-0.1em}
&\hspace*{-0.1em}\mathbf{0}_{w\times a}\hspace*{-0.1em}
&\hspace*{-0.1em} I_{w}\hspace*{-0.3em}
\end{array}%
\right]\hspace*{-0.1em}
\right),
\\
\Lambda_{1}\hspace*{-0.1em} & \hspace*{-0.1em}\triangleq\hspace*{-0.1em} &\hspace*{-0.1em}
\mathrm{Col}(L_{8},L_{2},L_{3}), \mathcal{G}_{1}\hspace*{-0.2em}\triangleq\hspace*{-0.2em}
\Lambda_{2}AL_{1}\hspace*{-0.2em}-\hspace*{-0.2em}\Lambda_{2}L_{1}\hspace*{-0.2em}+\hspace*{-0.2em}\Lambda_{2}B\Sigma_{1}\Lambda_{1},
\\
\mathcal{G}_{2}\hspace*{-0.1em} & \hspace*{-0.1em}\triangleq\hspace*{-0.1em} &\hspace*{-0.1em}
CL_{1}\hspace*{-0.2em}+\hspace*{-0.2em}D\Sigma_{1}\Lambda_{1},
\mathcal{G}_{3}\hspace*{-0.2em}\triangleq\hspace*{-0.2em}\mathrm{Col}(\Pi_{px}L_{8}\hspace*{-0.2em}+\hspace*{-0.2em}\Pi_{pm}L_{2},L_{2}),
\\
\Lambda_{2}\hspace*{-0.1em} & \hspace*{-0.1em}\triangleq\hspace*{-0.1em} &\hspace*{-0.1em}
\left[
\begin{array}{cc}
\hspace*{-0.3em}I_{n}\hspace*{-0.1em}
&\hspace*{-0.1em} \mathbf{0}_{n\hspace*{-0.1em}\times\hspace*{-0.1em}\psi}\hspace*{-0.3em}
\end{array}%
\right], R_{1}\hspace*{-0.2em}\triangleq\hspace*{-0.2em}\mathrm{Col}(L_{4},L_{5}, R_{1,3},L_{6}
\hspace*{-0.2em}+\hspace*{-0.3em} R_{1,3}),\\
R_{2}\hspace*{-0.1em} & \hspace*{-0.1em}\triangleq\hspace*{-0.1em} &\hspace*{-0.1em}
\mathrm{Col}(-\hspace*{-0.2em}L_{4},-\hspace*{-0.2em}L_{5}, R_{2,3},L_{7}\hspace*{-0.2em}-\hspace*{-0.2em}L_{5}
\hspace*{-0.2em}-\hspace*{-0.2em}\Lambda_{2}L_{1}),
\\
R_{3}\hspace*{-0.1em} & \hspace*{-0.1em}\triangleq\hspace*{-0.1em} &\hspace*{-0.1em}
\mathrm{Col}(L_{4},L_{5},\mathbf{0}_{n\hspace*{-0.1em}\times\hspace*{-0.1em}\bar{n}},L_{6}),
R_{4}\hspace*{-0.2em}\triangleq\hspace*{-0.2em}\mathrm{Col}(L_{4},L_{5},\mathbf{0}_{n\hspace*{-0.1em}\times\hspace*{-0.1em}
\bar{n}},L_{7}), \\
R_{5}\hspace*{-0.1em} & \hspace*{-0.1em}\triangleq\hspace*{-0.1em} &\hspace*{-0.1em}
\mathrm{Col}(\mathbf{0}_{n\times\bar{n}},\mathbf{0}_{n\times\bar{n}},\Lambda_{2}L_{1}\hspace*{-0.2em}-\hspace*{-0.2em}L_{4},
L_{6}\hspace*{-0.3em}-\hspace*{-0.2em}L_{4}),  R_{7}\hspace*{-0.2em}\triangleq\hspace*{-0.2em} R_{2}
\hspace*{-0.3em}-\hspace*{-0.3em} R_{6}, \\
R_{6}\hspace*{-0.1em} & \hspace*{-0.1em}\triangleq\hspace*{-0.1em} &\hspace*{-0.1em}
\mathrm{Col}(\mathbf{0}_{n\hspace*{-0.1em}\times\hspace*{-0.1em}\bar{n}},
\mathbf{0}_{n\hspace*{-0.1em}\times\hspace*{-0.1em}\bar{n}},L_{5}\hspace*{-0.3em}-\hspace*{-0.3em}\Lambda_{2}L_{1},
L_{7}\hspace*{-0.2em}-\hspace*{-0.2em}L_{5}), R_{8}\hspace*{-0.2em}\triangleq\hspace*{-0.2em} R_{1}
\hspace*{-0.3em}-\hspace*{-0.3em} R_{5}, \\
R_{9}\hspace*{-0.1em} & \hspace*{-0.1em}\triangleq\hspace*{-0.1em} &\hspace*{-0.1em}
\mathrm{Col}(\Lambda_{2}L_{1}\hspace*{-0.2em}-\hspace*{-0.2em}L_{4},\Lambda_{2}L_{1}\hspace*{-0.2em}+\hspace*{-0.2em}L_{4}
\hspace*{-0.2em}-\hspace*{-0.2em}L_{6}), \bar{n}\hspace*{-0.2em}\triangleq\hspace*{-0.2em}z
\hspace*{-0.2em}+\hspace*{-0.2em}a\hspace*{-0.2em}+\hspace*{-0.2em}w\hspace*{-0.2em}+\hspace*{-0.2em}5n,
\\
R_{10}\hspace*{-0.1em} & \hspace*{-0.1em}\triangleq\hspace*{-0.1em} &\hspace*{-0.1em}
\mathrm{Col}(L_{5}\hspace*{-0.2em}-\hspace*{-0.2em}\Lambda_{2}L_{1},L_{5}\hspace*{-0.2em}+\hspace*{-0.2em}\Lambda_{2}L_{1}
\hspace*{-0.2em}-\hspace*{-0.2em}L_{7}), L_{1}\hspace*{-0.2em}\triangleq\hspace*{-0.2em}\left[
\begin{array}{cc}
\hspace*{-0.3em}I_{z}\hspace*{-0.1em}
&\hspace*{-0.1em} \mathbf{0}_{z\hspace*{-0.1em}\times\hspace*{-0.1em}(\bar{n}-z)}\hspace*{-0.3em}
\end{array}%
\right]\hspace*{-0.2em},\\
R_{1,3}\hspace*{-0.1em} & \hspace*{-0.1em}\triangleq\hspace*{-0.1em} &\hspace*{-0.1em}
\Lambda_{2}AL_{1}\hspace*{-0.2em}+\hspace*{-0.2em}\Lambda_{2}B\Sigma_{1}\Lambda_{1}\hspace*{-0.2em}-\hspace*{-0.2em}L_{4},
L_{2}\hspace*{-0.2em}\triangleq\hspace*{-0.2em}\left[
\begin{array}{ccc}
\hspace*{-0.3em}\mathbf{0}_{a\hspace*{-0.1em}\times\hspace*{-0.1em}z}\hspace*{-0.1em}
&\hspace*{-0.1em}I_{a}\hspace*{-0.1em}
&\hspace*{-0.1em} \mathbf{0}_{a\hspace*{-0.1em}\times\hspace*{-0.1em}(\bar{n}\hspace*{-0.1em}-\hspace*{-0.1em}
z\hspace*{-0.1em}-\hspace*{-0.1em}a)}\hspace*{-0.3em}
\end{array}%
\right]\hspace*{-0.2em},\\
 R_{2,3}\hspace*{-0.1em} & \hspace*{-0.1em}\triangleq\hspace*{-0.1em} &\hspace*{-0.1em}
L_{5}\hspace*{-0.2em}-\hspace*{-0.2em}\Lambda_{2}AL_{1}\hspace*{-0.2em}-\hspace*{-0.2em}\Lambda_{2}B\Sigma_{1}\Lambda_{1},
L_{3}\hspace*{-0.2em}\triangleq\hspace*{-0.2em}
\left[
\begin{array}{ccc}
\hspace*{-0.3em}\mathbf{0}_{w\hspace*{-0.1em}\times\hspace*{-0.1em}(z\hspace*{-0.1em}+\hspace*{-0.1em}a})\hspace*{-0.1em}
&\hspace*{-0.1em}I_{w}\hspace*{-0.1em}
&\hspace*{-0.1em} \mathbf{0}_{w\hspace*{-0.1em}\times\hspace*{-0.1em}5n}\hspace*{-0.3em}
\end{array}%
\right]\hspace*{-0.2em},\\
L_{j}\hspace*{-0.1em} & \hspace*{-0.1em}\triangleq\hspace*{-0.1em} &\hspace*{-0.1em}
\left[
\begin{array}{cccc}
\hspace*{-0.3em}\mathbf{0}_{n\hspace*{-0.1em}\times\hspace*{-0.1em}(z\hspace*{-0.1em}+\hspace*{-0.1em}a\hspace*{-0.1em}+
\hspace*{-0.1em}w})\hspace*{-0.1em}
&\hspace*{-0.1em}\mathbf{0}_{n\hspace*{-0.1em}\times\hspace*{-0.1em}(j\hspace*{-0.1em}-\hspace*{-0.1em}4)n}\hspace*{-0.1em}
&\hspace*{-0.1em} I_{n}\hspace*{-0.1em}
&\hspace*{-0.1em} \mathbf{0}_{n\hspace*{-0.1em}\times\hspace*{-0.1em}(8\hspace*{-0.1em}-\hspace*{-0.1em}j)n}\hspace*{-0.3em}
\end{array}%
\right]\hspace*{-0.2em}, j\hspace*{-0.2em}\in\hspace*{-0.2em}\mathds{N}_{[4,8]}.
\end{eqnarray*}
 $W_{1,\jmath}$ denotes the $\jmath$-th row of the weight matrix $W_{1}$. Then an ellipsoid
$\mathcal{E}_{\hspace*{-0.1em}P_{1}}(x^{*}\hspace*{-0.1em})\hspace*{-0.2em}\triangleq\hspace*{-0.2em}
\{x\hspace*{-0.2em}\in\hspace*{-0.2em}\mathds{R}^{n}|\|x\hspace*{-0.2em}-\hspace*{-0.2em}x^{*}\|^{2}_{P_{1}}
\hspace*{-0.3em}\leq\hspace*{-0.3em}1\}$, where $P_{1}$ denotes the upper left block of
$P$ with the proper dimension, implies an inner approximation of RoA when $\alpha(0)\hspace*{-0.2em}=\hspace*{-0.2em}0$.
\end{theorem}
\textbf{Proof}\
We recall the neural-feedback loops \eqref{aug} and incorporate the dynamic variable
$\alpha(\theta_{j,q})$ to the closed-loop stability analysis. For
$k\hspace*{-0.2em}\in\hspace*{-0.2em}[\theta_{j,q},\theta_{j\hspace*{-0.1em}+\hspace*{-0.1em}1,q}\hspace*{-0.2em}-\hspace*{-0.2em}1]$
with $j\hspace*{-0.2em}\in\hspace*{-0.2em}\mathds{N}_{[0,h(q)]}$ and $q\hspace*{-0.2em}\in\hspace*{-0.2em}\mathds{N}$, we select the value
function with
\begin{equation}
\label{lya1}
\begin{array}{rcl}
\mathcal{W}(\eta(k),k)& \hspace*{-0.2em}=\hspace*{-0.2em} &
V(\eta(k))\hspace*{-0.2em}+\hspace*{-0.2em}H(\eta(k),k)\hspace*{-0.2em}+\hspace*{-0.2em}\alpha(k),
\end{array}
\end{equation}
in which $V(\eta(k))\hspace*{-0.3em}\triangleq\hspace*{-0.3em}\|\eta(k)\hspace*{-0.3em}-\hspace*{-0.3em}\eta^{*}\|_{P}^{2}$ with
$p\hspace*{-0.2em}\succ\hspace*{-0.2em}0$, $\alpha(k)$ in \eqref{as3} is nonnegative in view of Remark 1. The looped
function satisfies $H(\eta(k),k)\hspace*{-0.3em}\triangleq\hspace*{-0.3em}\sum_{\kappa=1}^{3}H_{\kappa}(k)$ with the form
\begin{eqnarray}
H_{1}(k)\hspace*{-0.1em}& \hspace*{-0.1em}\triangleq\hspace*{-0.1em} &\hspace*{-0.1em}
2\chi_{1}^{\top}(k)R\chi_{2}(k), \\
H_{2}(k)\hspace*{-0.1em}& \hspace*{-0.1em}\triangleq\hspace*{-0.1em} &\hspace*{-0.1em}
(\theta_{j+1,q}\hspace*{-0.3em}-\hspace*{-0.3em}k)\hspace*{-0.3em}\left(\scalebox{1}{\(\sum\limits\)}_{s=\theta_{j,q}}^{k}\|y(s)\|_{T_{1}}^{2}
\hspace*{-0.3em}-\hspace*{-0.3em}
\|y(k)\|_{T_{1}}^{2}\hspace*{-0.2em}\right)\hspace*{-0.3em},\\
H_{3}(k)\hspace*{-0.1em}& \hspace*{-0.1em}\triangleq\hspace*{-0.1em} &\hspace*{-0.1em}
(k\hspace*{-0.3em}-\hspace*{-0.3em}\theta_{j,q})\hspace*{-0.3em}\left(\|y(\theta_{j+1,q})
\|_{T_{2}}^{2}\hspace*{-0.3em}-\hspace*{-0.3em}\scalebox{1}{\(\sum\limits\)}_{s=k}^{\theta_{j+1,q}}
\|y(s)\|_{T_{2}}^{2}\hspace*{-0.2em}\right)\hspace*{-0.3em},
\end{eqnarray}
where $R\hspace*{-0.2em}\in\hspace*{-0.2em}\mathds{R}^{4n\hspace*{-0.1em}\times\hspace*{-0.1em}4n}$, $T_{\imath}\hspace*{-0.2em}\succ\hspace*{-0.2em}0$ with $\imath\hspace*{-0.2em}\in\hspace*{-0.2em}\mathds{N}_{[1,2]}$, and
\begin{eqnarray*}
\chi_{1}\hspace*{-0.1em}& \hspace*{-0.1em}\triangleq\hspace*{-0.1em} &\hspace*{-0.1em}
\mathrm{Col}\left((k\hspace*{-0.3em}-\hspace*{-0.3em}\theta_{j,q})m_{0},
x_{k}\hspace*{-0.3em}-\hspace*{-0.3em} x_{\theta_{j,q}},\scalebox{1}{\(\sum\limits\)}_{s=\theta_{j,q}}^{k}\hspace*{-0.3em}
x_{s}\hspace*{-0.3em}-\hspace*{-0.3em}x_{\theta_{j,q}}\right), \\
\chi_{2}\hspace*{-0.1em}& \hspace*{-0.1em}\triangleq\hspace*{-0.1em} &\hspace*{-0.1em}
\mathrm{Col}\left((\theta_{j+1,q}\hspace*{-0.3em}-\hspace*{-0.3em}k)m_{0},x_{\theta_{j\hspace*{-0.1em}+\hspace*{-0.1em}1,q}}
\hspace*{-0.3em}-\hspace*{-0.3em} x_{k},\scalebox{1}{\(\sum\limits\)}_{s=k}^{\theta_{j\hspace*{-0.1em}+\hspace*{-0.1em}1,q}}
x_{s}\hspace*{-0.3em}-\hspace*{-0.3em} x_{\theta_{j+1,q}}\right),\\
m_{0}\hspace*{-0.1em}& \hspace*{-0.1em}\triangleq\hspace*{-0.1em} &\hspace*{-0.1em}
\mathrm{Col}(x_{\theta_{j,q}},x_{\theta_{j\hspace*{-0.1em}+\hspace*{-0.1em}1,q}}),
x_{k}\hspace*{-0.2em}\triangleq\hspace*{-0.2em}x(k)\hspace*{-0.2em}-\hspace*{-0.2em}x^{*},
x_{\theta_{j,q}}\hspace*{-0.2em}\triangleq\hspace*{-0.2em}x(\theta_{j,q})\hspace*{-0.2em}-\hspace*{-0.2em}x^{*}, \\
x_{s}\hspace*{-0.1em}& \hspace*{-0.1em}\triangleq\hspace*{-0.1em} &\hspace*{-0.1em}
x(s)\hspace*{-0.2em}-\hspace*{-0.2em}x^{*}, x_{\theta_{j\hspace*{-0.1em}+\hspace*{-0.1em}1,q}}
\hspace*{-0.2em}\triangleq\hspace*{-0.2em}x(\theta_{j\hspace*{-0.1em}+\hspace*{-0.1em}1,q})\hspace*{-0.2em}-\hspace*{-0.2em}x^{*},
y(s)\hspace*{-0.2em}=\hspace*{-0.2em}x_{s\hspace*{-0.1em}+\hspace*{-0.1em}1}\hspace*{-0.2em}-\hspace*{-0.2em}x_{s}.
\end{eqnarray*}
Accordingly, the forward difference for each item in \eqref{lya1} can be calculated by
$\Delta\alpha(k)\hspace*{-0.2em}=\hspace*{-0.2em}\alpha(k\hspace*{-0.2em}+\hspace*{-0.2em}1)\hspace*{-0.2em}-\hspace*{-0.2em}\alpha(k)$, and
\begin{eqnarray}
 \Delta V\hspace*{-0.1em}& \hspace*{-0.1em}=\hspace*{-0.1em} &\hspace*{-0.1em}
\zeta^{\top}(k)((AL_{1}\hspace*{-0.3em}+\hspace*{-0.3em}B\Sigma_{1}\Lambda_{1})^{\hspace*{-0.2em}\top}\hspace*{-0.2em}
P(AL_{1}\hspace*{-0.3em}+\hspace*{-0.3em}B\Sigma_{1}\Lambda_{1}) \notag\\
\hspace*{-0.6em} \hspace*{-0.3em}& \hspace*{-0.6em}\hspace*{-0.6em} &\hspace*{-0.3em}
-L_{1}\hspace*{-0.2em}^{\hspace*{-0.2em}\top}PL_{1})\zeta(k), \notag \\
\Delta H_{1}\hspace*{-0.1em}& \hspace*{-0.1em}=\hspace*{-0.1em} &\hspace*{-0.1em}
\zeta^{\top}(k)\mathrm{Sym}(R_{1}^{\hspace*{-0.2em}\top}
RR_{2}\hspace*{-0.2em}-\hspace*{-0.2em}R_{5}^{\hspace*{-0.2em}\top}
RR_{6}\hspace*{-0.2em}+\hspace*{-0.2em}(k\hspace*{-0.2em}-\hspace*{-0.2em}\theta_{j,q})
R_{3}^{\hspace*{-0.2em}\top}RR_{7} \notag\\
\hspace*{-0.6em} \hspace*{-0.3em}& \hspace*{-0.6em}\hspace*{-0.6em} &\hspace*{-0.3em}
+(\theta_{j+1,q}\hspace*{-0.2em}-\hspace*{-0.2em}k)
R_{8}^{\hspace*{-0.2em}\top}RR_{4})\zeta(k), \notag \\
\Delta H_{2}\hspace*{-0.1em}& \hspace*{-0.1em}=\hspace*{-0.1em} &\hspace*{-0.1em}
(\theta_{j\hspace*{-0.1em}+\hspace*{-0.1em}1,q}\hspace*{-0.3em}-\hspace*{-0.3em}k\hspace*{-0.3em}-\hspace*{-0.3em}1)
\zeta^{\hspace*{-0.1em}\top}\hspace*{-0.3em}(k)\mathcal{G}^{\top}_{1}T_{1}\mathcal{G}_{1}\zeta(k)
\hspace*{-0.3em}-\hspace*{-0.3em}\scalebox{1}{\(\sum\limits\)}_{s=\theta_{j,q}}^{k-1}\hspace*{-0.2em}\|y(s)\|_{T_{1}}^{2}, \notag \\
\Delta H_{3}\hspace*{-0.1em}& \hspace*{-0.1em}=\hspace*{-0.1em} &\hspace*{-0.1em}
(k\hspace*{-0.3em}+\hspace*{-0.3em}1\hspace*{-0.3em}-\hspace*{-0.3em}\theta_{j,q})
\zeta^{\hspace*{-0.1em}\top}\hspace*{-0.3em}(k)\mathcal{G}^{\top}_{1}\hspace*{-0.2em}T_{2}\mathcal{G}_{1}\zeta(k)\hspace*{-0.3em}-\hspace*{-0.3em}
\scalebox{1}{\(\sum\limits\)}_{s=k}^{\theta_{j\hspace*{-0.1em}+\hspace*{-0.1em}1,q}-\hspace*{-0.1em}1}\hspace*{-0.2em}\|y(s)\|_{T_{2}}^{2}, \label{81}
\end{eqnarray}
in which $\zeta(k)\hspace*{-0.2em}\triangleq\hspace*{-0.2em}\mathrm{Col}(\eta_{k},m_{k},\omega_{k},x_{\theta_{j,q}},x_{\theta_{j
\hspace*{-0.1em}+\hspace*{-0.1em}1,q}},
\sum_{s=\theta_{j,q}}^{k}\hspace*{-0.3em}x_{s}/(k\hspace*{-0.3em}-\theta_{j,q}\hspace*{-0.2em}+\hspace*{-0.2em}1),
\sum_{s=k}^{\theta_{j+1,q}}\hspace*{-0.3em}x_{s}/(\theta_{j+1,q}\hspace*{-0.3em}-\hspace*{-0.3em}k\hspace*{-0.2em}+\hspace*{-0.2em}1),x_{k_{q}})
\hspace*{-0.3em}\in\hspace*{-0.3em}\mathds{R}^{\bar{n}}$ with $\eta_{k}\hspace*{-0.2em}\triangleq\hspace*{-0.2em}\eta(k)
\hspace*{-0.2em}-\hspace*{-0.2em}\eta^{*}$, $m_{k}\hspace*{-0.2em}\triangleq\hspace*{-0.2em}m(k)
\hspace*{-0.2em}-\hspace*{-0.2em}m^{*}$, $\omega_{k}\hspace*{-0.2em}\triangleq\hspace*{-0.2em}\omega(k)
\hspace*{-0.2em}-\hspace*{-0.2em}\omega^{*}$, and $x_{k_{q}}\hspace*{-0.2em}\triangleq\hspace*{-0.2em}x(k_{q})
\hspace*{-0.2em}-\hspace*{-0.2em}x^{*}$. By virtue of the nonlinearity isolation of $\pi_{\mathrm{DNN}}$ in \eqref{dnn3}, we have
\begin{equation}
\label{uu}
\begin{array}{rcl}
\bar{u}(k)\hspace*{-0.2em}-\hspace*{-0.2em}\bar{u}^{*}& \hspace*{-0.1em}=\hspace*{-0.1em} &
\Sigma_{1}\mathrm{Col}(x_{k_{q}},m_{k},\omega_{k})\hspace*{-0.2em}=\hspace*{-0.2em}\Sigma_{1}\Lambda_{1}\zeta(k),
\end{array}
\end{equation}
which is utilized for deriving $\Delta V$ in \eqref{81}. Then, in the light of the summation inequality, we can
further relax the summation terms of $\Delta H_{2}(k)$ and $\Delta H_{3}(k)$ by
\begin{eqnarray}
\hspace*{-0.3em}-
\scalebox{1}{\(\sum\limits\)}_{s=\theta_{j,q}}^{k-1}\|y(s)\|_{T_{1}}^{2}\hspace*{-0.1em}& \hspace*{-0.1em}\leq\hspace*{-0.1em} &\hspace*{-0.1em}
(k\hspace*{-0.2em}-\hspace*{-0.2em}\theta_{j,q})\zeta^{\top}(k)N_{1}^{\top}\mathcal{T}_{1}^{-1}N_{1}\zeta(k) \notag \\
\hspace*{-0.3em}\hspace*{-0.1em}& \hspace*{-0.1em}\hspace*{-0.1em} &\hspace*{-0.1em}
+\zeta^{\top}(k)\mathrm{Sym}(N_{1}R_{9})\zeta(k), \label{82}\\
\hspace*{-0.3em}-
\scalebox{1}{\(\sum\limits\)}_{s=k}^{\theta_{j+1,q}-\hspace*{-0.1em}1}\|y(s)\|_{T_{2}}^{2}
\hspace*{-0.1em}& \hspace*{-0.1em}\leq\hspace*{-0.1em} &\hspace*{-0.1em}
(\theta_{j+1,q}\hspace*{-0.2em}-\hspace*{-0.2em}k)\zeta^{\top}(k)N_{2}^{\top}\mathcal{T}_{2}^{-1}N_{2}\zeta(k) \notag \\
\hspace*{-0.3em}\hspace*{-0.1em}& \hspace*{-0.1em}\hspace*{-0.1em} &\hspace*{-0.1em}
+\zeta^{\top}(k)\mathrm{Sym}(N_{2}R_{10})\zeta(k), \label{83}
\end{eqnarray}
with $\mathcal{T}_{\imath}\hspace*{-0.2em}\triangleq\hspace*{-0.2em}\mathrm{Diag}(T_{\imath},3T_{\imath})$. Recall the
dynamic parameter $\alpha(\theta_{j,q})$ in \eqref{as3} and its nonnegativity attribute, we can derive
\begin{eqnarray}
\alpha(\theta_{j+1,q})\hspace*{-0.2em}-\hspace*{-0.2em}\alpha(\theta_{j,q})\hspace*{-0.1em}& \hspace*{-0.1em}=\hspace*{-0.1em} &\hspace*{-0.1em}
-\hspace*{-0.2em}\mu\alpha(\theta_{j,q})\hspace*{-0.2em}+\hspace*{-0.2em}\|\zeta(k)\|_{\mathcal{Q}}^{2}\hspace*{-0.2em}\leq\hspace*{-0.2em}
\|\zeta(k)\|_{\mathcal{Q}}^{2}. \label{84}
\end{eqnarray}
Moreover, combining \eqref{dnn3} with \eqref{dnnqc} leads to
\begin{eqnarray}
\zeta^{\top}(k)\mathcal{G}_{3}^{\top}\mathcal{M}_{\mathrm{DNN}}\mathcal{G}_{3}\zeta(k)\hspace*{-0.2em}\geq\hspace*{-0.2em}0. \label{85}
\end{eqnarray}
Hence, by virtue of \eqref{iqc}, \eqref{81}, and \eqref{82}-\eqref{85} in conjunction with the Lyapunov function $\mathcal{W}(\eta(k),k)$ in
\eqref{lya1}, we obtain
\begin{eqnarray}
&&\Delta \mathcal{W}(\eta(k),k)\hspace*{-0.2em}+\hspace*{-0.2em}\alpha(\theta_{j+1,q})\hspace*{-0.2em}-\hspace*{-0.2em}
\alpha(\theta_{j,q})\hspace*{-0.2em}-\hspace*{-0.2em}\Delta \alpha(k) \notag \\
&\hspace*{-0.8em}\hspace*{-0.8em}&+\|r(k)\hspace*{-0.2em}-\hspace*{-0.2em}r^{*}\|_{M_{\Theta}}^{2}
\hspace*{-0.2em}+\hspace*{-0.2em}\zeta^{\top}(k)\mathcal{G}_{3}^{\top}
\mathcal{M}_{\mathrm{DNN}}\mathcal{G}_{3}\zeta(k)\hspace*{-0.2em}<\hspace*{-0.2em} 0, \label{86}
\end{eqnarray}
which implies that the following inequality
\begin{eqnarray}
\zeta^{\top}(k)\left(\mathcal{G}\hspace*{-0.2em}+\hspace*{-0.2em}
\frac{k\hspace*{-0.2em}-\hspace*{-0.2em}\theta_{j,q}}{\vartheta}(\vartheta\amalg_{1}\hspace*{-0.2em}+\hspace*{-0.1em}\vartheta N_{1}^{\top}\mathcal{T}_{1}^{-1}N_{1})\right.&\hspace*{-0.1em}\hspace*{-0.1em}& \notag \\
+\left.\frac{\theta_{j+1,q}\hspace*{-0.2em}-\hspace*{-0.2em}k}{\vartheta}
(\vartheta\amalg_{2}\hspace*{-0.2em}+\hspace*{-0.1em}\vartheta N_{2}^{\top}\mathcal{T}_{2}^{-1}N_{2})\right)\zeta(k)&\hspace*{-0.1em}<\hspace*{-0.1em}& 0, \label{87}
\end{eqnarray}
holds for $k\hspace*{-0.2em}\in\hspace*{-0.2em}[\theta_{j,q},\theta_{j\hspace*{-0.1em}+\hspace*{-0.1em}1,q}\hspace*{-0.2em}-\hspace*{-0.2em}1]$.
For the coefficients $(k\hspace*{-0.2em}-\hspace*{-0.2em}\theta_{j,q})/\vartheta$ and
$(\theta_{j+1,q}\hspace*{-0.2em}-\hspace*{-0.2em}k)/\vartheta$ belonging to $\mathds{R}_{[0,1)}$,
we can derive that $\mathcal{G}\hspace*{-0.2em}+\hspace*{-0.3em}\vartheta\amalg_{\imath}\hspace*{-0.2em}+\hspace*{-0.1em}\vartheta N_{\imath}^{\top}\mathcal{T}_{\imath}^{-1}N_{\imath}\hspace*{-0.2em}\prec\hspace*{-0.2em}0$ holds for
$\imath\hspace*{-0.2em}\in\hspace*{-0.2em}\mathds{N}_{[1,2]}$. Hence, the first condition in \eqref{ma1} can be established by virtue of Schur complement, which is convex with regard to the decision variables. Note that the sixth term on the left side of \eqref{86} will
be greater than zero, allowing for the local sector quadratic constraint \eqref{dnnqc}. We then 
sum the remainder of \eqref{86} from
$k\hspace*{-0.2em}=\hspace*{-0.2em}\theta_{j,q}$ to $\theta_{j\hspace*{-0.1em}+\hspace*{-0.1em}1,q}\hspace*{-0.2em}-\hspace*{-0.2em}1$,
which leads to 
\begin{eqnarray}
& \hspace*{-0.1em}\hspace*{-0.1em}&
\scalebox{1}{\(\sum\limits\)}_{k=\theta_{j,q}}^{\theta_{j\hspace*{-0.1em}+\hspace*{-0.1em}1,q}\hspace*{-0.1em}-\hspace*{-0.1em}1}
\left(\Delta\mathcal{W}(\eta(k),k)\hspace*{-0.2em}+\hspace*{-0.2em}\alpha(\theta_{j+1,q})\hspace*{-0.2em}-\hspace*{-0.2em}
\alpha(\theta_{j,q})\hspace*{-0.2em}-\hspace*{-0.2em}\Delta\alpha(k)\right)\notag \\
& \hspace*{-0.1em}\hspace*{-0.1em}&
+\scalebox{1}{\(\sum\limits\)}_{k=\theta_{j,q}}^{\theta_{j\hspace*{-0.1em}+\hspace*{-0.1em}1,q}\hspace*{-0.1em}-\hspace*{-0.1em}1}
\|r(k)\hspace*{-0.2em}-\hspace*{-0.2em}r^{*}\|_{M_{\Theta}}^{2} \notag \\
& \hspace*{-0.1em}=\hspace*{-0.1em}&
V(\eta(\theta_{j+1,q}))\hspace*{-0.2em}+\hspace*{-0.2em}(\vartheta\hspace*{-0.3em}-\hspace*{-0.3em}1)
\alpha(\theta_{j\hspace*{-0.1em}+\hspace*{-0.1em}1,q})\hspace*{-0.3em}-\hspace*{-0.3em}V(\eta(\theta_{j,q}))
\hspace*{-0.3em}-\hspace*{-0.3em}(\vartheta\hspace*{-0.3em}-\hspace*{-0.3em}1)\alpha(\theta_{j,q})\notag \\
& \hspace*{-0.1em}\hspace*{-0.1em}&
+\hspace*{-0.1em}
\scalebox{1}{\(\sum\limits\)}_{k=\theta_{j,q}}^{\theta_{j\hspace*{-0.1em}+\hspace*{-0.1em}1,q}\hspace*{-0.1em}-\hspace*{-0.1em}1}
\|r(k)\hspace*{-0.2em}-\hspace*{-0.2em}r^{*}\|^{2}_{M_{\Theta}}
\hspace*{-0.2em}\leq\hspace*{-0.2em}-
\hbar_{1}\|\zeta(\theta_{j,q})\|^{2}\hspace*{-0.3em}<\hspace*{-0.2em}0,
\label{j1}
\end{eqnarray}
holding for the constant sampling interval $\vartheta\in\mathds{N}_{[\vartheta_{l},\vartheta_{u}]}$ and
the looped function $H(x(\theta_{j\hspace*{-0.1em}+\hspace*{-0.1em}1,q}),\theta_{j\hspace*{-0.1em}+\hspace*{-0.1em}1,q})
=H(x(\theta_{j,q}),\theta_{j,q})$. We generalize the time interval from
$[\theta_{j,q},\theta_{j\hspace*{-0.1em}+\hspace*{-0.1em}1,q}\hspace*{-0.2em}-\hspace*{-0.2em}1]$ to $[0,k_{\aleph}]$ with
$\mathcal{\aleph}\hspace*{-0.2em}\in\hspace*{-0.2em}\mathds{N}_{\hspace*{-0.1em}>\hspace*{-0.1em}1}$.
Let $k_{0}\hspace*{-0.2em}=\hspace*{-0.2em}0$ and recap
$\theta_{h(q)\hspace*{-0.1em}+\hspace*{-0.1em}1,q}\hspace*{-0.2em}=\hspace*{-0.2em}k_{q}\hspace*{-0.2em}+\hspace*{-0.2em}
\vartheta(h(q)\hspace*{-0.2em}+\hspace*{-0.2em}1)\hspace*{-0.2em}=\hspace*{-0.2em}k_{q\hspace*{-0.1em}+\hspace*{-0.1em}1}$,
based on the existence of $\hbar_{1}\hspace*{-0.2em}\in\hspace*{-0.2em}\mathds{R}_{>0}$ in \eqref{j1}, we further obtain
\begin{eqnarray}
& \hspace*{-0.1em}\hspace*{-0.1em}&
\scalebox{1}{\(\sum\limits\)}_{q=0}^{\aleph-1}
\scalebox{1}{\(\sum\limits\)}_{j=0}^{h(q)}\left(V(\eta(\theta_{j\hspace*{-0.1em}+\hspace*{-0.1em}1,q})
\hspace*{-0.2em}+\hspace*{-0.2em}(\vartheta\hspace*{-0.2em}-\hspace*{-0.2em}1)\alpha(\theta_{j\hspace*{-0.1em}+\hspace*{-0.1em}1,q})
\hspace*{-0.2em}-\hspace*{-0.2em}V(\eta(\theta_{j,q})) \right. \notag \\
& \hspace*{-0.1em}\hspace*{-0.1em}& \left.-(\vartheta\hspace*{-0.2em}-\hspace*{-0.2em}1)\alpha(\theta_{j,q})\hspace*{-0.2em}+\hspace*{-0.2em}
\scalebox{1}{\(\sum\limits\)}_{k=\theta_{j,q}}^{\theta_{j\hspace*{-0.1em}+\hspace*{-0.1em}1,q}\hspace*{-0.1em}-\hspace*{-0.1em}1}
\|r(k)\hspace*{-0.2em}-\hspace*{-0.2em}r^{*}\|^{2}_{M_{\Theta}}\right) \notag \\
& \hspace*{-0.1em}=\hspace*{-0.1em}&
V(\eta(k_{\aleph}))\hspace*{-0.2em}+\hspace*{-0.2em}(\vartheta\hspace*{-0.2em}-\hspace*{-0.2em}1)\alpha(\theta_{k_{\aleph}})
\hspace*{-0.2em}-\hspace*{-0.2em}V(\eta(0))\hspace*{-0.2em}-\hspace*{-0.2em}
(\vartheta\hspace*{-0.2em}-\hspace*{-0.2em}1)\alpha(0) \notag \\
& \hspace*{-0.1em}\hspace*{-0.1em}&
+\scalebox{1}{\(\sum\limits\)}_{q=0}^{\aleph-1}
\scalebox{1}{\(\sum\limits\)}_{j=0}^{h(q)}
\scalebox{1}{\(\sum\limits\)}_{k=\theta_{j,q}}^{\theta_{j\hspace*{-0.1em}+\hspace*{-0.1em}1,q}\hspace*{-0.1em}-\hspace*{-0.1em}1}
\|r(k)\hspace*{-0.2em}-\hspace*{-0.2em}r^{*}\|^{2}_{M_{\Theta}} \label{r2} \\
& \hspace*{-0.1em}\leq\hspace*{-0.1em}&-
\scalebox{1}{\(\sum\limits\)}_{q=0}^{\aleph-1}
\scalebox{1}{\(\sum\limits\)}_{j=0}^{h(q)}
\hbar_{1}\|\zeta(\theta_{j,q})\|^{2}. \label{r3}
\end{eqnarray}
Note that the last term of \eqref{r2} equals to $\scalebox{1}{\(\sum\limits\)}_{k=0}^{k_{\aleph}}\|r(k)\hspace*{-0.2em}-\hspace*{-0.2em}r^{*}\|^{2}_{M_{\Theta}}$,
which is no less than zero, allowing for the definition of IQC in \eqref{iqc}. By virtue of \eqref{r2} and \eqref{r3}, we obtain 
\begin{eqnarray}
& \hspace*{-0.1em}\hspace*{-0.1em}&
V(\eta(k_{\aleph}))\hspace*{-0.2em}+\hspace*{-0.2em}(\vartheta\hspace*{-0.2em}-\hspace*{-0.2em}1)\alpha(\theta_{k_{\aleph}})
\hspace*{-0.2em}-\hspace*{-0.2em}V(\eta(0))\hspace*{-0.2em}-\hspace*{-0.2em}
(\vartheta\hspace*{-0.2em}-\hspace*{-0.2em}1)\alpha(0) \notag \\
& \hspace*{-0.1em}\leq\hspace*{-0.1em}&-
\scalebox{1}{\(\sum\limits\)}_{k=0}^{k_{\aleph}}
\hbar_{1}\|\zeta(k)\|^{2}, \label{r4}
\end{eqnarray}
where the last term equals to the counterpart of \eqref{r3}. Hence, the asymptotical stability of
$V(\eta(k))\hspace*{-0.2em}+\hspace*{-0.2em}(\vartheta\hspace*{-0.2em}-\hspace*{-0.2em}1)\alpha(k)$ is ensured, that is,
$\eta(k)$ and $\alpha(k)$ converge to $\eta^{*}$ and the origin, respectively,
as $\hspace*{-0.1em}k\hspace*{-0.1em}$ tends to $\hspace*{-0.1em}\infty\hspace*{-0.1em}$. Besides, the second inequality of \eqref{ma1}
validates the prerequisite of local quadratic constraint of $\pi_{\mathrm{DNN}}$ \eqref{dnnqc}. We intuitively obtain
$W_{1,\jmath}P_{1}^{-\hspace*{-0.1em}1}\hspace*{-0.1em}W^{\top}_{1,\jmath}\hspace*{-0.2em}\leq\hspace*{-0.2em}
(\bar{p}_{1,\jmath}\hspace*{-0.2em}-\hspace*{-0.2em}p^{*}_{1,\jmath})^{2},\jmath\hspace*{-0.2em}\in\hspace*{-0.2em}\mathds{N}_{[1,a_{1}]}$ by
utilizing Schur complement, which ensures that the sublevel set
$\mathcal{E}_{\hspace*{-0.1em}P_{1}}(x^{*}\hspace*{-0.1em})$ is contained in the set $\mathcal{R}
\hspace*{-0.2em}\triangleq\hspace*{-0.2em}
\{x\hspace*{-0.2em}\in\hspace*{-0.2em}\mathds{R}^{n}|\|W_{1}\hspace*{-0.1em}(x\hspace*{-0.2em}-\hspace*{-0.2em}x^{*})\|^{2}
\hspace*{-0.3em}\leq\hspace*{-0.3em}(\bar{p}_{1}\hspace*{-0.2em}-\hspace*{-0.2em}p^{*}_{1})\hspace*{-0.1em}\}$
in view of the constrained quadratic \citep*[Lemma 1]{Hindi1998Analysis} combined with the IBP technique.
Thus, $\hspace*{-0.1em}p_{1}\hspace*{-0.3em}\in\hspace*{-0.3em}[\underline{p}_{1}
\hspace*{-0.1em},\hspace*{-0.1em}\overline{p}_{1}\hspace*{-0.1em}]$ holds if
$\hspace*{-0.1em}x(k)\hspace*{-0.3em}\in\hspace*{-0.3em}\mathcal{E}_{\hspace*{-0.1em}P_{1}}(x^{*}\hspace*{-0.1em})$,
and the local property is reasonable for the event-triggered $\hspace*{-0.1em}\pi_{\mathrm{DNN}}\hspace*{-0.1em}$.

Recalling
$\vartheta\hspace*{-0.2em}\geq\hspace*{-0.2em}1$ and the nonnegativity of $\alpha(k_{q})$ in the event-triggered scheme
\eqref{ets1} and letting $\alpha(0)\hspace*{-0.2em}=\hspace*{-0.2em}0$ as specified in \eqref{as3}, the inequality \eqref{r4} leads to
$V(\eta(k_{\aleph}))\hspace*{-0.2em}-\hspace*{-0.2em}V(\eta(0))\hspace*{-0.2em}\leq\hspace*{-0.2em}-
\scalebox{1}{\(\sum\limits\)}_{k=0}^{k_{\aleph}}
\hbar_{1}\|\zeta(k)\hspace*{-0.2em}-\hspace*{-0.2em}\zeta^{*}\|^{2}$, which implies that $\eta(k)\hspace*{-0.2em}\in\hspace*{-0.2em}
\mathcal{E}_{\hspace*{-0.1em}P}(\eta^{*}\hspace*{-0.1em})\hspace*{-0.2em}\triangleq\hspace*{-0.2em}
\{\eta\hspace*{-0.2em}\in\hspace*{-0.2em}\mathds{R}^{n}|\|\eta\hspace*{-0.2em}-\hspace*{-0.2em}\eta^{*}\|^{2}_{P}
\hspace*{-0.3em}\leq\hspace*{-0.3em}1\}$ holds for $\eta(0)\hspace*{-0.2em}\in\hspace*{-0.2em}
\mathcal{E}_{\hspace*{-0.1em}P}(\eta^{*}\hspace*{-0.1em})$.
In view of the initial value $\xi(0)\hspace*{-0.3em}=\hspace*{-0.3em}0$ of virtual filter \eqref{fil}, we obtain that
$\eta(0)\hspace*{-0.2em}\in\hspace*{-0.2em}
\mathcal{E}_{\hspace*{-0.1em}P}(\eta^{*}\hspace*{-0.1em})$ yields $x(0)\hspace*{-0.2em}\in\hspace*{-0.2em}
\mathcal{E}_{\hspace*{-0.1em}P_{1}}(x^{*}\hspace*{-0.1em})$, acting as an inner approximation of robust RoA defined by \eqref{rroa}, where
$P_{1}$ denotes the upper left block of $P$. The proof is completed.

The feasible solution of \eqref{ma1} captures an inner approximation of robust
RoA in \eqref{rroa}, since the volume of
$\mathcal{E}_{\hspace*{-0.1em}P_{1}}(x^{*}\hspace*{-0.1em})$ is proportional to the determinant of $P_{1}$, described by $\mathrm{Det}(P_{1})\hspace*{-0.2em}$
\citep*[Subsection 2.1]{Hindi1998Analysis}. For computing the \emph{largest} inner
approximation of robust RoA, we can formulate the following optimization problem
\begin{eqnarray}
\min\limits_{P\succ0,T_{\imath}\succ0,\Xi_{\imath}\succ0,\gamma\geq0,R,N_{\imath}}\log(\mathrm{Det}(P_{1})), \text{s.t.}\ \eqref{ma1} \ \text{holds},
\label{roa1}
\end{eqnarray}
which is convex about all decision variables defined in Theorem \ref{th1}.
\begin{remark}
\label{opt1}
 Note that a particular uncertainty $\Theta$
implies the existence of a class of time-domain IQCs depicted by a virtual filter $\Phi_{\Theta}$ in conjunction with
a \emph{weight} matrix $\hspace*{-0.1em}M_{\Theta}$, extracting from the constraint set $\mathcal{M}_{\Theta}$ with
$M_{\Theta}\hspace*{-0.3em}\in\hspace*{-0.3em}\mathcal{M}_{\Theta}\hspace*{-0.3em}\subseteq\hspace*{-0.3em}\mathds{S}^{\kappa}\hspace*{-0.1em}$.
Therefore, the convex optimization \eqref{roa1} can also regard $\hspace*{-0.1em}M_{\Theta}\hspace*{-0.1em}$ as an additional decision
variable, leading to less conservatism about calculating the triggering parameters of \eqref{as2}, \hspace*{-0.1em}compared to the scenario of fixed $\hspace*{-0.2em}M_{\Theta}$, \hspace*{-0.1em}which covers more flexibility of stability analysis by LMIs
in the sense of IQCs \hspace*{-0.2em}\citep*[Subsection 3.2]{Lessard2016Analysis}. Moreover, the upper bound $\hspace*{-0.1em}\overline{p}_{1}\hspace*{-0.1em}$ of the first-layer preactivation of $\hspace*{-0.1em}\pi_{\mathrm{DNN}}\hspace*{-0.1em}$ \eqref{dnn} affects the size of inner approximation of robust RoA explicitly. The local sector bound $\mathrm{Sec}[\rho,\sigma]\hspace*{-0.1em}$ is
sharpened by decreasing the value of $\overline{p}_{1}\hspace*{-0.3em}-\hspace*{-0.3em}p_{1}^{*}$,
which improves the robust RoA approximation in terms of reducing the convex relaxation error of $\hspace*{-0.1em}\pi_{\mathrm{DNN}}\hspace*{-0.1em}$. However, it narrows the size of constrained set $\mathcal{R}$ as
discussed in Appendix A, which contains the ellipsoid $\hspace*{-0.1em}\mathcal{E}_{\hspace*{-0.1em}P_{1}}(x^{*}\hspace*{-0.1em})\hspace*{-0.1em}$ \citep*[\hspace*{-0.1em}Lemma \hspace*{-0.1em}1]{Hindi1998Analysis}, hence to affect RoA inner approximations.
In contrast, enlarging $\overline{p}_{1}\hspace*{-0.3em}-\hspace*{-0.3em}p_{1}^{*}$
yields a larger constrained set $\hspace*{-0.1em}\mathcal{R}\hspace*{-0.1em}$ containing
$\hspace*{-0.1em}\mathcal{E}_{\hspace*{-0.1em}P_{1}}(x^{*}\hspace*{-0.1em})\hspace*{-0.1em}$ and implies
a looser local sector bound $\mathrm{Sec}[\rho,\sigma]\hspace*{-0.1em}$.
For further reducing conservatism, we can parameterize $\overline{p}_{1}\hspace*{-0.3em}-\hspace*{-0.3em}p_{1}^{*}
\hspace*{-0.3em}=\hspace*{-0.3em}\varsigma\mathbf{1}_{a_{1}}\hspace*{-0.2em}$
with $\underline{\varsigma}\hspace*{-0.2em}\leq\hspace*{-0.2em}\varsigma\hspace*{-0.2em}\leq\hspace*{-0.2em}\bar{\varsigma}$, in which
$\hspace*{-0.1em}\bar{\varsigma}\hspace*{-0.1em}$ is the largest value preserving the feasibility of condition \eqref{ma1}.
Hence, assessing $\varsigma$ within the gridded interval $[\underline{\varsigma},\bar{\varsigma}]$ leads to the \emph{largest} inner approximation of robust RoA.
Additionally, different from the time-triggered scheme \citep*{Yin2022Stability}, the interval $\hspace*{-0.1em}\vartheta\hspace*{-0.1em}$ of event-triggered logic \eqref{ets1} also affects the RoA estimation. Large
$\hspace*{-0.1em}\vartheta\hspace*{-0.1em}$ is beneficial for reducing computation burden, but the value of
$\vartheta$ (or $\hspace*{-0.1em}\vartheta_{u}\hspace*{-0.1em}$) is correlated to the feasible solution of \eqref{roa1}. Within
this context, the \emph{largest} inner approximation of robust RoA is
captured by adjusting $\hspace*{-0.1em}\vartheta\hspace*{-0.1em}$ through $[\vartheta_{l},\hspace*{-0.1em}\vartheta_{u}]$, which yields a tradeoff
between the communication burden and the stability metric. We
can synthesize the \emph{best}\hspace*{-0.1em} event-triggered neural control policy \eqref{aaug} by achieving the \emph{largest} RoA approximation.
\end{remark}
\begin{remark}
\label{306}
We make the discussions about the novelty and comparison of Theorem \ref{th1} in what follows, since fewer results have concentrated on
the event-triggered neural control design problems. First,
the convex optimization \eqref{roa1} incorporates the event-triggered transmission \eqref{ets1} into the neural-feedback loops \eqref{aug}
for assessing the stability metric by robust RoA estimation. Remark \ref{opt1} leads to the \emph{best} event-triggered logic corresponding to the \emph{largest} inner approximation of robust RoA, which builds upon
the strict stability criterion \eqref{ma1}, such that both the better adaptivity to uncertainty impacts by virtue of
$\hspace*{-0.1em}\pi_{\mathrm{DNN}}\hspace*{-0.1em}$ and the more efficient communication transmission induced by event-triggered logic
can be achieved synchronously. Second, albeit the Lipschitz continuity assumption on $\hspace*{-0.1em}\pi_{\mathrm{DNN}}\hspace*{-0.2em}$
\citep*[Lemma 4.2]{Jin2018Control} can lead to its quadratic constraint, associated with the forward dynamics of closed-loop system
\citep*[\hspace*{-0.1em}Proposition 1]{Talukder2023Robust}, to support stability analysis, the identification of Lipschitz bound also
needs an extra calculation step, such as the $\mathrm{LipSDP}$ procedure
\citep*[Theorem 2]{Fazlyab2019Efficient} and the maximum absolute row sum approach
\citep*[Equation (16)]{Talukder2023Robust}, which can increase computation burden especially for a 
large-scale $\hspace*{-0.1em}\pi_{\mathrm{DNN}}\hspace*{-0.1em}$. As an alternative, the local sector-bounded attribute of nonlinear activation
function $\varphi$ can represent the quadratic constraint of $\hspace*{-0.1em}\pi_{\mathrm{DNN}}$, and the nonlinear isolation technique
\eqref{dnn3} can integrate the activation $m_{i}$ of each layer of $\hspace*{-0.1em}\pi_{\mathrm{DNN}}$ into an augmented vector $\zeta$ to
perform stability analysis of ETC systems. Moreover, the dimensions of decision variables in Theorem \ref{th1} are
distinctly less than the counterparts calculated by Kronecker product \citep*{Talukder2023Robust}, which also reduces 
computation burden. Third, the stability verification performed in Theorem \hspace*{-0.1em}\ref{th1}\hspace*{-0.1em} relies on auxiliary looped functions, which implies less conservatism as shown in Remark \ref{loop}. Recalling the theoretical
derivations as aforementioned, the condition $H(\theta_{j\hspace*{-0.1em}+\hspace*{-0.1em}1,q})
\hspace*{-0.3em}=\hspace*{-0.3em}H(\theta_{j,q})\hspace*{-0.3em}=\hspace*{-0.3em}0$ holds. Compared to the
results for linear systems \citep*{Seuret2012A,Wang2022Model},
the nonlinearity of $\hspace*{-0.1em}\pi_{\mathrm{DNN}}\hspace*{-0.1em}$ herein is relaxed to develop stability criterion in the sense
of convex optimization, which generalizes the methods to the scenario of nonlinear systems and provides
strict theoretical guarantees in the complete time domain based on the IQC of uncertainty impact and the summation inequality.
\end{remark}
\begin{figure}[htbp]
\centerline{
\includegraphics[width=8.5cm]{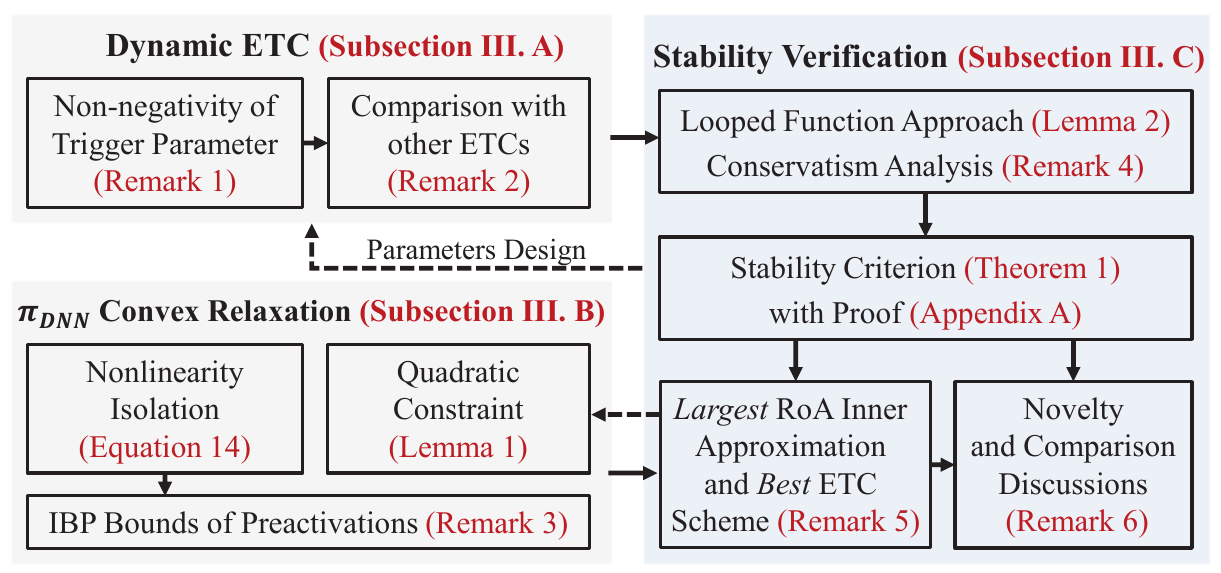} \vspace{-1ex}}
\caption{The architecture of event-triggered neural control in Section \ref{sec3}.}
\label{F2}
\end{figure}

\section{Self-Triggered Neural Control Policy}
\label{sec4}

\subsection{Self-Triggered Scheme}
\label{stt}

The core of self-triggered
communication transmission lies in specifying a function $\mho(x(k_{q}))$ to determine the next transmission instant $k_{q+1}$, namely,
\begin{eqnarray}
k_{q+1}\hspace*{-0.2em}=\hspace*{-0.2em}k_{q}\hspace*{-0.2em}+\hspace*{-0.1em}\mho(x(k_{q}),s_{k}), \label{s1}
\end{eqnarray}
where $s_{k}\hspace*{-0.2em}=\hspace*{-0.2em}k_{q+1}\hspace*{-0.2em}-\hspace*{-0.2em}k_{q}$ with $s_{k}\hspace*{-0.2em}\in\hspace*{-0.2em}\mathds{N}_{[1,\bar{s}]},
\bar{s}\hspace*{-0.2em}\in\hspace*{-0.2em}\mathds{N}_{>1}$. Different from event-triggered schemes using the current sampled data, 
self-triggered schemes depend on a predicted state $x(k_{q}\hspace*{-0.2em}+\hspace*{-0.2em}s_{k})$ to determine the next transmission
instant. For specifying the term $\mho(x(k_{q}),s_{k})$ in \eqref{s1}, we deploy the following condition
\begin{eqnarray}
&&\check{\epsilon}_{1}\|x(k_{q}\hspace*{-0.3em}+\hspace*{-0.3em}s_{k})\hspace*{-0.3em}-\hspace*{-0.3em}x^{*}\|^{2}_{
\check{\Xi}_{1}}
\hspace*{-0.3em}+\hspace*{-0.3em}\check{\epsilon}_{2}\|x(k_{q})\hspace*{-0.3em}-\hspace*{-0.3em}x^{*}\|^{2}_{\check{\Xi}_{1}}
\hspace*{-0.3em}-\hspace*{-0.3em} \|e(s_{k})\|^{2}_{\check{\Xi}}\hspace*{-0.3em}\geq\hspace*{-0.3em} 0, \label{st2}
\end{eqnarray}
where $\check{\epsilon}_{1},\check{\epsilon}_{2}\hspace*{-0.2em}\in\hspace*{-0.2em}\mathds{R}_{[0,1]}$ are discount parameters, 
$\check{\Xi}_{1},\check{\Xi}_{2}\hspace*{-0.2em}\in\hspace*{-0.2em}\mathds{S}^{n}_{\succ0}$ are weight matrices, and the divergence
between $x(k_{q})$ at the latest transmission instant $k_{q}$ and $x(k_{q}\hspace*{-0.2em}+\hspace*{-0.2em}s_{k})$ at the
predicted time instant $k_{q}\hspace*{-0.2em}+\hspace*{-0.2em}s_{k}$ can be defined by
$e(s_{k})\hspace*{-0.2em}\triangleq x(k_{q}\hspace*{-0.2em}+\hspace*{-0.2em}s_{k})\hspace*{-0.2em}-\hspace*{-0.2em}
x(k_{q})$. Hence, the condition \eqref{st2} is recapped by
\begin{eqnarray}
\mathcal{S}(x(k_{q}),s_{k})&\hspace*{-0.1em}=\hspace*{-0.1em}&\left[
\begin{array}{c}
\hspace*{-0.3em}x(k_{q}\hspace*{-0.2em}+\hspace*{-0.2em}s_{k})\hspace*{-0.2em}-\hspace*{-0.2em}x^{*}\hspace*{-0.3em}  \\
\hspace*{-0.3em}x(k_{q})\hspace*{-0.2em}-\hspace*{-0.2em}x^{*}\hspace*{-0.3em}
\end{array}%
\right]^{\hspace*{-0.2em}\top}\hspace*{-0.4em} \left[
\begin{array}{cc}
\hspace*{-0.3em}\check{\epsilon}_{1}\check{\Xi}_{1}\hspace*{-0.2em}-\hspace*{-0.2em}\check{\Xi}_{2}
\hspace*{-0.1em}&\hspace*{-0.1em}\check{\Xi}_{2}\hspace*{-0.1em}\\
\hspace*{-0.3em}\star\hspace*{-0.1em}&\hspace*{-0.1em}\check{\epsilon}_{2}\check{\Xi}_{1}\hspace*{-0.2em}-\hspace*{-0.2em}
\check{\Xi}_{2}\hspace*{-0.1em}
\end{array}%
\right] \notag\\
&\hspace*{-0.8em}\hspace*{-0.8em}&\times\left[
\begin{array}{c}
\hspace*{-0.3em}x(k_{q}\hspace*{-0.2em}+\hspace*{-0.2em}s_{k})\hspace*{-0.2em}-\hspace*{-0.2em}x^{*}\hspace*{-0.3em}  \\
\hspace*{-0.3em}x(k_{q})\hspace*{-0.2em}-\hspace*{-0.2em}x^{*}\hspace*{-0.3em}
\end{array}%
\right]\hspace*{-0.2em}\geq\hspace*{-0.2em} 0.
\label{st3}
\end{eqnarray}
The time instant $k_{q}\hspace*{-0.2em}+\hspace*{-0.2em}s_{k}$ reveals the next transmission on the premise of
$\mathcal{S}(x(k_{q}),s_{k})\hspace*{-0.2em}\geq\hspace*{-0.2em}0$. The ZOH is utilized to maintain the sampled state
$x(k_{q}\hspace*{-0.2em}+\hspace*{-0.2em}s_{k})$ within $[k_{q+1},k_{q+2}\hspace*{-0.2em}-\hspace*{-0.2em}1]$, when it is transmitted to the
controller $\pi_{\mathrm{DNN}}$. Meanwhile, the
self-triggered scheme is used to determine the next transmission time instant
$k_{q+2}$.
The term $\mho(x(k_{q}),s_{k})$ is with the form
\begin{equation}
\label{sts1}
\begin{array}{rcl}
\hspace*{-0.8em}\mho(x(k_{q}),s_{k})& \hspace*{-0.3em}=\hspace*{-0.3em} & \max_{s_{k}\in\mathds{N}}
\{s_{k}\hspace*{-0.2em}\in\hspace*{-0.2em}\mathds{N}_{\geq1}| \mathcal{S}(x(k_{q}),s_{k}) \hspace*{-0.2em}\geq\hspace*{-0.2em}0 \}.
\end{array}
\end{equation}
Hence, in view of the self-triggered scheme, the next transmission time instant $k_{q+1}$ is predicted by the current measurement
at time instant $k_{q}$, such that continuous monitoring of system state is unnecessary.
\begin{remark}
\hspace*{-0.3em}The static self-triggered scheme \eqref{s1}-\eqref{sts1} is akin to the results
\citep*{Wan2021Dynamic,Wang2022Model,Wildhagen2023Data}. Here, we only state the comparisons and discussions.
First, compared to \citep*[Subsection IV.A]{Wang2022Model}, the lifting technique
is unsuitable for the system \eqref{aug} with nonlinear $\hspace*{-0.1em}\pi_{\mathrm{DNN}}\hspace*{-0.1em}$, while the deployment of ZOH and the pretrained controller can still lead to $\hspace*{-0.1em}x(k_{q}\hspace*{-0.2em}+\hspace*{-0.2em}s_{k})\hspace*{-0.1em}$ based on
\eqref{aaug}, and the extension of \citep*{Wildhagen2023Data}
to the nonlinear situation is promising.
Second, the self-triggered policy \citep*[Theorem 1]{Wan2021Dynamic} depends heavily on convex relaxations to build the correlations between
$x(k_{q})$ and $x(k)$, whereas, the comparison lemma is unavailable for the discrete-time scenario directly. Note that the
self-triggered condition \eqref{st3} implies 
less conservatism and more conciseness, since an intermediate variable $x(k)$ is excluded intuitively.
Moreover, different from \citep*[\hspace*{-0.1em}Subsection 2.2\hspace*{-0.1em}]{Cao2023Self}, the network-induced delay during the state transmission is not incorporated into the self-triggered scheme as aforementioned, but it leaves the room for future research.   Third, compared to the event-triggered scheme in the
Subsection \ref{301}, the event-based condition \eqref{ets1} boils down to the self-triggered one \eqref{st2}
by letting $\alpha\hspace*{-0.2em}=\hspace*{-0.2em}0$, $g\hspace*{-0.2em}=\hspace*{-0.2em}1$, and
$j\vartheta\hspace*{-0.2em}=\hspace*{-0.2em}s_{k}$, while both of them depends on the current transmission $x(k_{q})$.
\end{remark}

We shall perform the closed-loop stability verification of
the self-triggered neural-feedback loops \eqref{aug} and \eqref{aaug}, allowing for the self-triggered strategy in
Subsection \ref{stt} and the convex relaxation of $\pi_{\mathrm{DNN}}$ in Subsection \ref{302}, in what follows.

\subsection{Closed-Loop Stability Analysis}
\label{ss302}

We search for a
looped function $\overline{W}(\eta,k):\mathds{R}^{\bar{n}}\times
\mathds{N}_{[k_{q},k_{q\hspace*{-0.1em}+\hspace*{-0.1em}1}\hspace*{-0.1em}-1]}\rightarrow\mathds{R}$, such that
$\overline{W}(\eta(k_{q\hspace*{-0.1em}+\hspace*{-0.1em}1}),k_{q\hspace*{-0.1em}+\hspace*{-0.1em}1})
\hspace*{-0.3em}=\hspace*{-0.3em}\overline{W}(\eta(k_{q}),k_{q})$ holds for
$s_{k}\hspace*{-0.2em}\in\hspace*{-0.2em}\mathds{N}_{[1,\bar{s}]}$. Based on the quadratic constraint \eqref{dnnqc} and
the nonlinearity isolation \eqref{dnn3} of $\pi_{\mathrm{DNN}}$, we perform the controller-dynamic association
for closed-loop stability analysis under the self-triggered scheme \eqref{s1} and \eqref{sts1}.
The following lemma provides a tool for stability analysis using a looped function for the scenario of self-triggered scheme.
\begin{lemma}
\label{lf1}
\hspace*{-0.4em}
For scalars $\check{\mu}_{1}\hspace*{-0.2em}<\hspace*{-0.2em}\check{\mu}_{2}$ with $\check{\mu}_{1},\check{\mu}_{2}\hspace*{-0.2em}\in\hspace*{-0.2em}\mathds{R}_{>0}$, and a
differentiable value function $\overline{V}(\eta):\mathds{R}^{\varsigma}\rightarrow\mathds{R}_{\geq0}$
for $\eta\hspace*{-0.2em}\in\hspace*{-0.2em}\mathds{R}^{\varsigma}$, satisfying
$\check{\mu}_{1}\|\eta\|^{2}_{2}\hspace*{-0.2em}\leq\hspace*{-0.2em}
\overline{V}(\eta)\hspace*{-0.2em}\leq\hspace*{-0.2em}\check{\mu}_{2}\|\eta\|^{2}_{2}$, the following two statements are equivalent. \newline
(i) The existence of
looped function $\overline{W}$ ensures that
$\Delta\overline{\mathcal{V}}(k)\hspace*{-0.2em}\triangleq\hspace*{-0.2em}
\overline{\mathcal{V}}(k\hspace*{-0.2em}+\hspace*{-0.2em}1)\hspace*{-0.2em}-\hspace*{-0.2em}\overline{\mathcal{V}}(k)
\hspace*{-0.2em}<\hspace*{-0.2em}0$ holds for
$k\hspace*{-0.2em}\in\hspace*{-0.2em}[k_{q},k_{q\hspace*{-0.1em}+\hspace*{-0.1em}1}\hspace*{-0.2em}-\hspace*{-0.2em}1]$, non-zero
$\eta(k_{q})$ and $\{k_{q}\}$, where $\overline{\mathcal{V}}\hspace*{-0.1em}(k)\hspace*{-0.2em}\triangleq\hspace*{-0.2em}
\overline{V}\hspace*{-0.1em}(\eta(k))\hspace*{-0.2em}+\hspace*{-0.2em}\overline{W}(\eta(k),k)$. \newline
(ii) The increment $\Delta \overline{V}\hspace*{-0.3em}\triangleq\hspace*{-0.3em}\overline{V}\hspace*{-0.1em}(\eta(\hspace*{-0.1em}
k_{q\hspace*{-0.1em}+\hspace*{-0.1em}1}\hspace*{-0.1em})\hspace*{-0.1em})
\hspace*{-0.3em}-\hspace*{-0.3em}\overline{V}\hspace*{-0.2em}
(\eta(\hspace*{-0.1em}k_{q}\hspace*{-0.1em})\hspace*{-0.1em})\hspace*{-0.3em}<\hspace*{-0.3em}0$ holds with
$\eta(k_{q})\hspace*{-0.3em}\neq\hspace*{-0.3em}0$ for a sequence $\{k_{q}\}$.
\end{lemma}
\textbf{Proof}\
The proof is similar to the counterpart of Lemma \ref{lf}, and we omit it here due to the limited space.
\begin{theorem}
\label{th2}
\hspace*{-0.4em}
For specified scalars $\check{\epsilon}_{1},\check{\epsilon}_{2}\hspace*{-0.2em}\in\hspace*{-0.2em}\mathds{R}_{[0,1]}$ and a   positive integer
$\bar{s}\hspace*{-0.2em}>\hspace*{-0.2em}1$, the neural-feedback loops \eqref{aug}
asymptotically converge to the equilibrium
point $\hspace*{-0.1em}(\hspace*{-0.1em}\eta^{*}\hspace*{-0.1em},\hspace*{-0.1em}u^{*}\hspace*{-0.1em},\hspace*{-0.1em}\omega^{*}\hspace*{-0.1em},
\hspace*{-0.1em}\nu^{*}\hspace*{-0.1em},\hspace*{-0.1em}p^{*}\hspace*{-0.1em},\hspace*{-0.1em}m^{*}\hspace*{-0.1em})$ under the self-triggered communication scheme
\eqref{s1} and \eqref{sts1}, if there exist matrices $\overline{P}\hspace*{-0.2em}\succ\hspace*{-0.2em}0$,
$\overline{T}_{\imath}\hspace*{-0.2em}\succ\hspace*{-0.2em}0$,
$\check{\Xi}_{\imath}\hspace*{-0.3em}\succ\hspace*{-0.2em}0$, $\hspace*{-0.1em}\overline{R},\hspace*{-0.1em}\overline{N}_{\imath}$,
a vector $\gamma\hspace*{-0.3em}\in\hspace*{-0.3em}\mathds{R}^{\hspace*{-0.1em}a}$ with
element-wise $\hspace*{-0.2em}\gamma_{i}\hspace*{-0.3em}\geq\hspace*{-0.3em}0$, and scalars $\bar{\lambda}_{\imath}\hspace*{-0.3em}>\hspace*{-0.3em}0$ as the decision variables, such that the conditions
\begin{eqnarray}
\left[
\begin{array}{cc}
\hspace*{-0.3em}\tilde{\mathcal{G}}\hspace*{-0.1em}&\hspace*{-0.1em}\eth_{1}(\overline{N}_{\imath},\hspace*{-0.2em}\overline{R})\hspace*{-0.2em}\\
\hspace*{-0.3em}\star\hspace*{-0.1em}&\hspace*{-0.1em}-\eth_{2}(\overline{\mathcal{T}}_{\imath},\hspace*{-0.2em}\bar{\lambda}_{\imath})\hspace*{-0.2em}
\end{array}%
\right]\hspace*{-0.1em} & \hspace*{-0.1em}\prec\hspace*{-0.1em} & 0, \left[
\begin{array}{cc}
\hspace*{-0.3em}(\bar{p}_{1,\jmath}\hspace*{-0.2em}-\hspace*{-0.2em}\bar{p}_{1,\jmath}^{*})^{2}\hspace*{-0.2em}&\hspace*{-0.2em}\left[
\begin{array}{cc}
\hspace*{-0.3em}W_{1,\jmath}\hspace*{-0.1em}
&\hspace*{-0.1em}\mathbf{0}_{1\times\psi}\hspace*{-0.3em}
\end{array}
\right]\hspace*{-0.2em}\\
\hspace*{-0.3em}\star\hspace*{-0.2em}&\hspace*{-0.2em}\overline{P}\hspace*{-0.2em}
\end{array}%
\right]\hspace*{-0.1em}\succeq
\hspace*{-0.1em}0,\label{ma2}
\end{eqnarray}
hold for $\underline{p}_{1}\hspace*{-0.3em}=\hspace*{-0.3em}-\overline{p}_{1}$,
$\imath\hspace*{-0.2em}\in\hspace*{-0.2em}\mathds{N}_{[1,2]}$, $\jmath\hspace*{-0.2em}\in\hspace*{-0.2em}\mathds{N}_{[1,a_{1}]}$, and $i\hspace*{-0.2em}\in\hspace*{-0.2em}\mathds{N}_{[1,a]}$,
where $\overline{\mathcal{T}}_{\imath}\hspace*{-0.2em}\triangleq\hspace*{-0.2em}\mathrm{Diag}(\overline{T}_{\imath},3\overline{T}_{\imath})$,  
\begin{eqnarray*}
\tilde{\mathcal{G}}\hspace*{-0.1em} & \hspace*{-0.1em}\triangleq\hspace*{-0.1em} &\hspace*{-0.1em}
(A\overline{L}_{1}\hspace*{-0.3em}+\hspace*{-0.3em}B\Sigma_{1}\overline{\Lambda}_{1})^{\hspace*{-0.2em}\top}\hspace*{-0.1em}
\overline{P}(A\overline{L}_{1}\hspace*{-0.3em}+\hspace*{-0.3em}B\Sigma_{1}\overline{\Lambda}_{1})\hspace*{-0.2em}+\hspace*{-0.2em}
(\bar{s}\hspace*{-0.2em}-\hspace*{-0.2em}1)\overline{\mathcal{G}}_{1}^
{\hspace*{-0.2em}\top}\overline{T}_{1}\overline{\mathcal{G}}_{1}\hspace*{-0.2em}+\\
\hspace*{-0.1em} & \hspace*{-0.1em}\hspace*{-0.1em} &\hspace*{-0.1em}
(\bar{s}\hspace*{-0.2em}+\hspace*{-0.2em}1)\overline{\mathcal{G}}_{1}^
{\hspace*{-0.2em}\top}\overline{T}_{2}\overline{\mathcal{G}}_{1}\hspace*{-0.2em}+\hspace*{-0.2em}\bar{s}\bar{\lambda}_{1}
\overline{R}_{7}^{\top}\overline{R}_{7}\hspace*{-0.2em}+\hspace*{-0.2em}\bar{s}\bar{\lambda}_{2}
\overline{R}_{4}^{\top}\overline{R}_{4}\hspace*{-0.2em}+\hspace*{-0.2em}\overline{\mathcal{G}}_{2}^{\top}M_{\Theta}
\overline{\mathcal{G}}_{2}\hspace*{-0.2em}+\\
\hspace*{-0.1em} & \hspace*{-0.1em}\hspace*{-0.1em} &\hspace*{-0.1em}
\overline{\mathcal{G}}_{3}^{\top}\hspace*{-0.2em}\mathcal{M}_{\mathrm{DNN}}
\overline{\mathcal{G}}_{3}\hspace*{-0.2em}+\hspace*{-0.2em}\mathrm{Sym}(\overline{R}_{1}^{\top}\hspace*{-0.2em}\overline{R}
\overline{R}_{2}
\hspace*{-0.2em}-\hspace*{-0.3em}\overline{R}_{5}^{\top}\hspace*{-0.2em}\overline{R}
\overline{R}_{6}\hspace*{-0.3em}+\hspace*{-0.3em}
\overline{N}_{1}\hspace*{-0.1em}\overline{R}_{9}\hspace*{-0.3em}+\hspace*{-0.3em}
\overline{N}_{2}\hspace*{-0.1em}\overline{R}_{10})\\
\hspace*{-0.1em} & \hspace*{-0.1em}\hspace*{-0.1em} &\hspace*{-0.1em}
-\overline{R}_{11}^{\top}\widetilde{\Xi}\overline{R}_{11}\hspace*{-0.3em}-\hspace*{-0.3em}
\overline{L}_{1}^{\top}\overline{P}\overline{L}_{1}, \\
\eth_{1}\hspace*{-0.1em} & \hspace*{-0.1em}\triangleq\hspace*{-0.1em} &\hspace*{-0.1em}
\left[
\begin{array}{cccc}
\hspace*{-0.3em}\bar{s}\overline{N}_{1}
&\bar{s}\overline{N}_{2}
& \bar{s}\overline{R}_{3}^{\top}\overline{R}
& \bar{s}\overline{R}_{8}^{\top}\overline{R}\hspace*{-0.3em}
\end{array}%
\right]\hspace*{-0.2em}, \overline{\mathcal{G}}_{2}\hspace*{-0.2em}\triangleq\hspace*{-0.2em}C\overline{L}_{1}
\hspace*{-0.2em}+\hspace*{-0.2em}D\Sigma_{1}\overline{\Lambda}_{1},\\
\eth_{2}\hspace*{-0.1em} & \hspace*{-0.1em}\triangleq\hspace*{-0.1em} &\hspace*{-0.1em}
\mathrm{Diag}(\bar{s}\overline{\mathcal{T}}_{1},\bar{s}\overline{\mathcal{T}}_{2},
\bar{s}\bar{\lambda}_{1}I_{4n},\bar{s}\bar{\lambda}_{2}I_{4n}),
\overline{R}_{7}\hspace*{-0.3em}\triangleq
\hspace*{-0.3em}\overline{R}_{2}\hspace*{-0.3em}-\hspace*{-0.3em}\overline{R}_{6}, \\
\bar{\mathcal{G}}_{1}\hspace*{-0.1em} & \hspace*{-0.1em}\triangleq\hspace*{-0.1em} &\hspace*{-0.1em}
\Lambda_{2}A\overline{L}_{1}\hspace*{-0.3em}-\hspace*{-0.3em}\Lambda_{2}\overline{L}_{1}\hspace*{-0.3em}
+\hspace*{-0.3em}\Lambda_{2}B\Sigma_{1}\overline{\Lambda}_{1},
\overline{\Upsilon}_{3}\hspace*{-0.3em}\triangleq\hspace*{-0.3em}\mathrm{Col}
(\overline{L}_{4},\overline{L}_{5},\mathbf{0}_{n\hspace*{-0.1em}\times
\hspace*{-0.1em}\tilde{n}},\overline{L}_{6}),\\
\bar{\mathcal{G}}_{3}\hspace*{-0.1em} & \hspace*{-0.1em}\triangleq\hspace*{-0.1em} &\hspace*{-0.1em}
\mathrm{Col}(\Pi_{px}\overline{L}_{4}\hspace*{-0.2em}+\hspace*{-0.2em}\Pi_{pm}\overline{L}_{2},\overline{L}_{2}),
\overline{\Upsilon}_{4}\hspace*{-0.3em}\triangleq\hspace*{-0.3em}\mathrm{Col}(\overline{L}_{4},\overline{L}_{5},
\mathbf{0}_{n\hspace*{-0.1em}\times
\hspace*{-0.1em}\tilde{n}},\overline{L}_{7}),
\\
\overline{R}_{1}\hspace*{-0.1em} & \hspace*{-0.1em}\triangleq\hspace*{-0.1em} &\hspace*{-0.1em}
\mathrm{Col}(\overline{L}_{4},\overline{L}_{5},\overline{R}_{12},\overline{L}_{6}
\hspace*{-0.3em}+\hspace*{-0.3em}\overline{R}_{12}),
\overline{R}_{8}\hspace*{-0.3em}\triangleq\hspace*{-0.3em}
\overline{R}_{1}\hspace*{-0.3em}-\hspace*{-0.3em}\overline{R}_{5},\\
\overline{R}_{2}\hspace*{-0.1em} & \hspace*{-0.1em}\triangleq\hspace*{-0.1em} &\hspace*{-0.1em}
\mathrm{Col}(-\hspace*{-0.2em}\overline{L}_{4},-\hspace*{-0.2em}\overline{L}_{5},\overline{R}_{13},
\overline{L}_{7}\hspace*{-0.3em}-\hspace*{-0.3em}\overline{L}_{5}
\hspace*{-0.3em}-\hspace*{-0.3em}\Lambda_{2}\overline{L}_{1}),
\overline{R}_{11}\hspace*{-0.3em}\triangleq\hspace*{-0.3em}\mathrm{Col}(\overline{L}_{5},\overline{L}_{4}),
\\
\overline{R}_{5}\hspace*{-0.1em} & \hspace*{-0.1em}\triangleq\hspace*{-0.1em} &\hspace*{-0.1em}
\mathrm{Col}(\mathbf{0}_{n\hspace*{-0.1em}\times\hspace*{-0.1em}\tilde{n}},\mathbf{0}_{n\hspace*{-0.1em}\times\hspace*{-0.1em}\tilde{n}},
\Lambda_{2}\overline{L}_{1}\hspace*{-0.3em}-\hspace*{-0.3em}\overline{L}_{4},
\overline{L}_{6}\hspace*{-0.3em}-\hspace*{-0.3em}\overline{L}_{4}), \overline{L}_{1}\hspace*{-0.3em}\triangleq\hspace*{-0.3em}\left[
\begin{array}{cc}
\hspace*{-0.3em}I_{z}\hspace*{-0.1em}
&\hspace*{-0.1em} \mathbf{0}_{z\hspace*{-0.1em}\times\hspace*{-0.1em}(\tilde{n}-z)}\hspace*{-0.3em}
\end{array}%
\right]\hspace*{-0.3em},\\
\overline{R}_{6}\hspace*{-0.1em} & \hspace*{-0.1em}\triangleq\hspace*{-0.1em} &\hspace*{-0.1em}
\mathrm{Col}(\mathbf{0}_{n\hspace*{-0.1em}\times\hspace*{-0.1em}\tilde{n}},\mathbf{0}_{n\hspace*{-0.1em}\times\hspace*{-0.1em}\tilde{n}},
\overline{L}_{5}\hspace*{-0.3em}-\hspace*{-0.3em}\Lambda_{2}\overline{L}_{1},
\overline{L}_{7}\hspace*{-0.3em}-\hspace*{-0.3em}\overline{L}_{5}),
\widetilde{\Xi}\hspace*{-0.3em}\triangleq\hspace*{-0.3em}\left[
\begin{array}{cc}
\hspace*{-0.3em}\mathcal{X}_{1}\hspace*{-0.1em}
&\hspace*{-0.1em}\mathcal{X}_{2}\hspace*{-0.3em}
\end{array}%
\right]\hspace*{-0.3em},\\
\overline{R}_{9}\hspace*{-0.1em} & \hspace*{-0.1em}\triangleq\hspace*{-0.1em} &\hspace*{-0.1em}
\mathrm{Col}(\Lambda_{2}\overline{L}_{1}\hspace*{-0.3em}-\hspace*{-0.3em}\overline{L}_{4},
\Lambda_{2}\overline{L}_{1}\hspace*{-0.3em}+\hspace*{-0.3em}\overline{L}_{4}\hspace*{-0.3em}-\hspace*{-0.3em}\overline{L}_{6}),
\tilde{n}\hspace*{-0.2em}\triangleq\hspace*{-0.2em}z
\hspace*{-0.2em}+\hspace*{-0.2em}a\hspace*{-0.2em}+\hspace*{-0.2em}w\hspace*{-0.2em}+\hspace*{-0.2em}4n,\\
\overline{R}_{10}\hspace*{-0.1em} & \hspace*{-0.1em}\triangleq\hspace*{-0.1em} &\hspace*{-0.1em}
\mathrm{Col}(\overline{L}_{5}\hspace*{-0.3em}-\hspace*{-0.3em}\Lambda_{2}\overline{L}_{1},
\overline{L}_{5}\hspace*{-0.3em}+\hspace*{-0.3em}\Lambda_{2}\overline{L}_{1}\hspace*{-0.3em}-\hspace*{-0.3em}\overline{L}_{7}),
\overline{\Lambda}_{1}\hspace*{-0.3em}\triangleq\hspace*{-0.3em}\mathrm{Col}(\overline{L}_{4},\overline{L}_{2},\overline{L}_{3}), \\
\overline{R}_{12}\hspace*{-0.1em} & \hspace*{-0.1em}\triangleq\hspace*{-0.1em} &\hspace*{-0.1em}\Lambda_{2}A\overline{L}_{1}\hspace*{-0.2em}+\hspace*{-0.2em}\Lambda_{2}B\Sigma_{1}\overline{\Lambda}_{1},
\overline{R}_{13}\hspace*{-0.3em}\triangleq\hspace*{-0.3em}\overline{L}_{5}\hspace*{-0.3em}-\hspace*{-0.3em}\Lambda_{2}A\overline{L}_{1}
\hspace*{-0.3em}-\hspace*{-0.3em}\Lambda_{2}B\Sigma_{1}\overline{\Lambda}_{1},
\\
\overline{L}_{2}\hspace*{-0.1em} & \hspace*{-0.1em}\triangleq\hspace*{-0.1em} &\hspace*{-0.1em}
\left[
\begin{array}{ccc}
\hspace*{-0.3em}\mathbf{0}_{a\hspace*{-0.1em}\times\hspace*{-0.1em}z}
&I_{a}& \mathbf{0}_{a\hspace*{-0.1em}\times\hspace*{-0.1em}(\tilde{n}\hspace*{-0.1em}-\hspace*{-0.1em}
z\hspace*{-0.1em}-\hspace*{-0.1em}a)}\hspace*{-0.3em}
\end{array}%
\right]\hspace*{-0.2em}, \overline{L}_{3}\hspace*{-0.2em}\triangleq\hspace*{-0.2em}
\left[
\begin{array}{ccc}
\hspace*{-0.3em}\mathbf{0}_{w\hspace*{-0.1em}\times\hspace*{-0.1em}(z\hspace*{-0.1em}+\hspace*{-0.1em}a})
&I_{w}
& \mathbf{0}_{w\hspace*{-0.1em}\times\hspace*{-0.1em}4n}\hspace*{-0.3em}
\end{array}%
\right]\hspace*{-0.2em},\\
\overline{L}_{j}\hspace*{-0.1em} & \hspace*{-0.1em}\triangleq\hspace*{-0.1em} &\hspace*{-0.1em}
\left[
\begin{array}{cccc}
\hspace*{-0.3em}\mathbf{0}_{n\hspace*{-0.1em}\times\hspace*{-0.1em}(z\hspace*{-0.1em}+\hspace*{-0.1em}a\hspace*{-0.1em}+
\hspace*{-0.1em}w})
&\mathbf{0}_{n\hspace*{-0.1em}\times\hspace*{-0.1em}(j\hspace*{-0.1em}-\hspace*{-0.1em}4)n}
& I_{n}
& \mathbf{0}_{n\hspace*{-0.1em}\times\hspace*{-0.1em}(7\hspace*{-0.1em}-\hspace*{-0.1em}j)n}\hspace*{-0.3em}
\end{array}%
\right]\hspace*{-0.2em}, j\hspace*{-0.2em}\in\hspace*{-0.2em}\mathds{N}_{[4,7]},\\
\mathcal{X}_{1}\hspace*{-0.1em} & \hspace*{-0.1em}\triangleq\hspace*{-0.1em} &\hspace*{-0.1em}
\mathrm{Col}(\check{\epsilon}_{1}\check{\Xi}_{1}\hspace*{-0.2em}-\hspace*{-0.2em}\check{\Xi}_{2},\check{\Xi}^{\top}_{2}),
\mathcal{X}_{2}\hspace*{-0.2em}\triangleq\hspace*{-0.2em}\mathrm{Col}(\check{\Xi}_{2},
\check{\epsilon}_{2}\check{\Xi}_{1}\hspace*{-0.2em}-\hspace*{-0.2em}\check{\Xi}_{2}), 
\end{eqnarray*}
and the other terms embedded in \eqref{ma2} are consistent to the counterparts in Theorem \ref{th1},
Then the ellipsoid
$\mathcal{E}_{\hspace*{-0.1em}\overline{P}_{1}}(x^{*}\hspace*{-0.1em})$, where $\overline{P}_{1}$ is
the upper left block of $\overline{P}$ with the proper dimension, implies an inner approximation of RoA.
\end{theorem}
\textbf{Proof}\ We recall the neural-feedback loops \eqref{aug} with an augmented state $\eta(k)$, and incorporate the self-triggered transmission
scheme \eqref{s1} and \eqref{sts1} to the closed-loop stability analysis.
For $k\hspace*{-0.2em}\in\hspace*{-0.2em}[k_{q},k_{q\hspace*{-0.1em}+\hspace*{-0.1em}1}\hspace*{-0.3em}-\hspace*{-0.3em}1]$ with
$s_{k}\hspace*{-0.3em}=\hspace*{-0.3em}k_{q+1}\hspace*{-0.3em}-\hspace*{-0.3em}k_{q}$ and
$s_{k}\hspace*{-0.3em}\in\hspace*{-0.3em}\mathds{N}_{[1,\bar{s}]}$, we select a value function with the form
\begin{equation}
\label{lya2}
\begin{array}{rcl}
\overline{\mathcal{V}}(\eta(k),k)& \hspace*{-0.2em}=\hspace*{-0.2em} &
\overline{V}(\eta(k))\hspace*{-0.2em}+\hspace*{-0.2em}\overline{W}(\eta(k),k),
\end{array}
\end{equation}
where $\overline{V}\hspace*{-0.1em}(\hspace*{-0.1em}\eta(k)\hspace*{-0.1em})\hspace*{-0.3em}=\hspace*{-0.3em}
\|\eta(k)\hspace*{-0.3em}-\hspace*{-0.3em}\eta^{*}\hspace*{-0.1em}\|_{\overline{P}}^{2}$ with
$\overline{P}\hspace*{-0.3em}\succ\hspace*{-0.3em}0$, and $\overline{W}(\eta(k),k)$ is
a looped function $\overline{W}(\eta(k),k)\hspace*{-0.3em}\triangleq\hspace*{-0.3em}\sum_{\kappa=1}^{3}\hspace*{-0.3em}\overline{W}_{\hspace*{-0.3em}\kappa}(k)$ with 
\begin{eqnarray}
\overline{W}_{\hspace*{-0.2em}1}(k)\hspace*{-0.1em}& \hspace*{-0.1em}\triangleq\hspace*{-0.1em} &\hspace*{-0.1em}
2\overline{\chi}_{1}^{\top}(k)\overline{R}\overline{\chi}_{2}(k), \\
\overline{W}_{\hspace*{-0.2em}2}(k)\hspace*{-0.1em}& \hspace*{-0.1em}\triangleq\hspace*{-0.1em} &\hspace*{-0.1em}
(k_{q+1}\hspace*{-0.3em}-\hspace*{-0.3em}k)\hspace*{-0.3em}\left(\scalebox{1}{\(\sum\limits\)}_{s=k_{q}}^{k}\|y(s)\|_{\overline{T}_{1}}^{2}
\hspace*{-0.3em}-\hspace*{-0.3em}
\|y(k)\|_{\overline{T}_{1}}^{2}\right)\hspace*{-0.3em},\\
\overline{W}_{\hspace*{-0.2em}3}(k)\hspace*{-0.1em}& \hspace*{-0.1em}\triangleq\hspace*{-0.1em} &\hspace*{-0.1em}
(k\hspace*{-0.3em}-\hspace*{-0.3em}k_{q})\hspace*{-0.3em}\left(\|y(k_{q+1})
\|_{\overline{T}_{2}}^{2}\hspace*{-0.3em}-\hspace*{-0.3em}\scalebox{1}{\(\sum\limits\)}_{s=k}^{k_{q+1}}
\|y(s)\|_{\overline{T}_{2}}^{2}\right)\hspace*{-0.3em},
\end{eqnarray}
where $\overline{R}\hspace*{-0.2em}\in\hspace*{-0.2em}\mathds{R}^{4n\hspace*{-0.1em}\times\hspace*{-0.1em}4n}$, $\overline{T}_{\imath}\hspace*{-0.2em}\succ\hspace*{-0.2em}0$ with $\imath\hspace*{-0.2em}\in\hspace*{-0.2em}\mathds{N}_{[1,2]}$, and
\begin{eqnarray*}
\overline{\chi}_{1}\hspace*{-0.1em}& \hspace*{-0.1em}\triangleq\hspace*{-0.1em} &\hspace*{-0.1em}
\mathrm{Col}\left((k\hspace*{-0.3em}-\hspace*{-0.3em}k_{q})\overline{m}_{0},
x_{k}\hspace*{-0.3em}-\hspace*{-0.3em} x_{k_{q}},\scalebox{1}{\(\sum\limits\)}_{s=k_{q}}^{k}
x_{s}\hspace*{-0.3em}-\hspace*{-0.3em}x_{k_{q}}\right), \\
\hspace*{-0.6em} \overline{\chi}_{2}\hspace*{-0.1em}& \hspace*{-0.1em}\triangleq\hspace*{-0.1em} &\hspace*{-0.1em}
\mathrm{Col}\left((k_{q+1}\hspace*{-0.3em}-\hspace*{-0.3em}k)m_{0},x_{k_{q\hspace*{-0.1em}+\hspace*{-0.1em}1}}
\hspace*{-0.3em}-\hspace*{-0.3em} x_{k},\scalebox{1}{\(\sum\limits\)}_{s=k}^{k_{q\hspace*{-0.1em}+\hspace*{-0.1em}1}}
x_{s}\hspace*{-0.3em}-\hspace*{-0.3em} x_{k_{q+1}}\right),\\
m_{0}\hspace*{-0.1em}& \hspace*{-0.1em}\triangleq\hspace*{-0.1em} &\hspace*{-0.1em}
\mathrm{Col}(x_{k_{q}},x_{k_{q\hspace*{-0.1em}+\hspace*{-0.1em}1}}),
x_{k}\hspace*{-0.2em}\triangleq\hspace*{-0.2em}x(k)\hspace*{-0.2em}-\hspace*{-0.2em}x^{*},
x_{k_{q}}\hspace*{-0.2em}\triangleq\hspace*{-0.2em}x(k_{q})\hspace*{-0.2em}-\hspace*{-0.2em}x^{*}, \\
\hspace*{-0.6em} x_{s}\hspace*{-0.1em}& \hspace*{-0.1em}\triangleq\hspace*{-0.1em} &\hspace*{-0.1em}
x(s)\hspace*{-0.2em}-\hspace*{-0.2em}x^{*}, x_{k_{q\hspace*{-0.1em}+\hspace*{-0.1em}1}}
\hspace*{-0.2em}\triangleq\hspace*{-0.2em}x(k_{q\hspace*{-0.1em}+\hspace*{-0.1em}1})\hspace*{-0.2em}-\hspace*{-0.2em}x^{*},
y(s)\hspace*{-0.2em}=\hspace*{-0.2em}x_{s\hspace*{-0.1em}+\hspace*{-0.1em}1}\hspace*{-0.2em}-\hspace*{-0.2em}x_{s}.
\end{eqnarray*}
Accordingly, the forward difference for each item in \eqref{lya2} can be calculated by
\begin{eqnarray}
 \Delta \overline{V}\hspace*{-0.3em}& \hspace*{-0.1em}=\hspace*{-0.1em} &\hspace*{-0.1em}
\overline{\zeta}^{\top}(k)((A\overline{L}_{1}\hspace*{-0.3em}+\hspace*{-0.3em}B\Sigma_{1}\overline{\Lambda}_{1})^{\hspace*{-0.2em}\top}
\hspace*{-0.2em}
\overline{P}(A\overline{L}_{1}\hspace*{-0.3em}+\hspace*{-0.3em}B\Sigma_{1}\overline{\Lambda}_{1}) \notag\\
 \hspace*{-0.1em}& \hspace*{-0.1em}\hspace*{-0.1em} &\hspace*{-0.1em}
-\overline{L}_{1}\hspace*{-0.2em}^{\hspace*{-0.2em}\top}\overline{P}\overline{L}_{1})\overline{\zeta}(k), \notag \\
\Delta \overline{W}_{\hspace*{-0.2em}1}\hspace*{-0.1em}& \hspace*{-0.1em}=\hspace*{-0.1em} &\hspace*{-0.1em}
\overline{\zeta}^{\top}(k)\mathrm{Sym}(\overline{R}_{1}^{\hspace*{-0.2em}\top}
\overline{R}\overline{R}_{2}\hspace*{-0.2em}-\hspace*{-0.2em}\overline{R}_{1}^{\hspace*{-0.2em}\top}
\overline{R}\overline{R}_{6}\hspace*{-0.2em}+\hspace*{-0.2em}(k\hspace*{-0.2em}-\hspace*{-0.2em}k_{q})
\overline{R}_{3}^{\hspace*{-0.2em}\top}\overline{R}\overline{R}_{7} \notag\\
 \hspace*{-0.1em}& \hspace*{-0.1em}\hspace*{-0.1em} &\hspace*{-0.1em}
+(k_{q+1}\hspace*{-0.2em}-\hspace*{-0.2em}k)
\overline{R}_{8}^{\hspace*{-0.2em}\top}\overline{R}\overline{R}_{4})\overline{\zeta}(k), \notag \\
\Delta \overline{W}_{\hspace*{-0.2em}2}\hspace*{-0.1em}& \hspace*{-0.1em}=\hspace*{-0.1em} &\hspace*{-0.1em}
(k_{q\hspace*{-0.1em}+\hspace*{-0.1em}1}\hspace*{-0.3em}-\hspace*{-0.3em}k\hspace*{-0.3em}-\hspace*{-0.3em}1)
\overline{\zeta}^{\hspace*{-0.1em}\top}\hspace*{-0.3em}(k)\overline{\mathcal{G}}^{\top}_{1}\overline{T}_{1}\overline{\mathcal{G}}_{1}
\overline{\zeta}(k)
\hspace*{-0.3em}-\hspace*{-0.3em}\scalebox{1}{\(\sum\limits\)}_{s=k_{q}}^{k-1}\hspace*{-0.2em}\|y(s)\|_{\overline{T}_{1}}^{2}, \notag \\
\Delta \overline{W}_{\hspace*{-0.2em}3}\hspace*{-0.1em}& \hspace*{-0.1em}=\hspace*{-0.1em} &\hspace*{-0.1em}
(k\hspace*{-0.3em}+\hspace*{-0.3em}1\hspace*{-0.3em}-\hspace*{-0.3em}k_{q})
\overline{\zeta}^{\hspace*{-0.1em}\top}\hspace*{-0.3em}(k)\overline{\mathcal{G}}^{\top}_{1}
\hspace*{-0.2em}\overline{T}_{2}\overline{\mathcal{G}}_{1}\overline{\zeta}(k)\hspace*{-0.3em}-\hspace*{-0.3em}
\scalebox{1}{\(\sum\limits\)}_{s=k}^{k_{q\hspace*{-0.1em}+\hspace*{-0.1em}1}-\hspace*{-0.1em}1}\hspace*{-0.2em}
\|y(s)\|_{\overline{T}_{2}}^{2}, \label{91}
\end{eqnarray}
in which $\overline{\zeta}(k)\hspace*{-0.2em}\triangleq\hspace*{-0.2em}\mathrm{Col}(\eta_{k},m_{k},\omega_{k},
x_{k_{q}},x_{k_{q+1}},\sum_{s=k_{q}}^{k}\hspace*{-0.3em}x_{s}/(k\hspace*{-0.3em}-\hspace*{-0.3em}k_{q}\hspace*{-0.2em}+\hspace*{-0.2em}1),
\sum_{s=k}^{k_{q\hspace*{-0.1em}+\hspace*{-0.1em}1}}\hspace*{-0.3em}x_{s}/\hspace*{-0.1em}(k_{q\hspace*{-0.1em}+\hspace*{-0.1em}1}
\hspace*{-0.3em}-\hspace*{-0.3em}k\hspace*{-0.2em}+\hspace*{-0.2em}1))
\hspace*{-0.3em}\in\hspace*{-0.3em}\mathds{R}^{\tilde{n}}$, with $\eta_{k},m_{k},\omega_{k},x_{k_{q}}$ prescribed as in \eqref{81} and
$x_{k_{q+1}}\hspace*{-0.3em}\triangleq\hspace*{-0.2em}
x(k_{q\hspace*{-0.1em}+\hspace*{-0.1em}1}\hspace*{-0.1em})\hspace*{-0.3em}-\hspace*{-0.3em}x^{*}$. By virtue of the nonlinearity
isolation of $\pi_{\mathrm{DNN}}$ in \eqref{dnn3}, we have
\begin{equation}
\label{uu2}
\begin{array}{rcl}
\bar{u}(k)\hspace*{-0.2em}-\hspace*{-0.2em}\bar{u}^{*}& \hspace*{-0.2em}=\hspace*{-0.2em} &
\Sigma_{1}\mathrm{Col}(x_{k_{q}},m_{k},\omega_{k})\hspace*{-0.2em}=\hspace*{-0.2em}\Sigma_{1}\overline{\Lambda}_{1}\zeta(k),
\end{array}
\end{equation}
which is utilized for deriving $\Delta \overline{V}$ in \eqref{91}. Then, in the light of the summation inequality \citep*[Corollary 3]{Chen2017Novel},
we can further relax the summation terms of $\Delta \overline{W}_{\hspace*{-0.2em}2}(k)$ and $\Delta \overline{W}_{\hspace*{-0.2em}3}(k)$ by
\begin{eqnarray}
-\scalebox{1}{\(\sum\limits\)}_{s=k_{q}}^{k-1}\|y(s)\|_{\overline{T}_{1}}^{2}\hspace*{-0.1em}& \hspace*{-0.1em}\leq\hspace*{-0.1em} &\hspace*{-0.1em}
(k\hspace*{-0.2em}-\hspace*{-0.2em}k_{q})\overline{\zeta}^{\top}(k)\overline{N}_{1}^{\top}\overline{\mathcal{T}}_{1}^{-1}\overline{N}_{1}
\overline{\zeta}(k) \notag \\
\hspace*{-0.1em}& \hspace*{-0.1em}\hspace*{-0.1em} &\hspace*{-0.1em}
+\overline{\zeta}^{\top}(k)\mathrm{Sym}(\overline{N}_{1}\overline{R}_{9})\overline{\zeta}(k), \label{92}\\
-
\scalebox{1}{\(\sum\limits\)}_{s=k}^{k_{q+1}-\hspace*{-0.1em}1}\|y(s)\|_{\overline{T}_{2}}^{2}
\hspace*{-0.1em}& \hspace*{-0.1em}\leq\hspace*{-0.1em} &\hspace*{-0.1em}
(k_{q+1}\hspace*{-0.2em}-\hspace*{-0.2em}k)\overline{\zeta}^{\top}(k)\overline{N}_{2}^{\top}\overline{\mathcal{T}}_{2}^{-1}
\overline{N}_{2}\overline{\zeta}(k) \notag \\
\hspace*{-0.1em}& \hspace*{-0.1em}\hspace*{-0.1em} &\hspace*{-0.1em}
+\overline{\zeta}^{\top}(k)\mathrm{Sym}(\overline{N}_{2}\overline{R}_{10})\overline{\zeta}(k), \label{93}
\end{eqnarray}
with $\overline{\mathcal{T}}_{\imath}\hspace*{-0.2em}\triangleq\hspace*{-0.2em}\mathrm{Diag}(\overline{T}_{\imath},3\overline{T}_{\imath})$.
Moreover, combining \eqref{dnn3} with \eqref{dnnqc} for the convex relaxation of $\pi_{\mathrm{DNN}}$ leads to
\begin{eqnarray}
\overline{\zeta}^{\top}(k)\overline{\mathcal{G}}_{3}^{\top}\mathcal{M}_{\mathrm{DNN}}
\overline{\mathcal{G}}_{3}\zeta(k)\hspace*{-0.2em}\geq\hspace*{-0.2em}0. \label{95}
\end{eqnarray}
Hence, by virtue of \eqref{iqc}, \eqref{91}, and \eqref{92}-\eqref{95} combined with the self-triggered scheme \eqref{s1} and \eqref{sts1}, we obtain
\begin{eqnarray}
&\hspace*{-0.1em}\hspace*{-0.1em}&\Delta \overline{\mathcal{V}}(\eta(k),k)\hspace*{-0.2em}+\hspace*{-0.2em}\|r(k)\hspace*{-0.2em}-\hspace*{-0.2em}r^{*}\|_{M_{\Theta}}^{2}
\hspace*{-0.2em}+\hspace*{-0.2em}\overline{\zeta}^{\hspace*{-0.2em}\top}\hspace*{-0.2em}(k)\overline{\mathcal{G}}_{3}^{\top}
\mathcal{M}_{\mathrm{DNN}}\overline{\mathcal{G}}_{3}\overline{\zeta}(k) \notag \\
&\hspace*{-0.1em}\hspace*{-0.1em}&-
\mathcal{S}(x(k_{q}),s_{k})\hspace*{-0.2em}<\hspace*{-0.2em} 0, \label{96}
\end{eqnarray}
holding for $k\hspace*{-0.2em}\in\hspace*{-0.2em}[k_{q},k_{q\hspace*{-0.1em}+\hspace*{-0.1em}1}\hspace*{-0.3em}-\hspace*{-0.3em}1]$, which implies that
\begin{eqnarray}
&\hspace*{-0.1em}\hspace*{-0.1em}&\overline{\zeta}^{\hspace*{-0.2em}\top}\hspace*{-0.2em}(k)\hspace*{-0.2em}
\left(\overline{\mathcal{G}}\hspace*{-0.2em}+\hspace*{-0.2em}
(k\hspace*{-0.2em}-\hspace*{-0.2em}k_{q})\overline{\amalg}_{1}\hspace*{-0.2em}+\hspace*{-0.2em}
(k_{q+1}\hspace*{-0.2em}-\hspace*{-0.2em}k)\overline{\amalg}_{2}\hspace*{-0.2em}+\hspace*{-0.2em}(k\hspace*{-0.2em}-\hspace*{-0.2em}k_{q})
\overline{N}_{1}^{\top}\overline{\mathcal{T}}_{1}^{-1}\overline{N}_{1}
\right. \notag \\
&\hspace*{-0.1em}\hspace*{-0.1em}&+\left.(k_{q+1}\hspace*{-0.2em}-\hspace*{-0.2em}k)
\overline{N}_{2}^{\top}\overline{\mathcal{T}}_{2}^{-1}\overline{N}_{2}
\right)\overline{\zeta}(k)\hspace*{-0.2em}<\hspace*{-0.2em} 0, \label{97}
\end{eqnarray}
is satisfied with the terms defined as
\begin{eqnarray*}
\overline{\mathcal{G}}\hspace*{-0.1em} & \hspace*{-0.1em}\triangleq\hspace*{-0.1em} &\hspace*{-0.1em}
(A\overline{L}_{1}\hspace*{-0.3em}+\hspace*{-0.3em}B\Sigma_{1}\overline{\Lambda}_{1})^{\hspace*{-0.2em}\top}\hspace*{-0.1em}
\overline{P}(A\overline{L}_{1}\hspace*{-0.3em}+\hspace*{-0.3em}B\Sigma_{1}\overline{\Lambda}_{1})
\hspace*{-0.2em}+\hspace*{-0.2em}
\overline{\mathcal{G}}_{1}^{\hspace*{-0.2em}\top}(\overline{T}_{2}\hspace*{-0.2em}-\hspace*{-0.2em}\overline{T}_{1})
\overline{\mathcal{G}}_{1}\\
\hspace*{-0.1em} & \hspace*{-0.1em}\hspace*{-0.1em} &\hspace*{-0.1em}
+\overline{\mathcal{G}}_{2}^{\top}M_{\Theta}
\overline{\mathcal{G}}_{2}\hspace*{-0.2em}+\hspace*{-0.2em}\overline{\mathcal{G}}_{3}^{\top}\hspace*{-0.2em}\mathcal{M}_{\mathrm{DNN}}
\overline{\mathcal{G}}_{3}\hspace*{-0.2em}+\hspace*{-0.2em}\mathrm{Sym}
(\overline{R}_{1}^{\top}\hspace*{-0.2em}\overline{R}\overline{R}_{2}
\hspace*{-0.2em}-\hspace*{-0.3em}\overline{R}_{5}^{\top}\hspace*{-0.2em}\overline{R}\overline{R}_{6}\\
\hspace*{-0.4em} & \hspace*{-0.5em}\hspace*{-0.5em} &\hspace*{-0.4em}
+\overline{N}_{1}\hspace*{-0.1em}\overline{R}_{9}\hspace*{-0.3em}+\hspace*{-0.3em}
\overline{N}_{2}\hspace*{-0.1em}\overline{R}_{10})
\hspace*{-0.3em}-\hspace*{-0.3em}\overline{R}_{11}^{\top}\widetilde{\Xi}\overline{R}_{11}\hspace*{-0.3em}-\hspace*{-0.3em}
\overline{L}_{1}^{\top}\overline{P}\overline{L}_{1}, \\
\overline{\amalg}_{1}\hspace*{-0.1em} & \hspace*{-0.1em}\triangleq\hspace*{-0.1em} &\hspace*{-0.1em}
\overline{\mathcal{G}}_{1}^{\top}\hspace*{-0.2em}\overline{\mathcal{T}}_{2}\overline{\mathcal{G}}_{1}\hspace*{-0.3em}+\hspace*{-0.3em}
\mathrm{Sym}(\overline{R}_{3}^{\top}\hspace*{-0.2em}\overline{R}\overline{R}_{7}),
\overline{\amalg}_{2}\hspace*{-0.2em}\triangleq\hspace*{-0.2em}
\overline{\mathcal{G}}_{1}^{\top}\hspace*{-0.2em}\overline{\mathcal{T}}_{1}\overline{\mathcal{G}}_{1}\hspace*{-0.3em}+\hspace*{-0.3em}
\mathrm{Sym}(\overline{R}_{8}^{\top}\hspace*{-0.2em}\overline{R}\overline{R}_{4}).
\end{eqnarray*}
Note that the conditions $0\hspace*{-0.2em}\leq\hspace*{-0.2em}k\hspace*{-0.2em}-\hspace*{-0.2em}k_{q}\hspace*{-0.2em}\leq\hspace*{-0.2em}
s_{k}\hspace*{-0.2em}\leq\hspace*{-0.2em}\bar{s}$ and
$0\hspace*{-0.2em}<\hspace*{-0.2em}k_{q+1}\hspace*{-0.2em}-\hspace*{-0.2em}k\hspace*{-0.2em}\leq
s_{k}\hspace*{-0.3em}\leq\hspace*{-0.3em}\bar{s}$ hold obviously within the time interval $[k_{q},k_{q\hspace*{-0.1em}+\hspace*{-0.1em}1}\hspace*{-0.3em}-\hspace*{-0.3em}1]$, thus
based on \citep*[Lemma 2]{Wan2021Dynamic}, there exist scalars $\overline{\lambda}_{\imath},\imath\hspace*{-0.2em}\in\hspace*{-0.2em}\mathds{N}_{[1,2]}$,
such that the inequalities
\begin{eqnarray}
(k\hspace*{-0.2em}-\hspace*{-0.2em}k_{q})\mathrm{Sym}(\overline{R}_{3}^{\top}\hspace*{-0.2em}
\overline{R}\overline{R}_{7})
\hspace*{-0.1em} & \hspace*{-0.1em}\leq\hspace*{-0.1em} &\hspace*{-0.1em}\bar{s}
(\overline{\lambda}_{1}^{-1}\overline{R}_{3}^{\top}\hspace*{-0.2em}\overline{R}\overline{R}^{\top}\hspace*{-0.2em}
\overline{R}_{3}\hspace*{-0.2em}+\hspace*{-0.2em}\overline{\lambda}_{1}\overline{R}_{7}^{\top}
\hspace*{-0.2em}\overline{R}_{7}),\\
(k_{q\hspace*{-0.1em}+\hspace*{-0.1em}1}\hspace*{-0.3em}-\hspace*{-0.2em}k)
\mathrm{Sym}(\overline{R}_{8}^{\top}\hspace*{-0.2em}\overline{R}\overline{R}_{4})
\hspace*{-0.1em} & \hspace*{-0.1em}\leq\hspace*{-0.1em} &\hspace*{-0.1em}\bar{s}
(\overline{\lambda}_{2}^{-1}\overline{R}_{8}^{\top}\hspace*{-0.2em}\overline{R}\overline{R}^{\top}\hspace*{-0.2em}
\overline{R}_{8}\hspace*{-0.2em}+\hspace*{-0.2em}\overline{\lambda}_{2}\overline{R}_{4}^{\top}\hspace*{-0.2em}
\overline{R}_{4}), \label{198}
\end{eqnarray}
hold obviously. In addition, allowing for the terms
$\overline{\mathcal{G}}_{1}^{\top}\hspace*{-0.2em}\overline{\mathcal{T}}_{\imath}\overline{\mathcal{G}}_{1}\hspace*{-0.2em}\succ0$ for $\imath\hspace*{-0.2em}\in\hspace*{-0.2em}\mathds{N}_{[1,2]}$ with $\overline{\mathcal{T}}_{\imath}\hspace*{-0.2em}\succ\hspace*{-0.2em}0$,
the conditions \eqref{97}-\eqref{198} yield
\begin{eqnarray}
&\hspace*{-0.1em}\hspace*{-0.1em}&
\left(\overline{\mathcal{G}}\hspace*{-0.3em}+\hspace*{-0.3em}
\bar{s}(\overline{\mathcal{G}}_{1}^{\top}\overline{\mathcal{T}}_{2}
\overline{\mathcal{G}}_{1}\hspace*{-0.3em}+\hspace*{-0.3em}\overline{\mathcal{G}}_{1}^{\top}\overline{\mathcal{T}}_{1}\overline{\mathcal{G}}_{1}
\hspace*{-0.3em}+\hspace*{-0.3em}\overline{\lambda}_{1}\hspace*{-0.2em}\overline{R}_{7}^{\top}
\hspace*{-0.2em}\overline{R}_{7}\hspace*{-0.3em}+\hspace*{-0.3em}\overline{\lambda}_{2}
\overline{R}_{4}^{\top}\hspace*{-0.2em}
\overline{R}_{4})
\right. \notag \\
&\hspace*{-0.1em}\hspace*{-0.1em}&\left.+\bar{s}(\overline{N}_{1}^{\top}\hspace*{-0.2em}
\overline{\mathcal{T}}_{1}^{-\hspace*{-0.1em}1}\overline{N}_{1}+
\overline{N}_{2}^{\top}\overline{\mathcal{T}}_{2}^{-\hspace*{-0.1em}1}\overline{N}_{2})\hspace*{-0.2em}+\hspace*{-0.2em}
\bar{s}\overline{\lambda}_{1}^{-1}\overline{R}_{3}^{\top}\hspace*{-0.2em}\overline{R}\overline{R}^{\top}\hspace*{-0.2em}
\overline{R}_{3} \right. \notag \\
&\hspace*{-0.1em}\hspace*{-0.1em}&\left.
+\bar{s}\overline{\lambda}_{2}^{-1}\overline{R}_{8}^{\top}\hspace*{-0.2em}\overline{R}\overline{R}^{\top}\hspace*{-0.2em}
\overline{R}_{8}
\right)\hspace*{-0.2em}\prec\hspace*{-0.2em} 0. \label{99}
\end{eqnarray}
Accordingly, by using Schur complement for \eqref{99}, the first LMI in \eqref{ma2} can be obtained, which
is convex for all decision variables. Besides, the feasibility of \eqref{99} implies that the condition \eqref{96} holds obviously.
Note that the third term on the left side of \eqref{96} is greater than zero, allowing for the local sector
quadratic constraint \eqref{dnnqc}. Then, $\mathcal{S}\hspace*{-0.1em}(\hspace*{-0.1em}x(k_{q})\hspace*{-0.1em},s_{k}\hspace*{-0.1em})\hspace*{-0.3em}<\hspace*{-0.3em}0$ holds during the
execution interval $[k_{q},k_{q\hspace*{-0.1em}+\hspace*{-0.1em}1}\hspace*{-0.3em}-\hspace*{-0.3em}1]$. We
sum the remainder of \eqref{96} from $k\hspace*{-0.2em}=\hspace*{-0.2em}k_{q}$ to $k_{q+1}$, leading to 
\begin{eqnarray}
& \hspace*{-0.1em}\hspace*{-0.1em}&
\scalebox{1}{\(\sum\limits\)}_{k=k_{q}}^{k_{q\hspace*{-0.1em}+\hspace*{-0.1em}1}\hspace*{-0.1em}-\hspace*{-0.1em}1}
\left(\Delta\overline{\mathcal{V}}(\eta(k),k)\hspace*{-0.2em}+\hspace*{-0.2em}
\|r(k)\hspace*{-0.2em}-\hspace*{-0.2em}r^{*}\|_{M_{\Theta}}^{2}\right) \notag \\
& \hspace*{-0.1em}=\hspace*{-0.1em}&
\overline{V}(\eta(k_{q+1}))\hspace*{-0.2em}-\hspace*{-0.2em}\overline{V}(\eta(k_{q}))\hspace*{-0.2em}+\hspace*{-0.2em}
\scalebox{1}{\(\sum\limits\)}_{k=k_{q}}^{k_{q\hspace*{-0.1em}+\hspace*{-0.1em}1}}
\|r(k)\hspace*{-0.2em}-\hspace*{-0.2em}r^{*}\|_{M_{\Theta}}^{2} \notag \\
& \hspace*{-0.1em}\leq\hspace*{-0.1em}&
-\hbar_{2}\|\overline{\zeta}(k_{q})\|^{2}\hspace*{-0.3em}<\hspace*{-0.2em}0,
\label{j2}
\end{eqnarray}
which is satisfied for the looped function
$\overline{W}(x(k_{q\hspace*{-0.1em}+\hspace*{-0.1em}1}),k_{q\hspace*{-0.1em}+\hspace*{-0.1em}1})
\hspace*{-0.3em}=\hspace*{-0.3em}\overline{W}(x(k_{q}),k_{q})$ and $\hspace*{-0.1em}\hbar_{2}\hspace*{-0.3em}\in\hspace*{-0.3em}
\mathds{R}_{\hspace*{-0.1em}>\hspace*{-0.1em}0}$. We generalize the time interval from $[k_{q},k_{q\hspace*{-0.1em}+\hspace*{-0.1em}1}\hspace*{-0.3em}-\hspace*{-0.3em}1]$
to $[0,k_{\aleph}]$ with
$\mathcal{\aleph}\hspace*{-0.2em}\in\hspace*{-0.2em}\mathds{N}_{\hspace*{-0.1em}>\hspace*{-0.1em}1}$.
Letting $k_{0}\hspace*{-0.3em}=\hspace*{-0.3em}0$ and allowing for \eqref{j2}, we further obtain
\begin{eqnarray*}
& \hspace*{-0.1em}\hspace*{-0.1em}&\hspace*{-0.1em}
\scalebox{1}{\(\sum\limits\)}_{q=0}^{\aleph-1}\hspace*{-0.2em}
\left(\overline{V}\hspace*{-0.2em}(\eta(k_{q\hspace*{-0.1em}+\hspace*{-0.1em}1}))
\hspace*{-0.2em}-\hspace*{-0.2em}\overline{V}\hspace*{-0.2em}(\eta(k_{q}))\right)\hspace*{-0.3em}+\hspace*{-0.3em}
\scalebox{1}{\(\sum\limits\)}_{q=0}^{\aleph-1}
\scalebox{1}{\(\sum\limits\)}_{k=k_{q}}^{k_{q\hspace*{-0.1em}+\hspace*{-0.1em}1}}
\|r(k)\hspace*{-0.3em}-\hspace*{-0.3em}r^{*}\hspace*{-0.1em}\|_{M_{\Theta}}^{2} \notag \\
& \hspace*{-0.1em}\hspace*{-0.1em}&\hspace*{-0.1em}=\hspace*{-0.1em}
\overline{V}\hspace*{-0.2em}(\eta(k_{\aleph})\hspace*{-0.1em})\hspace*{-0.3em}-\hspace*{-0.3em}
\overline{V}\hspace*{-0.2em}(\eta(0)\hspace*{-0.1em})\hspace*{-0.3em}+\hspace*{-0.3em}
\scalebox{1}{\(\sum\limits\)}_{k=0}^{k_{\aleph}}\|r(k)\hspace*{-0.3em}-\hspace*{-0.3em}
r^{*}\hspace*{-0.1em}\|_{M_{\Theta}}^{2} \notag\\
& \hspace*{-0.1em}\hspace*{-0.1em}&\hspace*{-0.1em}\leq\hspace*{-0.1em}
-\hspace*{-0.3em}\scalebox{1}{\(\sum\limits\)}_{k=0}^{k_{\aleph}}
\hbar_{2}\|\overline{\zeta}(k)\|^{2}\hspace*{-0.2em}.
\end{eqnarray*}
Note that $\scalebox{1}{\(\sum\limits\)}_{k=0}^{k_{\aleph}}\|r(k)\hspace*{-0.3em}-\hspace*{-0.3em}
r^{*}\hspace*{-0.1em}\|_{M_{\Theta}}^{2}\hspace*{-0.2em}\geq\hspace*{-0.2em}0$ holds based on the definition of
IQC in \eqref{iqc}, which yields
\begin{eqnarray}
\overline{V}(\eta(k_{\aleph}))\hspace*{-0.2em}-\hspace*{-0.2em}\overline{V}(\eta(0))
& \hspace*{-0.2em}\leq\hspace*{-0.2em}&-\hspace*{-0.3em}\scalebox{1}{\(\sum\limits\)}_{k=0}^{k_{\aleph}}
\hbar_{2}\|\overline{\zeta}(k)\|^{2}\hspace*{-0.2em}. \label{994}
\end{eqnarray}
Hence, the asymptotical stability of $\overline{V}(\eta(k))$ is ensured, namely, $\eta(k)$ converges to $\eta^{*}$ as
$k$ tends to $\hspace*{-0.1em}\infty\hspace*{-0.1em}$. Similar to the proof as aforementioned, the second inequality of
\eqref{ma2} shows the prerequisite of local quadratic constraint of $\pi_{\mathrm{DNN}}$ \eqref{dnnqc}, that is,
$\hspace*{-0.1em}p_{1}\hspace*{-0.3em}\in\hspace*{-0.3em}[\underline{p}_{1}
\hspace*{-0.1em},\hspace*{-0.1em}\overline{p}_{1}\hspace*{-0.1em}]$ holds if
$\hspace*{-0.1em}x(k)\hspace*{-0.3em}\in\hspace*{-0.3em}\mathcal{E}_{\hspace*{-0.1em}P_{1}}(x^{*}\hspace*{-0.1em})$, in view of the
constrained quadratic \citep*[Lemma 1]{Hindi1998Analysis} and the IBP specified in Remark \ref{ibp}. Thus,
the local property is reasonable for self-triggered $\pi_{\mathrm{DNN}}$.

We perform an inner-approximation of RoA based on the previous analysis. Recalling the inequality \eqref{994}, it is intuitive that
$\eta(k)\hspace*{-0.3em}\in\hspace*{-0.3em}
\mathcal{E}_{\hspace*{-0.1em}\overline{P}}(\eta^{*}\hspace*{-0.1em})
\hspace*{-0.3em}\triangleq\hspace*{-0.3em}
\{\eta\hspace*{-0.3em}\in\hspace*{-0.3em}\mathds{R}^{n}|\|\eta\hspace*{-0.2em}-\hspace*{-0.2em}\eta^{*}\|^{2}_{\overline{P}}
\hspace*{-0.3em}\leq\hspace*{-0.3em}1\}$ holds for an initial state
$\eta(0)\hspace*{-0.3em}\in\hspace*{-0.3em}\mathcal{E}_{\hspace*{-0.1em}P}(\eta^{*}\hspace*{-0.1em})$.
Based on the initial value $\xi(0)\hspace*{-0.3em}=\hspace*{-0.3em}0$ of virtual filter \eqref{fil},
we have that $\eta(0)\hspace*{-0.3em}\in\hspace*{-0.3em}
\mathcal{E}_{\hspace*{-0.1em}\overline{P}}(\eta^{*}\hspace*{-0.1em})$ yields $x(0)\hspace*{-0.3em}\in\hspace*{-0.3em}
\mathcal{E}_{\hspace*{-0.1em}\overline{P}_{1}}\hspace*{-0.2em}(x^{*}\hspace*{-0.1em})$, which is an
inner approximation of robust RoA defined by \eqref{rroa}, with $\overline{P}_{1}$ denoting the upper left block of $\overline{P}$.
The proof is completed.

Similar to \eqref{roa1}, the \emph{largest} RoA estimation can be obtained by solving the optimization problem
\begin{eqnarray}
\min\limits_{\overline{P}\succ0,\overline{T}_{\imath}\succ0,\check{\Xi}_{\imath}\succ0,\gamma\geq0,\overline{\lambda}_{\imath}\geq0,
\overline{R},\overline{N}_{\imath}}\hspace*{-0.2em}&&\log(\mathrm{Det}(\overline{P}_{1})\hspace*{-0.1em}), \notag \\
 &&\text{s.t.}\ \eqref{ma2} \ \text{holds},
\label{roa2}
\end{eqnarray}
which is convex about all decision variables defined in Theorem \ref{th2}.
\begin{remark}
\label{rrr8}
\hspace*{-0.2em} The upper bound $\overline{p}_{1}$ of the first-layer
peractivation of $\pi_{\mathrm{DNN}}$ \eqref{dnn} explicitly affects the size of inner approximation of robust RoA as discussed in
Remark \ref{opt1}. Different from the event-triggered scheme in Section \ref{sec3}, Theorem \ref{th2} captures the relation between
the upper bound of transmission interval $s_{k}$ and the feasibility of \eqref{roa2}. Larger $\bar{s}$ is conducive to reduce transmission times for improving communication efficiency, but the obtained RoA estimation may be more conservative at the cost of sacrificing a stability metric
$\mathrm{Det}(\overline{P}_{1})$. In addition, the feasibility of \eqref{roa2} reveals that multiple self-triggered policies satisfying
$\hspace*{-0.2em}s_{k}\hspace*{-0.2em}\leq\hspace*{-0.2em}\bar{s}\hspace*{-0.1em}$ will lead to the same inner approximation of robust RoA
$\mathcal{E}_{\hspace*{-0.1em}\overline{P}_{1}}\hspace*{-0.1em}(x^{*}\hspace*{-0.1em})$, which implies more generality and
flexibility. Considering a tradeoff between communication burden and stability assessment, it is potential to capture the \emph{largest} RoA approximation in view of determining the \emph{best} upper bound $\hspace*{-0.1em}\bar{s}\hspace*{-0.1em}$ of transmission intervals of self-triggered DNN control scheme \eqref{aaug}.
\end{remark}
\begin{remark}
We make discussions about the novelty of Theorem \ref{th2} in what follows, since fewer results have focused on designing 
self-triggered DNN control schemes. First, similar to Remark \ref{306}, the convex optimization \eqref{roa2} builds upon the
association of the self-triggered transmissions \eqref{sts1} and the neural-feedback loops \eqref{aug}, leading to both better
adaptivity to uncertainty impacts and more efficient
communication transmission. Besides, we establish a relationship between the upper bound of
self-triggered transmission intervals and the stability metric evaluated by $\mathrm{Det}(P_{1})$, which was not reported in the existing
literature. 
Second, we consider the local sector-bounded attribute of nonlinear $\pi_{\mathrm{DNN}}$ and
relax the nonlinearity to conduct the
controller-dynamics association. Apart from the discussions in Remark \ref{306}, Theorem \ref{th2} well generalizes the linear case
\citep*[Section IV]{Wang2022Model} to the nonlinear case, which also implies the potential of control synthesis of self-triggered system 
by virtue of
nonlinearity cancellation/relaxation techniques \citep*{Persis2023Learning}. Third, the LMI-based condition, verifying the closed-loop stability of self-triggered neural-feedback loops, is put forward to determine the triggering parameters in \eqref{st2}, where
looped functions are utilized to reduce conservatism. Particularly, the strict stability criterion \eqref{ma2} for self-triggered
logic \eqref{sts1} is proposed for the first time.
\end{remark}
\begin{figure}[htbp]
\centerline{
\includegraphics[width=8.5cm]{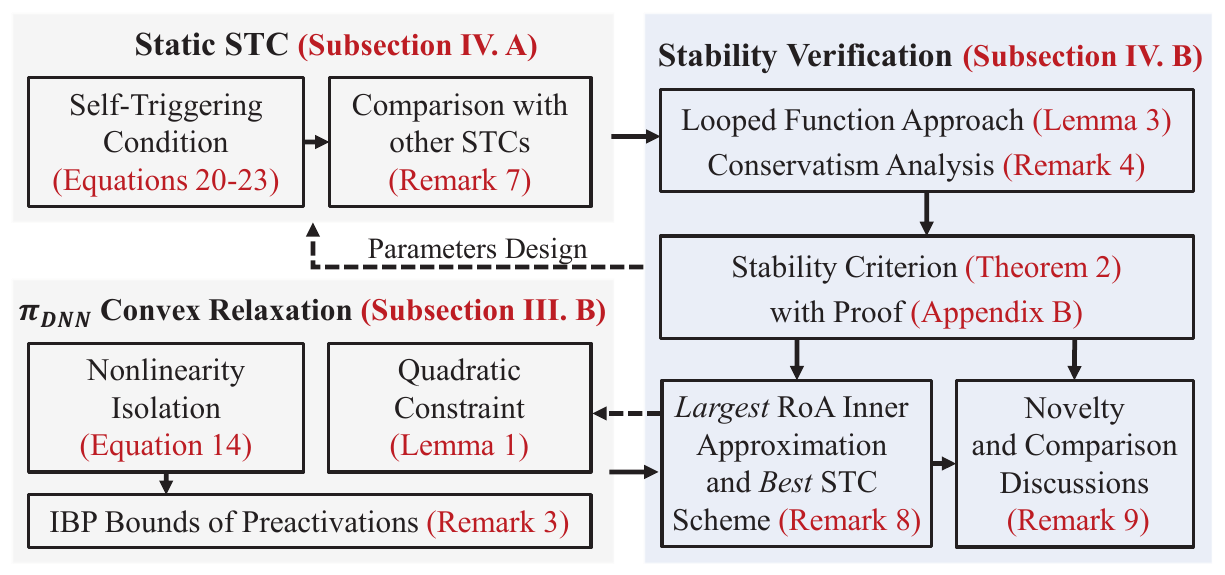} \vspace{-1ex}}
\caption{The architecture of self-triggered neural control in Section \ref{sec4}.}
\label{F3}
\end{figure}

\begin{remark}
We discuss the extensions of the technical results made above. First, considering the system without IQC filter \eqref{fil}, that is, $\hspace*{-0.2em}x(k\hspace*{-0.2em}+\hspace*{-0.2em}1)\hspace*{-0.3em}=\hspace*{-0.3em}
A_{\Gamma}x(k)\hspace*{-0.3em}+\hspace*{-0.3em}B_{\Gamma}u(k)\hspace*{-0.3em}+\hspace*{-0.3em}F_{\Gamma}\omega(k)$ holds only with
$\omega$ representing the bounded uncertainty/fault \citep*{Ma2021Sparse}, we still can perform stability
analyses for aperiodic-sampled neural-feedback loops as Theorems \ref{th1} and \ref{th2}, by virtue of replacing the matrices
$\hspace*{-0.1em}A\hspace*{-0.1em}$ and $\hspace*{-0.1em}B\hspace*{-0.1em}$ with $\hspace*{-0.1em}A_{\Gamma}\hspace*{-0.1em}$ and
$\hspace*{-0.1em}\left[
\begin{array}{cc}
\hspace*{-0.4em}B_{\Gamma}
&F_{\Gamma}\hspace*{-0.4em}
\end{array}%
\right]\hspace*{-0.1em}$, while adjusting the dimensions of $L_{\iota},\iota\hspace*{-0.3em}\in\hspace*{-0.3em}\mathds{N}_{[1,8]}$ and $\overline{L}_{\hslash},\hslash\hspace*{-0.3em}\in\hspace*{-0.3em}\mathds{N}_{[1,7]}$, respectively. Note that
the asymptotic stability degrades to the ultimate boundness, due to the existence of $\hspace*{-0.1em}\omega\hspace*{-0.1em}$
as \citep*[\hspace*{-0.2em}Appendix \hspace*{-0.1em}A\hspace*{-0.1em}]{Ma2021Sparse} \hspace*{-0.1em}and \citep*[\hspace*{-0.2em}Theorem \hspace*{-0.1em}1\hspace*{-0.1em}]{Ye2023Global},  which leads to a  robust invariant set as
  $\hspace*{-0.1em}\overline{\mathcal{R}}_{F}^{x}\hspace*{-0.3em}\triangleq\hspace*{-0.3em}
\left\{\hspace*{-0.1em}x_{0}\hspace*{-0.3em}\in\hspace*{-0.3em}\mathds{R}^{n}\hspace*{-0.1em}|\hspace*{-0.1em}
\|\chi(\infty,\hspace*{-0.1em}x_{0}\hspace*{-0.1em})\hspace*{-0.3em}-\hspace*{-0.3em}x^{*}\hspace*{-0.1em}\|
\hspace*{-0.3em}\leq\hspace*{-0.3em}J(\bar{\omega})\hspace*{-0.1em}\right\}$ with $\hspace*{-0.1em}J(\bar{\omega})\hspace*{-0.1em}$ denoting
a scalar function of the norm bound $\bar{\omega}$.
It is possible to explore inner approximations of RoA based on the positive invariance of reachable sets.
Second, apart from the LMI-based conditions herein, one can
utilize convex relaxation of $\hspace*{-0.1em}\pi_{\mathrm{DNN}}\hspace*{-0.1em}$
combined with sums of squares optimization
to determine inner approximations of RoA for a more general nonlinear systems, for example, the
polynomial systems \citep*[\hspace*{-0.1em}Algorithm \hspace*{-0.1em}1\hspace*{-0.1em}]{Iannelli2019Region}. Moreover, the auxiliary
looped-function approach herein can be leveraged in analyzing aperiodic-sampled neural-feedback loops based on the sums of
squares optimization, which is consistent with \citep*[\hspace*{-0.1em}Remark \hspace*{-0.1em}2\hspace*{-0.1em}]{Iannelli2019Region} and
illuminates the future directions. Third, we suppose that the controller is pretrained by either imitating the experimental
data pairs or approximating the model predictive control policy \citep*{Karg2020Efficient}, which is akin to use deterministic feedback control strategies to design aperiodic-sampled schemes \citep*{Wildhagen2023Data}. In practice, there exist
suboptimal demonstrations and DNN approximation errors for training $\hspace*{-0.1em}\pi_{\mathrm{DNN}}\hspace*{-0.1em}$ \citep*[Theorem 10]{Drummond2022Bounding}, which influence  
setpoint tracking stability. Note that the loop transformation technique combined with
the alternating direction method of multipliers algorithm, for example,
\citep*[Section V]{Yin2022Imitation} and \citep*[Section III]{Pauli2021Training},
reveal a roadmap of learning $\pi_{\mathrm{DNN}}$ with closed-loop performance guarantees, which can be generalized to 
aperiodic-sampled communication transmissions, yielding a codesign of triggering logic and learning strategy.
The relevant issues remain to be investigated in future works.
\end{remark}

\section{Numerical Simulation}
\label{sec5}

In this section, the Euler discretization of inverted pendulum \citep*[\hspace*{-0.1em}Example 1\hspace*{-0.1em}]{Persis2023Learning} is utilized to validate the effectiveness of our
theoretical derivations. The numerical simulation is performed using \texttt{Matlab} with \texttt{CVX} toolkit and
\texttt{MOSEK} solver. The codes are available at https://github.com/Renjie-Ma/Aperiodic-Sampled-DNN-Controller.git.

The inverted pendulum can be formulated by the interconnection of a nominal plant and an
uncertainty part
\begin{eqnarray*}
x_{1}(k\hspace*{-0.2em}+\hspace*{-0.3em}1)& = &
x_{1}(k)\hspace*{-0.2em}+\hspace*{-0.2em}T_{s}x_{2}(k),\\
x_{2}(k\hspace*{-0.2em}+\hspace*{-0.3em}1)& = &
\frac{T_{s}g}{l_{d}}x_{1}(k)\hspace*{-0.2em}+\hspace*{-0.2em}(1\hspace*{-0.3em}-\hspace*{-0.3em}
\frac{T_{s}\mu}{ml_{d}^{2}})x_{2}(k)\hspace*{-0.2em}+\hspace*{-0.3em}\frac{T_{s}}{ml_{d}^{2}}u(k)\\
& &
-\frac{T_{s}g}{l_{d}}\omega(k), \\
\omega(k)& = &
x_{1}(k)\hspace*{-0.2em}-\hspace*{-0.2em}\sin\left(x_{1}(k)\right)\hspace*{-0.2em}=\hspace*{-0.2em}\Theta(\nu(k))\hspace*{-0.1em},
\end{eqnarray*}
where $T_{s}$ is the sampling time interval, $g$ is the acceleration of gravity, $l_{d}$ is the distance from the base
to the center of mass of the balanced body, $\mu$ is the coefficient of rotational friction, and $m$ is the mass.
The system state is $x\hspace*{-0.2em}=\hspace*{-0.2em}\mathrm{Col}(x_{1},x_{2})$, where
$x_{1}\hspace*{-0.1em}$ and $\hspace*{-0.1em}x_{2}\hspace*{-0.1em}$ denote the angular position and the velocity, respectively.
We set $T_{s}\hspace*{-0.2em}=\hspace*{-0.2em}0.01\mathrm{s}$, $g\hspace*{-0.2em}=\hspace*{-0.2em}9.80\mathrm{m}\hspace*{-0.1em}/\hspace*{-0.1em}\mathrm{s}^{2}$,
$l_{d}\hspace*{-0.2em}=\hspace*{-0.2em}0.50\mathrm{m}$,
$\mu\hspace*{-0.2em}=\hspace*{-0.2em}0.05\mathrm{N ms}\hspace*{-0.1em}/\hspace*{-0.1em}\mathrm{rad}$,
and $m\hspace*{-0.2em}=\hspace*{-0.2em}0.15\mathrm{kg}$ for the numerical simulation.
We assume that $\hspace*{-0.1em}x_{1}\hspace*{-0.1em}(k)\hspace*{-0.3em}\in\hspace*{-0.3em}
[\underline{\varphi},\hspace*{-0.1em}\overline{\varphi}]\hspace*{-0.1em}$ with
$\hspace*{-0.1em}\overline{\varphi}\hspace*{-0.2em}=\hspace*{-0.3em}-\hspace*{-0.2em}
\underline{\varphi}\hspace*{-0.2em}=\hspace*{-0.2em}0.73\mathrm{rad}\hspace*{-0.1em}$ and
$\hspace*{-0.1em}u\hspace*{-0.1em}(k)\hspace*{-0.3em}\in\hspace*{-0.3em}
[\underline{u},\hspace*{-0.1em}\overline{u}]\hspace*{-0.1em}$ with
$\hspace*{-0.1em}\overline{u}\hspace*{-0.2em}=\hspace*{-0.3em}-\hspace*{-0.2em}
\underline{u}\hspace*{-0.2em}=\hspace*{-0.2em}0.7\mathrm{Nm}\hspace*{-0.1em}$ hold in practice. The uncertainty $\Theta$ is locally sector-bounded by $\mathrm{Sec}(\textbf{\emph{l}}_{s},\textbf{\emph{m}}_{s})$.
Based on the definition of off-by-one IQC \cite[\hspace*{-0.1em}Lemma \hspace*{-0.1em}8]{Lessard2016Analysis}, we determine the parameters
of virtual filter \eqref{fil} by $\hspace*{-0.1em}A_{\Phi}\hspace*{-0.3em}=\hspace*{-0.3em}0$,
$\hspace*{-0.1em}B_{\Phi}\hspace*{-0.3em}=\hspace*{-0.3em}-\textbf{\emph{l}}_{s}$,
$\hspace*{-0.1em}C_{\Phi}\hspace*{-0.3em}=\hspace*{-0.3em}\mathrm{Col}(1,\hspace*{-0.1em}0)$,
$F_{\Phi}\hspace*{-0.3em}=\hspace*{-0.3em}1$,
$\hspace*{-0.1em}D_{\Phi}\hspace*{-0.3em}=\hspace*{-0.3em}\mathrm{Col}(\textbf{\emph{l}}_{s},\hspace*{-0.1em}-\textbf{\emph{m}}_{s})$, and
$G_{\Phi}\hspace*{-0.3em}=\hspace*{-0.3em}\mathrm{Col}(-\hspace*{-0.1em}1,\hspace*{-0.1em}1)$. Note that the physical
constraints as aforementioned directly yield $\textbf{\emph{l}}_{s}\hspace*{-0.3em}=\hspace*{-0.3em}(\overline{\varphi}\hspace*{-0.2em}-\hspace*{-0.2em}
\sin(\overline{\varphi}))/\overline{\varphi}$ and
$\textbf{\emph{m}}_{s}\hspace*{-0.3em}=\hspace*{-0.3em}0$. \hspace*{-0.1em}Thus, \hspace*{-0.1em}we can determine the system matrices of \eqref{aug} accordingly.

The expert demonstrations can be obtained by solving an
explicit model predictive control policy with \texttt{MPT3} toolkit.
We fit the state-input data pairs by a DNN controller $\pi_{\mathrm{DNN}}$ \eqref{dnn}, which is parameterized
by two hidden layers  $\hspace*{-0.1em}a_{1}\hspace*{-0.3em}=\hspace*{-0.3em}32\hspace*{-0.1em}$ and $a_{2}\hspace*{-0.3em}=\hspace*{-0.3em}32$ with $\tanh$ activation function. The training process is implemented via \texttt{PyTorch} in the same way as in \citep*{Souza2023Event}. 
We assume that $\pi_{\mathrm{DNN}}$ is
bias-free, and the neural-feedback loop \eqref{aug} has a zero equilibrium as explained in Remark \ref{ibp}.

\begin{figure}[htbp]
\centerline{
\includegraphics[width=8.5cm]{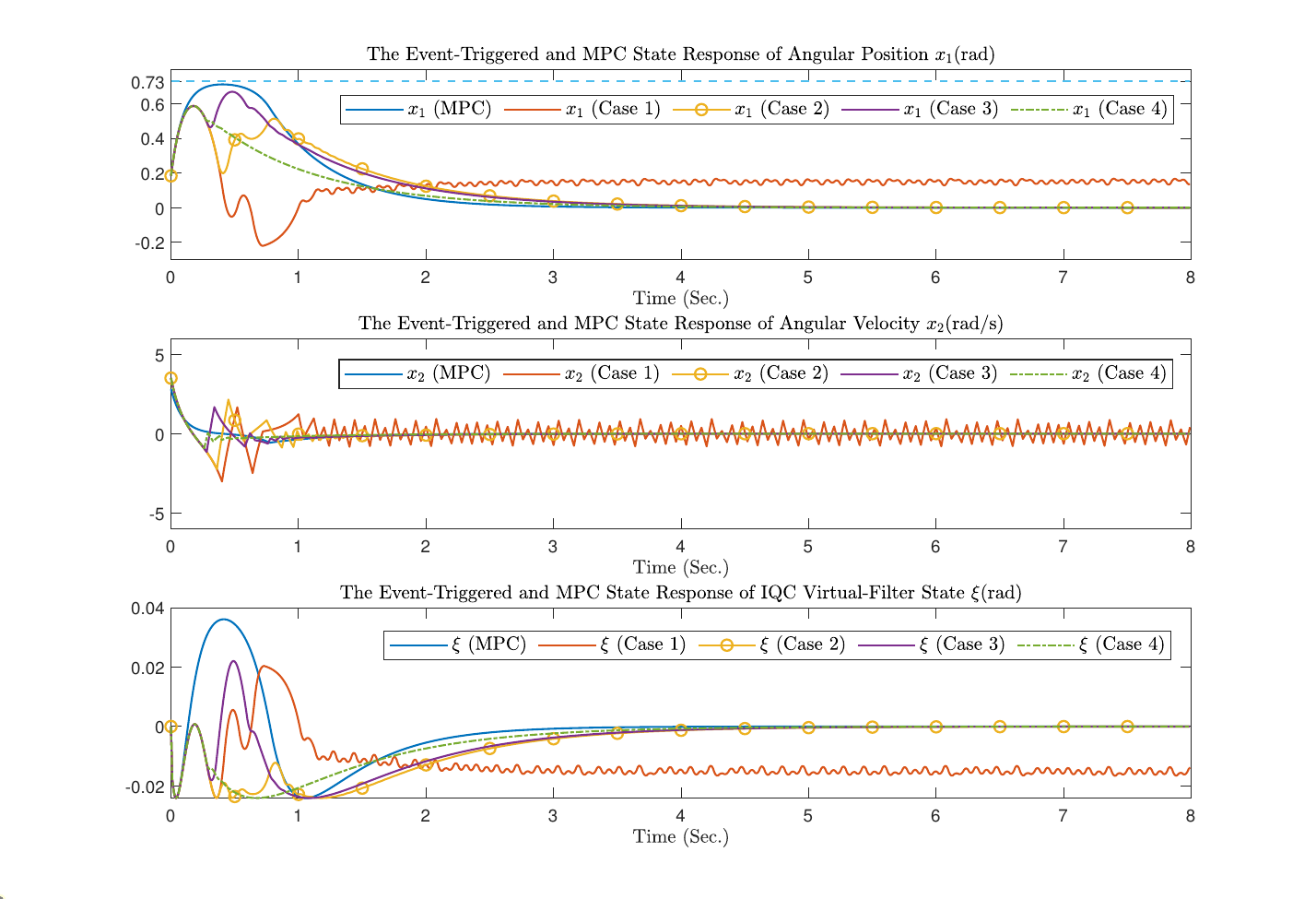} \vspace{-1ex}}
\caption{The state trajectories of event-triggered neural-feedback loops \eqref{aug} and MPC-feedback loops with the angular position $x_{1}$, the angular velocity $x_{2}$, and the IQC virtual filter state $\xi$.}
\label{F4}
\end{figure}

\begin{figure}[htbp]
\centerline{
\includegraphics[width=8.5cm]{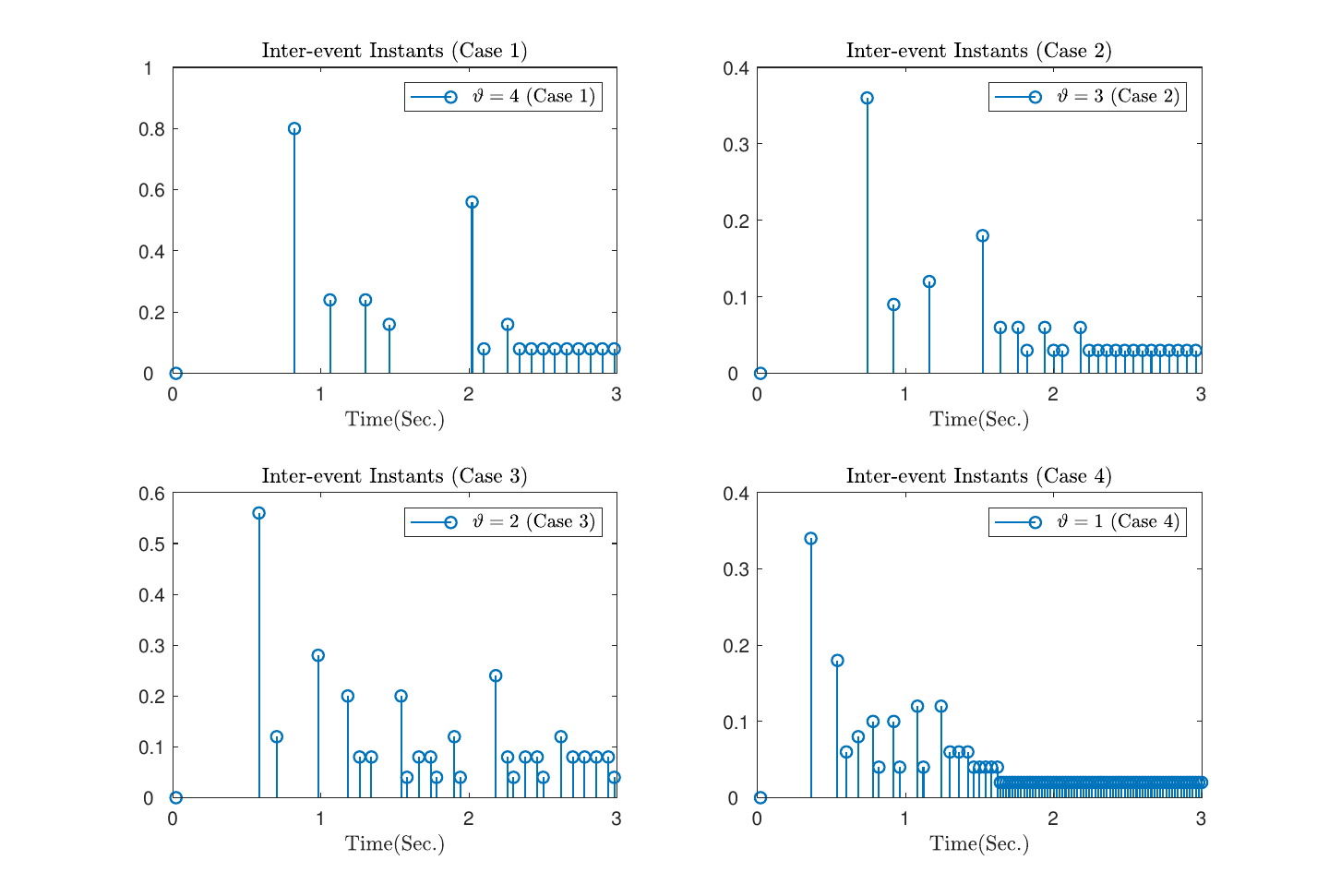} \vspace{-1ex}}
\caption{The triggering instants and intervals of \eqref{ets1} corresponding to different constant sampling intervals $\vartheta$.}
\label{F5}
\end{figure}

\noindent $\mathbf{1)}$ \emph{\textbf{Event-triggered scenario}}. Based on the pretrained $\pi_{\mathrm{DNN}}$, we compile the control policy in Subsection \ref{sec33}. The upper bound of sampling interval $\vartheta_{u}$ is set to be $5$ for ensuring the
feasibility of semidefinite programming \eqref{roa1}. We specify the initial state by $\eta(0)\hspace*{-0.2em}=\hspace*{-0.2em}\mathrm{Col}(0.19,3.5,0)$ and prescribe the scalars $\epsilon_{1}\hspace*{-0.2em}=\hspace*{-0.2em}0.003$
and $\epsilon_{2}\hspace*{-0.2em}=\hspace*{-0.2em}0.002$. Besides, the triggering parameters $\mu$ in \eqref{as3} and $g$ in \eqref{asl} are set to be $0.05$ and $500$, respectively, such that
the constraint $1\hspace*{-0.3em}-\hspace*{-0.3em}\mu\hspace*{-0.3em}-\hspace*{-0.3em}g^{-1}
\hspace*{-0.3em}\geq\hspace*{-0.3em}0$ holds as analyzed in Remark \ref{r1}. The state trajectories $\eta(k)$
of event-triggered neual-feedback loops \eqref{aug} under different
$\vartheta$ satisfying $1\hspace*{-0.2em}=\hspace*{-0.2em}\vartheta_{l}
\hspace*{-0.2em}\leq\hspace*{-0.2em}\vartheta\hspace*{-0.2em}\leq\hspace*{-0.2em}\vartheta_{u}$ can be illustrated in Fig. \ref{F4}, in view of feasible solutions of triggering gains
$\Xi_{1}$ and $\Xi_{2}$ in \eqref{as2}, where \emph{Case \hspace*{-0.1em}1}, \emph{Case \hspace*{-0.1em}2}, \emph{Case \hspace*{-0.1em}3}, and \emph{Case \hspace*{-0.1em}4} correspond to
$\vartheta\hspace*{-0.2em}=\hspace*{-0.2em}4,3,2,1$, respectively. The simulation results reveal that the sampling intervals and their related triggering intervals degrade the closed-loop tracking stability as $\vartheta$ becomes larger. There is a tradeoff between  communication efficiency and stability assurance, which is consistent with
the theoretical analysis made in Remark \ref{opt1}. We cut out the event-triggered 
times and intervals for different $\vartheta$ in Fig. \ref{F5}. 
For a complete time domain, the number of sampling times is $800$, and the numbers of transmission instants are $179$, $242$, $351$, and $739$, which imply that communication efficiencies are improved by
$77.63\%$, $69.75\%$, $56.13\%$, and $7.63\%$, respectively.
	
\begin{figure}[htbp]
\centerline{
\includegraphics[width=8.5cm]{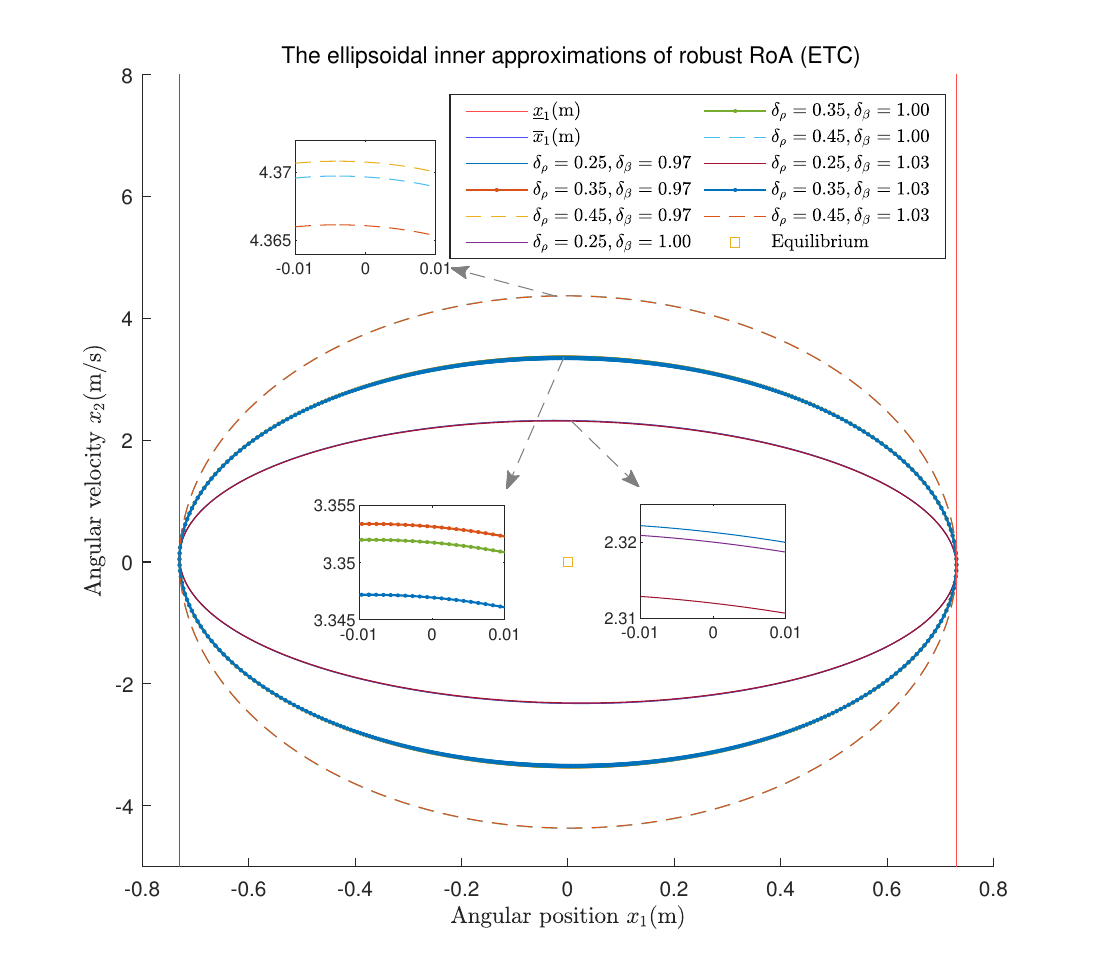} \vspace{-1ex}}
\caption{The ellipsoidal inner approximations of robust RoA with different bounds ($\delta_{\rho},\delta_{\beta}$) of first-layer DNN preactivation under a fixed event-triggered transmission scheme.}
\label{F6}
\end{figure}

We assess the sizes of ellipsoidal inner-approximations of robust RoA for event-triggered neural-feedback loops in
\eqref{aug}. First, we discuss the impacts induced by the local sector bound
attribute $\mathrm{Sec}[\rho,\sigma]\hspace*{-0.1em}$ of nonlinear DNN activation functions, which directly affects the feasible shape matrix $P$ via semidefinite programming
\eqref{roa1}. In Fig. \ref{F6}, we illustrate ellipsoidal inner approximations of robust RoA, introduced by Definition \ref{def2}, with
different bounds $(\delta_{\rho},\delta_{\beta})$ of the first-layer DNN preactivation.
We specify
$p_{1}\hspace*{-0.3em}\in\hspace*{-0.3em}[\underline{p}_{1}\hspace*{-0.1em},\hspace*{-0.1em}\overline{p}_{1}]$ with
$\overline{p}_{1}\hspace*{-0.3em}=\hspace*{-0.3em}\underline{p}_{1}\hspace*{-0.3em}=\hspace*{-0.3em}\delta_{\rho}\mathbf{1}_{a_{1}}$, in which
$\delta_{\rho}\hspace*{-0.2em}\in\hspace*{-0.2em}\mathds{R}_{(0,1)}$ and $\mathbf{1}\hspace*{-0.2em}\in\hspace*{-0.2em}\mathds{R}^{a_{1}}$ whose elements are all ones. 
Based on IBP technique stated in Remark \ref{ibp}, we can calculate
the term $\rho$ and its related $M_{\rho}$ in \eqref{dnnqc}. For comparisons, $\delta_{\rho}$ can be selected by $0.25$, $0.35$, and $0.45$
during the simulation. We prescribe the local upper bound $\sigma\hspace*{-0.2em}=\hspace*{-0.2em}\delta_{\beta}\mathbf{1}_{a_{1}}$, where
$\delta_{\beta}$ can be selected by $0.97$, $1.00$, and $1.03$, respectively.
Fig. \ref{F6} mirrors the results corresponding to several combinations of $\delta_{\rho}$ and
$\delta_{\beta}$. The volumes of ellipsoidal RoA estimations are different, according to simulation results, the best
choices leading to the \emph{largest} RoA estimation are $\delta_{\rho}\hspace*{-0.3em}=\hspace*{-0.3em}0.45$ and $\delta_{\beta}\hspace*{-0.3em}=\hspace*{-0.3em}0.97$. 
Besides, compared to the simulation results in \citep*[\hspace*{-0.1em}Fig. 2\hspace*{-0.1em}]{Souza2023Event}, where the event-triggered scheme located within the interior of a DNN controller, we potentially enlarge the
volumes while satisfying the physical constraint
on the angular position $x_{1}$. Hence, Fig. \ref{F6} is consistent with the
theoretical discussions in Remark \ref{opt1}.

\begin{figure}[htbp]
\centerline{
\includegraphics[width=8.5cm]{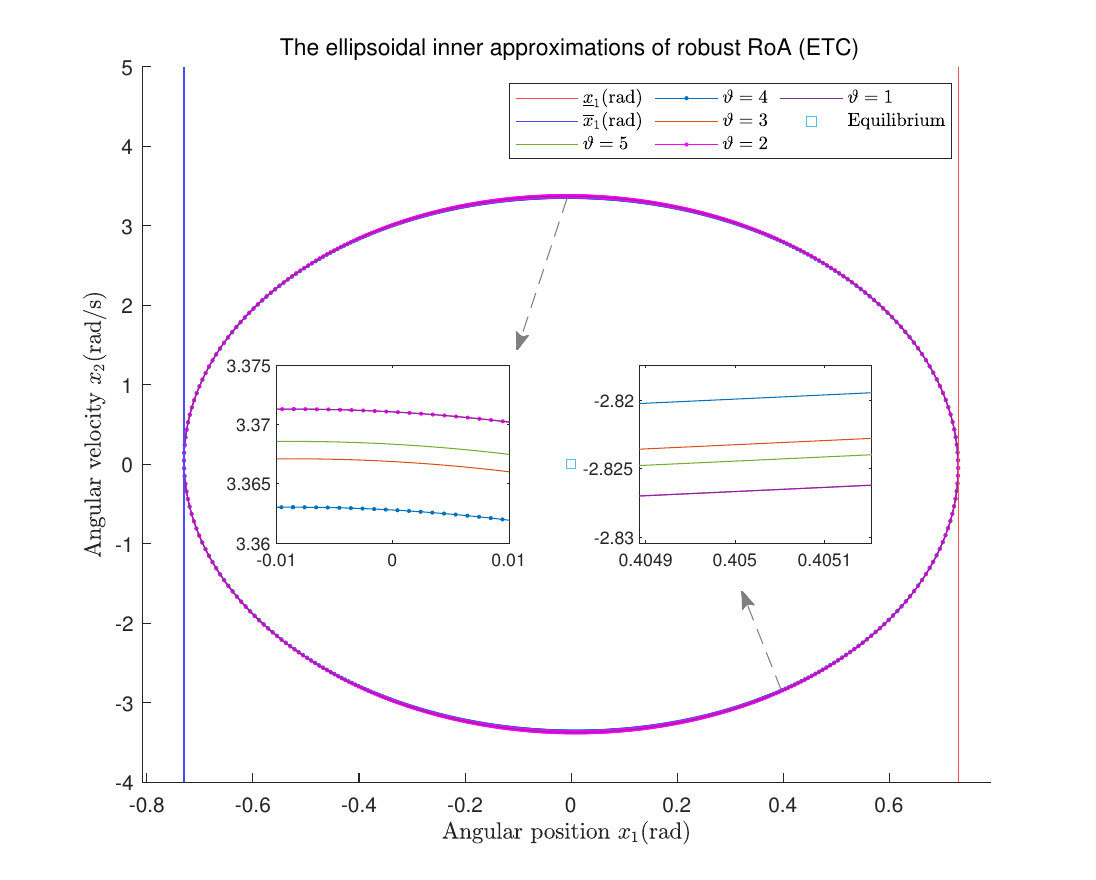} \vspace{-1ex}}
\caption{The ellipsoidal inner approximations of robust
RoA under different  sampling intervals $\vartheta$ of event-triggered transmissions.}
\label{F7}
\end{figure}

Besides, we also explore the impacts of sampling intervals on the
inner approximations of robust RoA. Within this context, we prescribe the parameters of local sector
bound of DNN by
$(\delta_{\rho},\delta_{\beta})\hspace*{-0.3em}=\hspace*{-0.3em}(0.35,1.00)$ for comparisons.
According to the simulation results illustrated in Fig. \ref{F7}, in which the RoA estimations with different $\vartheta$ satisfying
$1\hspace*{-0.2em}=\hspace*{-0.2em}\vartheta_{l}\hspace*{-0.2em}\leq\hspace*{-0.2em}
\vartheta\hspace*{-0.2em}\leq\hspace*{-0.2em}\vartheta_{u}$ are plotted, we obtain that the size of ellipsoidal approximation is decrescent when the sampling interval $\vartheta$ becomes larger. Albeit increasing the value
of $\vartheta$ is beneficial for reducing the
computation burden, the size of RoA inner approximation, acting as a stability metric, can be contractible, which also reveals a tradeoff as in Remark \ref{opt1}.

\begin{figure}[htbp]
\centerline{
\includegraphics[width=8.5cm]{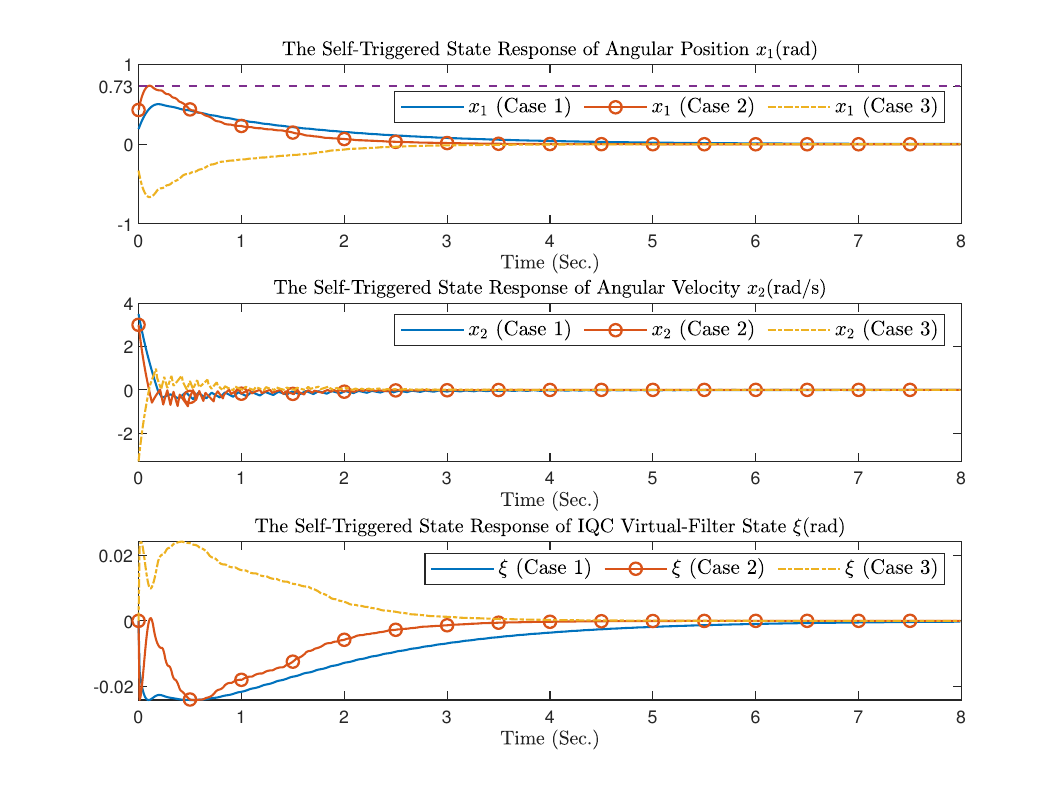} \vspace{-1ex}}
\caption{The state trajectories of self-triggered neural-feedback loops \eqref{aug} with the angular position $x_{1}$, the angular velocity $x_{2}$, and the IQC virtual filter state $\xi$.}
\label{F8}
\end{figure}

\begin{figure}[htbp]
\centerline{
\includegraphics[width=8.5cm]{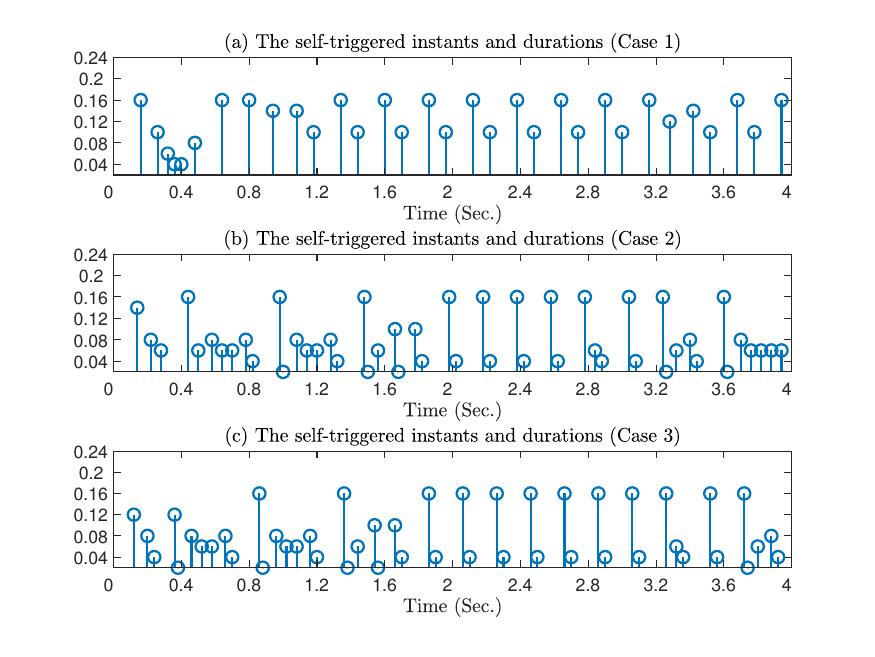} \vspace{-1ex}}
\caption{The self-triggered times and intervals of \eqref{sts1} corresponding to different initial states}
\label{F9}
\end{figure}

\noindent $\mathbf{2)}$ \emph{\textbf{Self-triggered scenario}}. Based on a pretrained $\pi_{\mathrm{DNN}}$, we execute the control policy  in Subsection \ref{ss302}. The self-triggered parameters are prescribed with
$\check{\epsilon}_{1}\hspace*{-0.2em}=\hspace*{-0.2em}0.8$, $\check{\epsilon}_{2}\hspace*{-0.2em}=\hspace*{-0.2em}0.6$, \hspace*{-0.1em}and $\bar{s}\hspace*{-0.2em}=\hspace*{-0.2em}10$. By solving the
semidefinite programming problem \eqref{roa2}, we obtain the feasible solutions $\check{\Xi}_{1}$ and $\check{\Xi}_{2}$ in \eqref{st3}. The closed-loop state trajectories $\eta(k)$ of self-triggered
neural-feedback loops \eqref{aug} under different
initial conditions are illustrated in
Fig. \ref{F8}, in which \emph{Case \hspace*{-0.1em}1}, \emph{Case \hspace*{-0.1em}2}, and \emph{Case \hspace*{-0.1em}3} correspond to
the initial states $\mathrm{Col}(0.19,3.5,0)$,
$\mathrm{Col}(0.43,3.0,0)$, and $\mathrm{Col}(-0.33,-3.3,0)$, respectively. The effectiveness of self-triggered DNN controller to stabilize the system \eqref{aug} is verified. We
illustrate self-triggered times and intervals of \eqref{sts1} in Fig. \ref{F9}.
The number of time steps is $800$, and the numbers 
of triggered times are $222$, $219$, and $218$, such that the communication efficiencies are 
$72.25\%$, $72.63\%$, and $72.75\%$, respectively. Thus, the self-triggered DNN controller is
resource-efficient while ensuring the setpoint tracking stability.

\begin{figure}[htbp]
\centerline{
\includegraphics[width=8.5cm]{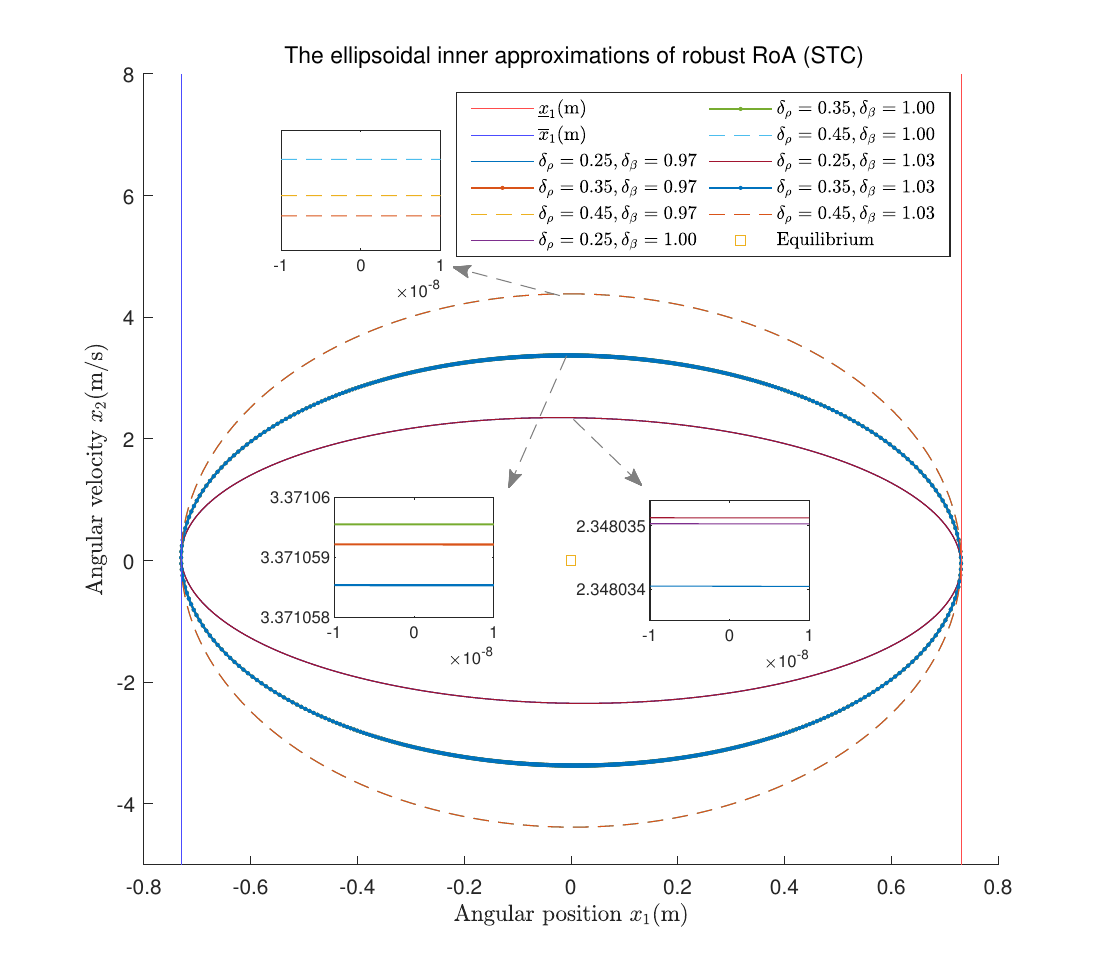} \vspace{-1ex}}
\caption{The ellipsoidal inner approximations of robust
RoA with different bounds ($\delta_{\rho},\delta_{\beta}$) of first-layer DNN preactivation under a fixed self-triggered transmission scheme.}
\label{F10}
\end{figure}

We then assess the sizes of ellipsoidal inner-approximations of robust RoA for the self-triggered neural-feedback loops \eqref{aug}. As we discussed in Remark \ref{rrr8}, the feasibility of ellipsoidal shape matrix $\overline{P}$ in \eqref{roa2}
leads to a same inner approximation of robust RoA $\mathcal{E}_{\hspace*{-0.1em}\overline{P}_{1}}\hspace*{-0.1em}(x^{*}\hspace*{-0.1em})$ for $s_{k}\hspace*{-0.2em}\leq\hspace*{-0.2em}\bar{s}$. We only
need to compare RoA estimations corresponding to different local sector bounds of DNNs. 
We illustrate the simulation results in Fig. \ref{F10}, based
on which, the \emph{largest} RoA estimation matches the
selection $\delta_{\rho}\hspace*{-0.2em}=\hspace*{-0.2em}0.45$ and $\delta_{\beta}=\hspace*{-0.2em}1.03$. Moreover, we also
notice that the volumes of ellipsoidal RoA estimations are increased, as $\delta_{\rho}$ becomes larger to some extent, and
are related with the preactivation bounds of DNNs.
The simulation results are consistent with the discussions in Subsection \ref{ss302}.

\section{Conclusion}
\label{sec6}

This paper has investigated aperiodic-sampled DNN control schemes for a class of uncertain systems, 
whose setpoint tracking stability can be strictly ensured. 
Firstly,
both event- and self-triggered schemes were
incorporated with neural-feedback loops to determine the communication transmission instants. 
Then, the quadratic constraint of DNN, based on the local sector-bounded specifications of activation functions, and
the auxiliary looped functions were utilized for deriving the stability guarantees, leading to convex programming problems, whose 
feasible solutions can be utilized to design the triggering parameters. 
Besides, we assessed the sizes of ellipsoidal RoA inner approximations, acting as a stability metric, which can be correlated with 
the triggering logics and the local attributes of DNN activation functions. 
 Finally, we provided a numerical example of an inverted pendulum to validate the effectiveness of theoretical derivations herein.

\begin{ack}                               
The authors sincerely thank Prof. Dr. Marios M. Polycarpou from the University of Cyprus and 
Prof. Dr. Luca Zaccarian from the LAAS-CNRS, University of Toulouse, CNRS for their valuable suggestions and 
fruitful discussions in the research. We also thank Mr. Su Zhang and Mr. Zeyu Wang for their 
contributions in numerical simulations. 
\end{ack}
 
\bibliographystyle{elsarticle-harv1} 
\bibliography{autosam}

\end{document}